\documentclass[useAMS,usenatbib,manuscript]{mn2e}

\usepackage{amssymb,amsmath,amsfonts}
\usepackage{graphicx}
\usepackage{epsfig,color} 
\usepackage{natbib}
\usepackage{lscape}
\usepackage[section] {placeins}

\title[Satellite abundances]{Satellite abundances around bright
  isolated galaxies}

\author[Wang et al.]{Wenting Wang$^{1,2,3}$, Simon D.M. White$^{2}$\\
  {}$^{1}$Key Laboratory for Research in Galaxies and Cosmology of Chinese Academy of Sciences, Max-Panck-Institute Partner\\
          Group, Shanghai Astronomical Observatory, Nandan Road 80, Shanghai 200030, China\\
  {}$^{2}$Max Planck Institut fur Astrophysik, Karl-Schwarzschild-Str. 1, 85741 Garching b. M\"unchen, Germany\\
  {}$^{3}$Graduate School of the Chinese Academy of Sciences, 19A, Yuquan Road, Beijing, China, 100080
}
\begin{document}

\date{Submitted to MNRAS}

\pagerange{\pageref{firstpage}--\pageref{lastpage}} \pubyear{2011}

\maketitle

\label{firstpage}

\begin{abstract}

We study satellite galaxy abundances by counting photometric galaxies
from the Eighth Data Release of the Sloan Digital Sky Survey (SDSS/DR8) 
around isolated bright primary galaxies from SDSS/DR7. We present results 
as a function of the luminosity, stellar
mass and colour of the satellites, and of the stellar mass and colour
of the primaries. For massive primaries ($\log M_\star/M_\odot >
11.1$) the luminosity and stellar mass functions of satellites with
$\log M_\star/M_\odot > 8$ are similar in shape to those of field
galaxies, but for lower mass primaries they are significantly steeper,
even accounting for exclusion effects due to the isolation criteria.
The steepening is particularly marked for the stellar mass function.
Satellite abundance increases strongly with primary stellar mass,
approximately in proportion to expected dark halo mass. For $\log
M_\star/M_\odot > 10.8$, red primaries have more satellites than blue
ones of the same stellar mass. The effect exceeds a factor of two at
$\log M_\star/M_\odot \sim 11.2$. Satellite galaxies are
systematically redder than field galaxies of the same stellar mass,
except around primaries with $\log M_\star/M_\odot <10.8$ where their
colours are similar or even bluer.  Satellites are also systematically
redder around more massive primaries. At fixed primary mass, they are
redder around red primaries. We select similarly isolated galaxies
from mock catalogues based on the galaxy formation simulations of
Guo et al. and analyze them in parallel with the SDSS
data. The simulation reproduces all the above trends qualitatively,
except for the steepening of the satellite luminosity and stellar mass
functions with decreasing primary mass. Model satellites, however, are
systematically redder than in the SDSS, particularly at low mass and
around low-mass primaries. Simulated haloes of a given mass have
satellite abundances that are independent of central galaxy colour,
but red centrals tend to have lower stellar masses, reflecting earlier
quenching of star formation by feedback. This explains the correlation
between satellite abundance and primary colour in the simulation. The
correlation between satellite colour and primary colour arises because
red centrals live in haloes which are more massive, older and more
gas-rich, so that satellite quenching is more efficient.

\end{abstract}


\section{Introduction}

Clustering studies provide insights into the formation and evolution
of galaxies that complement those coming from the joint distribution
of intrinsic properties -- mass, size, morphology, gas content,
star-formation rate, nuclear activity, characteristic velocity and
metallicity. In particular, clustering studies connect galaxies to
their unseen dark matter haloes and indicate how these were
assembled. According to the current standard
$\mathrm{\Lambda}\text{CDM}$ paradigm, galaxies form as gas cools and
condenses at the centres of a hierarchically aggregating population of
dark matter haloes, as originally outlined by
\cite{1978MNRAS.183..341W}.  As smaller haloes fall into more massive
ones, their central galaxies become satellites of these new hosts,
occasionally merging at some later time into the new central galaxies
in their cores.  Thus, each halo contains a dominant galaxy at the
bottom of its potential well, and a set of satellites which were the
central galaxies of smaller progenitors.  Observational studies of
satellite populations provide a check on this picture, indicating how
central galaxy properties relate to halo mass, and how these
properties are modified when a halo falls into a bigger system.

The abundance of satellites, their spatial distribution and their
intrinsic properties are thus intimately bound up with halo merger
histories, which are themselves closely related to the underlying
cosmology. For example, the evolution of merger rates is sensitive to
the cosmic matter density, while the mass distribution of merging
objects depends on the linear power spectrum of initial density
fluctuations \citep{1993MNRAS.262..627L}. Thus, satellite properties
can, in principle, be used to constrain cosmological parameters. On
the other hand, the physical processes driving galaxy evolution have
strong effects on satellites. For example, their colours are affected
by stripping of the gas reservoirs which supply star formation, and
both gravitational and hydrodynamical processes can modify their
structure, changing their morphology and partially or even totally
disrupting them. Modern evolutionary models for the galaxy population
attempt to include such processes and can be tested by comparison with
the abundances, colours and spatial distributions of satellites.

The ``missing satellite problem'' is a particularly striking example
of how satellite galaxy studies can constrain cosmology and galaxy
evolution. The problem highlights an apparent mismatch between the large
number of self-bound subhaloes found in $\mathrm{\Lambda}\text{CDM}$
simulations of the formation of haloes like those of the Milky Way and
M~31, and the much smaller number of satellite galaxies observed
around these two Local Group galaxies
\citep[]{1999ApJ...522...82K,1999ApJ...524L..19M,
  2004ApJ...609..482K}. This discrepancy has traditionally been
addressed by claiming that photoionisation and supernova feedback
suppress cooling and star formation in low-mass haloes, so that only
the most massive Milky Way subhaloes were able to make stars, the rest
remaining dark \citep[e.g.][]{1993MNRAS.264..201K,
  2000ApJ...539..517B, 2002MNRAS.333..177B, 2002ApJ...572L..23S,
  2010MNRAS.402.1995M, 2011MNRAS.413..101G}. Recent analyses of the
kinematics of Galactic satellites suggest, however, that their dark
matter haloes are not dense enough to correspond to the most massive
subhaloes in a $\mathrm{\Lambda}\text{CDM}$ universe
\citep{2012MNRAS.422.1203B, 2011arXiv1111.6609F}. Such problems have
led a number of authors to invoke warm dark matter (WDM) to eliminate
subhaloes less massive than a few billion solar masses, and to reduce
the central density of subhaloes above this cut-off
\citep[e.g.][]{2000ApJ...535L..21M, 2000PhRvL..84.3760S,
  2000ApJ...544L..87Y,2001ApJ...556...93B,2009ApJ...700.1779Z,
  2012MNRAS.420.2318L}. Others claim that part of the problem may be
incompleteness of the observed satellite population
\citep[e.g.,][]{2004MNRAS.353..639W,2007ApJ...670..313S,2008ApJ...686..279K,
  2008ApJ...688..277T,2009AJ....137..450W}.

Two kinds of methods are now commonly used to compare galaxy clustering in
large redshift surveys to the predictions of high-resolution simulations of
cosmic structure formation. The Halo Occupation Distribution \citep[HOD;
  e.g.,][]{ 1998ApJ...494....1J, 1998ApJ...503...37J, 2000MNRAS.318.1144P,
  2000ApJ...543..503M, 2000MNRAS.318..203S, 2002ApJ...575..587B,
  2002PhR...372....1C, 2005ApJ...633..791Z} and the closely related
Conditional Luminosity Function \citep[CLF;][]{2003MNRAS.339.1057Y} approaches
determine the central and satellite galaxy populations of haloes as a function
of mass by optimizing the fit to abundance and clustering observations,
typically luminosity and correlation functions. In a non-parametric variant,
the observed abundance of central galaxies is matched directly to the
simulated abundance of haloes to obtain a monotonic relation between galaxy
luminosity and halo mass \citep[Abundance Matching,
  AM:][]{2004ApJ...614..533T, 2004MNRAS.353..189V, 2006ApJ...647..201C,
  2010ApJ...710..903M, 2010MNRAS.404.1111G}. This relation can be used to
populate both main and satellite subhaloes, if it is assumed to hold when
satellites first fall into more massive systems. By construction, these
methods fit observed luminosity functions perfectly. The same is true for
observed correlation functions for HOD and CLF, whereas these serve as a test
for AM. None of these schemes explains {\it why} haloes of given mass have
central galaxies with specific properties.

In contrast, semi-analytic models use simplified representations of the
relevant astrophysics to follow galaxy growth within the evolving
dark halo population, and so attempt to predict the detailed
properties of both central and satellite galaxies \citep[SAM;
  e.g.,][]{1991ApJ...379...52W, 1993MNRAS.264..201K, 1994MNRAS.271..781C,
  1999MNRAS.303..188K, 1999MNRAS.310.1087S, 2001MNRAS.328..726S,
  2005ApJ...631...21K}. Here, adjustable parameters correspond to the
efficiencies of poorly understood processes like star formation or AGN
feedback, so that the values derived from fitting to observation are
interesting in their own right. In recent years, ever more detailed
astrophysical models have been incorporated into ever larger and higher
resolution simulations of dark matter evolution, leading to increasingly
faithful representation of the observed galaxy population
\citep[e.g.][]{2005Natur.435..629S,2006MNRAS.365...11C,
  2006MNRAS.370..645B,2007MNRAS.375....2D,2011MNRAS.413..101G}.

Many studies of satellite galaxy populations have focused on the Local Group
(LG) because of the greater depth and detail with which nearby galaxies can be
studied (see \cite{2000ESASP.445...87G, 2001ApSSS.277..231G,
  2007ggnu.conf....3G,2011EAS....48..315G} and references therein).  Recent
work has been particularly concerned with comparing
$\mathrm{\Lambda}\text{CDM}$ predictions to the abundance and internal
structure of dwarf spheroidal galaxies and to the orbital and internal
properties of the Magellanic Clouds \citep[e.g.][]{2009ApJ...696.2179K,
  2011ApJ...733...62L, 2011MNRAS.417.1260F,  2011ApJ...738..102T, 
2011MNRAS.418..648S,2011MNRAS.415L..40B}. Such studies are 
limited by the fact that the LG contains only two large spirals, since
considerable scatter is expected among the satellite populations of similar
mass haloes \citep[e.g.][]{2010MNRAS.406..896B,2011MNRAS.413..101G}.

Beyond the LG, many observational studies of satellite populations have
estimated their luminosity functions and their radial
distribution around their primaries, but rather few have compared directly
with theoretical expectations.  Some studies have used redshift surveys to
investigate the projected number density profiles of satellites
\citep[e.g.][]{1991ApJ...379L...1V, 2005MNRAS.356.1045S,2006ApJ...647...86C}
usually fitting power laws $\Sigma(r)\sim r^\alpha$, and obtaining slopes
$\alpha$ between $-0.5$ and $-1.2$. The availability of redshifts for all
galaxies facilitates discrimination between satellites and background objects,
but, for most objects, satellites are detectable only one or two magnitudes
fainter than their primaries. An important application made possible by the
redshifts is the measurement of mean dynamical masses for haloes as a function
of central galaxy luminosity and colour \citep{1993ApJ...405..464Z,
  1997ApJ...478...39Z, 2003ApJ...598..260P,
  2005ApJ...635..982C,2007ApJ...654..153C, 2011MNRAS.410..210M}. The last of
these papers finds that, at given luminosity, red central galaxies have more
massive haloes than blue ones, but that this difference goes away if red and
blue primaries are compared at the same stellar mass.  We will return to this
issue below.

The abundance of satellites at magnitudes much fainter than their primaries is
most easily studied by combining a redshift survey with a photometric survey
which catalogues galaxies several magnitudes below the spectroscopic limit. In
the absence of redshifts, it is not possible, of course, to distinguish true
satellites from background galaxies. Results can therefore be obtained only by
stacking large samples of primaries so that a statistical substraction of the
background population is possible \citep{1969ArA.....5..305H,
  1987MNRAS.229..621P, 1994MNRAS.269..696L, 2004ApJ...617.1017S, 2012ApJ...746..138T}.  In
particular, \cite{1994MNRAS.269..696L} counted the number of faint images on
Schmidt survey plates around primaries of known redshift, using a
``bootstrap'' method to remove the background and fitting the projected
surface density to a power law $\Sigma(r_p)\sim r_p^{-\alpha}$, finding
$\alpha\sim 0.9$.  They showed that satellites are more abundant and are
concentrated to smaller radii around early-type primaries than around
late-types.

Most recent work has taken advantage of the enormous increase in data provided
by the Sloan Digital Sky Survey  \citep[SDSS;][]{2000AJ....120.1579Y}. \cite{2006MNRAS.372.1161W} used the
group catalogue of \cite{2007ApJ...671..153Y}, constructed from the SDSS
spectroscopic data, to study in detail how the properties of satellite
galaxies depend on the colour, luminosity, and morphology of the central
galaxy and on their inferred dark halo mass. They compared their observational
results with a mock redshift survey based on the SAM of
\cite{2006MNRAS.365...11C}, finding significant discrepancies. In particular,
the model overpredicted the number of faint satellites in massive haloes and
produced too many red satellites.  The fraction of blue central galaxies was
also too high at high luminosities.  \cite{2006MNRAS.372.1161W} argued that
the satellite problems most likely reflect an improper treatment of tidal
stripping or of the truncation of star formation, while the central problem
may reflect an overly simple treatment of dust or of AGN feedback. In
\cite{2011MNRAS.416.1197W} a mock catalog based on the more recent model of
\cite{2011MNRAS.413..101G} was compared with several nearby galaxy clusters as
well as with the group catalogue of \cite{2007ApJ...671..153Y}.  Discrepancies
were weaker than for the earlier model, but the predicted fraction of red
dwarf satellites remains higher than in the Virgo cluster or in the group
catalogue of \cite{2007ApJ...671..153Y}, although agreeing with the fractions
found in the Coma and Perseus clusters.

Studies of satellite galaxies based on both spectroscopic and
photometric data from the SDSS have been published recently by
\cite{2011AJ....142...13L}, by \cite{2011MNRAS.417..370G} and by
\cite{2012ApJ...746..138T,2012ApJ...751L...5T}. The first of these
investigated how the luminosity function and number density profile of
satellites depend on the colour and luminosity of their central
galaxy, finding the abundance of satellites to depend strongly on
primary luminosity, and the faint-end slope of their luminosity
function to be consistent with that of the field.  Using similar
datasets, \cite{2011MNRAS.417..370G} investigated the satellite
luminosity function and its dependence on primary luminosity, colour
and concentration.  Their satellite luminosity function estimates are
not well fit by Schechter functions, tending to be flat at bright
luminosities but very steep at faint luminosities, apparently at odds
with the conclusions of \cite{2011AJ....142...13L}. For the primary
magnitude range ($M_V=-21.25\pm 0.5$) the mean luminosity function of
\cite{2011MNRAS.417..370G}is similar in shape to that of the MW and
M31, but the abundance of satellites is about a factor two
lower. \cite{2012ApJ...746..138T} studied satellites of SDSS Luminous
Red Galaxies using, in particular, the deep Stripe 82 data finding a
luminosity function with a shallow faint end slope and a very
different shape from those of \cite{2011MNRAS.417..370G}. 
\cite{2012ApJ...751L...5T}  constructed radial number density
profiles for these same systems, concluding that they are well fitted 
by a NFW model \citep{1996ApJ...462..563N,1997ApJ...490..493N} on large scales 
while at small radii there is an excess  of satellites compared with the NFW profile, 
which can be well dscribed by a Sersic model.  Using data from the Galaxy and
Mass Assembly Survey \citep[GAMA;][]{2009A&G....50e..12D,
  2011MNRAS.413..971D}, \cite{2011MNRAS.417.1374P} also studied
satellite number density profiles and red fractions as functions of
projected separation and the masses of both satellite and primary,
arguing that their results favour removal of gas reservoirs as the
main mechanism quenching star formation in satellites. Finally,
\cite{2012ApJ...752...99N} use HST data from the Cosmological Evolution Survey (COSMOS) to
study similar issues for smaller samples of galaxies, but out to
$z\sim0.8$.

In the present paper we return to many of these questions, using the full SDSS
spectroscopic and photometric databases in conjunction with the galaxy
population simulations of \citet[][hereafter G11]{2011MNRAS.413..101G}. The
simulations allow us to compare expectations based on our current
understanding of galaxy formation in a $\Lambda$CDM universe with the observed
dependences of satellite luminosity, stellar mass and colour on primary galaxy
properties.  By using mock catalogues from the simulations, we are able to
explore how satellite populations relate to the dark matter haloes in which
they are embedded, to gain insight into the effect of physical processes
like quenching and tidal stripping on their observable properties, and to explore
how the observational criteria defining isolated primary galaxies impact the
clustering of other galaxies around them (see \cite{1976ApJ...205L.121F} for an
old example of the potential strength of such effects).

We describe the datasets we use and the selection criteria which
define our primary and satellite galaxy samples in
section~\ref{sec:data}.  In section~\ref{sec:method} we introduce our
background subtraction method.  We present our SDSS results and
compare them directly with the G11 simulation in
sections~\ref{sec:LFMF} and~\ref{sec:colour}. Further discussion and
comparison with previous work is given in our concluding section. An
appendix describes a variety of tests for systematics in the SDSS
photometric data and in the techniques we use to correct satellite
counts for contamination by foreground and background
galaxies. Throughout this paper, we convert observational to intrinsic
properties assuming a cosmology with $\Omega_m=0.25$,
$\Omega_\Lambda=0.75$ and $h=0.73$. We quote all masses in units of
$M_\odot$ rather than $h^{-1}M_\odot$.

\section{Data and Sample Selection}
\label{sec:data}
\subsection{Primary Selection}
\label{subsec:select} 
We wish to study the satellite populations of bright isolated galaxies out to
distances $\sim 0.5$~Mpc.  We begin by considering all galaxies brighter than
$r=16.6$ ($r$-band extinction corrected Petrosian magnitude) in the
spectroscopic galaxy catalogue of the New York University Value Added Galaxy
Catalog (NYU-VAGC)\footnote{http://sdss.physics.nyu.edu/vagc/} \citep{2011ApJS..193...29A}. 
This catalogue was built by \cite{2005AJ....129.2562B} on the basis of the 
seventh Data Release of the Sloan Digital Sky Survey
\citep[SDSS/DR7;][]{2009ApJS..182..543A}. This apparent magnitude limit provides
us with a parent catalogue of 145070 objects.  We select isolated galaxies
from this sample by requiring that there should be no companion in the
spectroscopic sample at $r_p<0.5$~Mpc and $|\Delta{z}|<1000$~km/s that is less
than a magnitude fainter in $r$ than the central object, and no companion at
$r_p<1$~Mpc and $|\Delta{z}|<1000$~km/s that is brighter than it. These
isolation criteria reduce our sample to 66285 objects.

The SDSS spectroscopic sample is incomplete, because observing efficiency
constraints made it impossible to put a fibre on all candidates or to
re-observe objects where an initial spectrum yielded an unreliable
redshift. The completeness varies with position on the sky and has a mean of
$91.5\%$ for our parent sample. Thus $\sim 10\%$ of our ``isolated'' galaxies
will not, in fact, be isolated according to our criteria, because their
companion was missed by the spectroscopic survey. To eliminate such systems we
apply an additional cut using the SDSS photometric data.  The photometric
redshift 2 catalogue \citep[photoz2;][]{2009MNRAS.396.2379C} on the SDSS
website provides redshift probability distributions for all galaxies in the
SDSS footprint down to apparent magnitude limits much fainter than we require.
These distributions are tabulated for 100 redshift bins, centered from
$z_1=0.03$ to $z_2=1.47$ with spacing $dz=1.44/99$. We find all the objects in
our candidate isolated galaxy list which have a companion in the photoz2
catalogue satisfying the above projected separation and magnitude difference
criteria, and we discard those where the companion has a photometrically
estimated redshift distribution compatible with the spectroscopic redshift of
the primary.  Our definition of ``compatible'' is that the probability for the
companion to have a redshift equal to or less than that of the primary exceeds
0.1. Apparent companions which fail this test usually do so because their
colours are too red to be consistent with a redshift as low as that of the
primary. After applying this additional cut, 41883 objects remain in our
isolated galaxy sample.

Finally, we exclude any object for which more than $20\%$ of a surrounding
disc of radius $r_p=1$Mpc lies outside the survey footprint.  Such objects
could have bright companions which are not included in the SDSS databases. To
evaluate these overlaps we made use of the set of ``spherical polygons''
provided on the NYU-VAGC website. These account both for the survey boundary
and for masked areas around bright stars. We generate a large number of points
uniformly and randomly over the 1~Mpc disc surrounding each galaxy and discard
any which lie outside the .survey boundary. A galaxy is eliminated from the
sample if more than 20\% of its points are discarded in this way. This last
cut removes about 1.5\% of our objects, leaving a final sample of 41271 bright
isolated galaxies. The set of randomly generated points surrounding each of
these ``primaries'' is kept for later use when estimating background
corrections to the counts of its faint companions (see below).

Figure~\ref{fig:prop_dis} compares the distributions of colour,
Petrosian half-light radius $R_{50}$, concentration $C=R_{90}/R_{50}$
and stellar surface mass density $\mu_\star= M_\star/2\pi R_{50}^2$
for our parent galaxy sample (all 145070 galaxies with $r<16.6$) and
for our 41271 isolated primaries, separated into bins of galaxy stellar 
mass.  These quantities were taken directly from the NYU-VAGC 
catalogue. The stellar masses were estimated by fitting stellar population 
synthesis models to the K-corrected galaxy colours assuming a 
\cite{2003PASP..115..763C} initial mass function as in 
\cite{2007AJ....133..734B}. The sensitivity of the stellar masses to 
assumptions underlying the estimation technique is explored in the 
Appendix of \cite{2009MNRAS.398.2177L}.
Each stellar mass bin is a factor of two wide, and we show data for
the five mass bins on which we will concentrate our analysis
throughout the rest of this paper. The numbers in the lower right of
the $R_{50}$ plots indicate the number of isolated primaries in each
mass bin.  Volume corrections have been applied to all the histograms
in this figure by calculating the total volume $V_{\rm max}$ of the
survey over which each individual galaxy would be brighter than the
flux limit, $r=16.6$, and accumulating counts weighted by $1/V_{\rm
  max}$. It turns out that our isolated primaries are slightly bluer
than the parent sample, particularly at low masses.  In addition, they
are slightly more concentrated than the parent sample.  Our selection
procedure appears to have no significant effect on the distributions
of the other two properties, and the same is true for the redshift
distributions which we do not show. To a very good approximation our
isolated primary galaxies are typical objects of their stellar mass,
although as we will see below, our selection has a strong influence on
their relation to their environment.  Vertical dashed lines in the
colour plots indicate the split we adopt when separating our primaries
into red and blue populations.  This split is slightly dependent on
stellar mass.

\begin{figure*}
\centerline{             \epsfig{figure=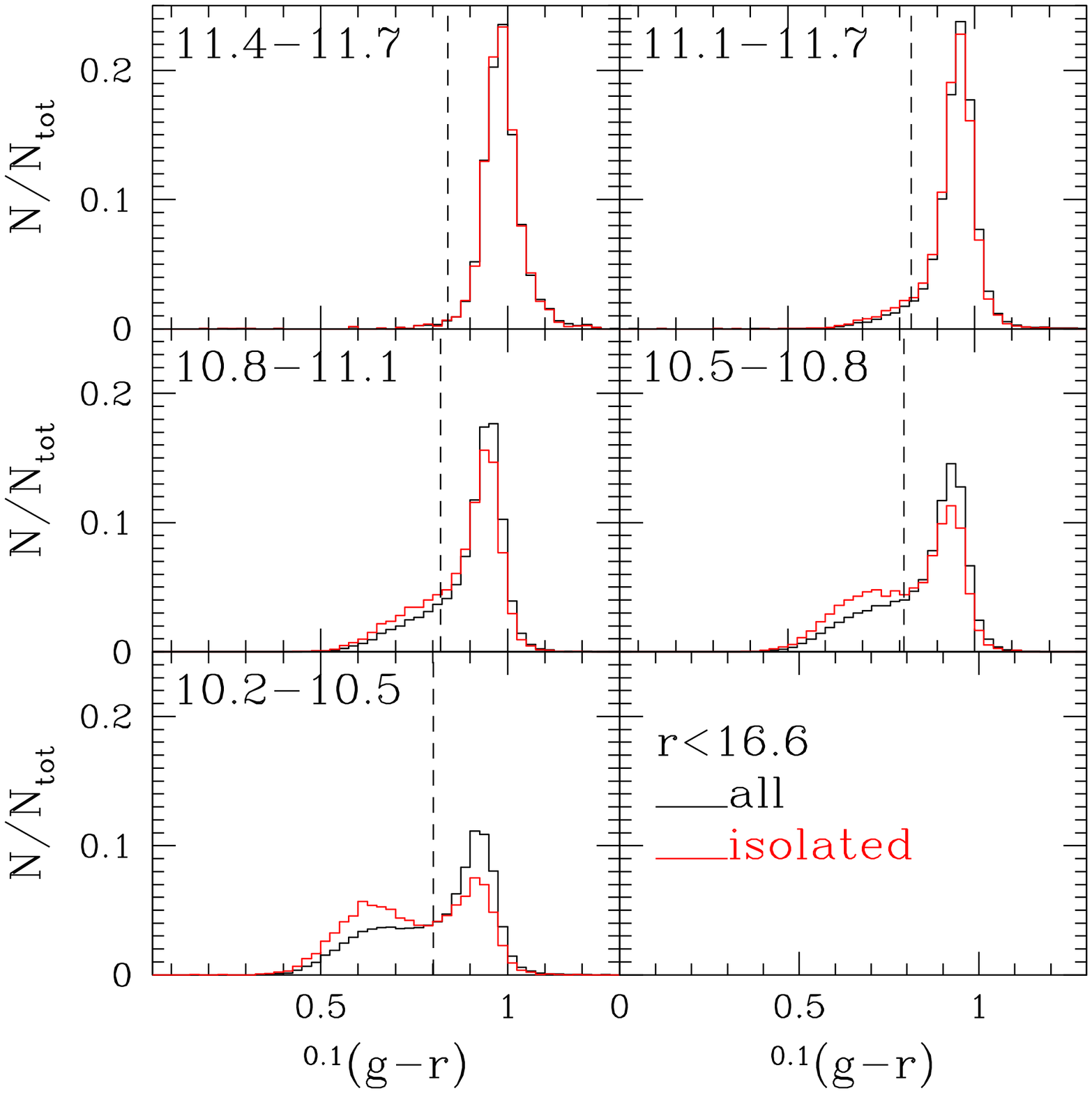,width=0.45\textwidth}
  \epsfig{figure=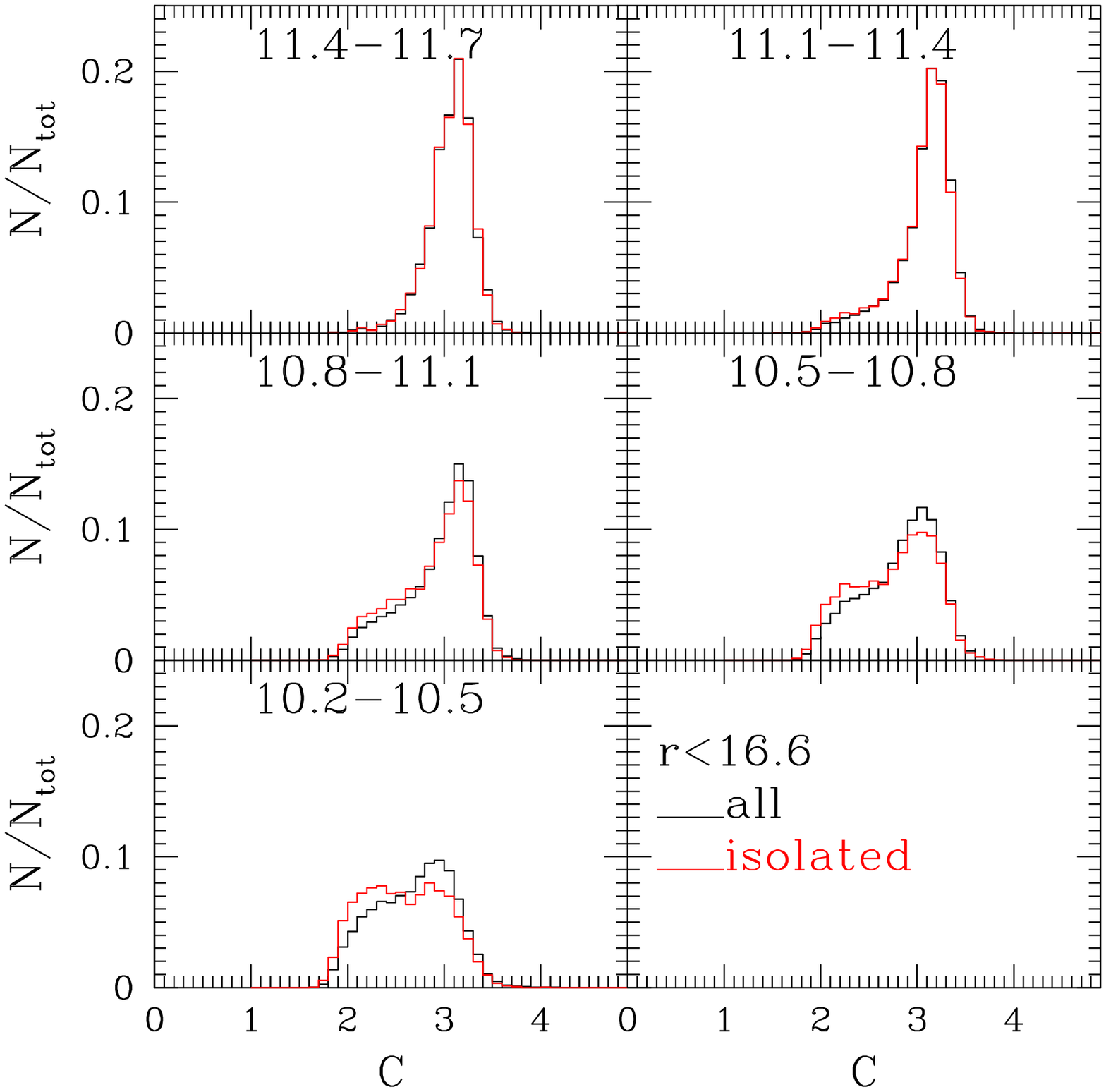,width=0.45\textwidth}}          \centerline{
  \epsfig{figure=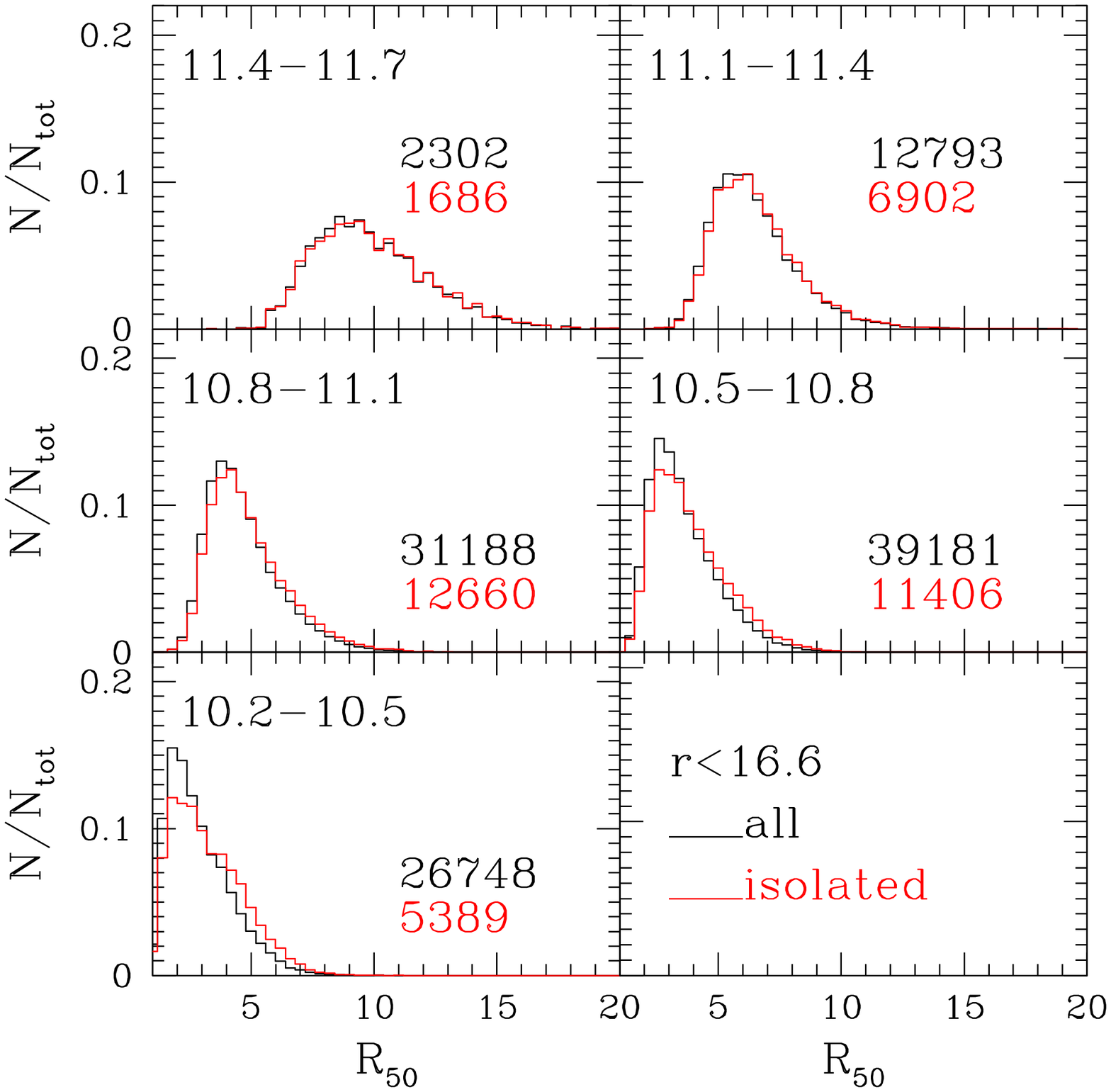,width=0.45\textwidth}
  \epsfig{figure=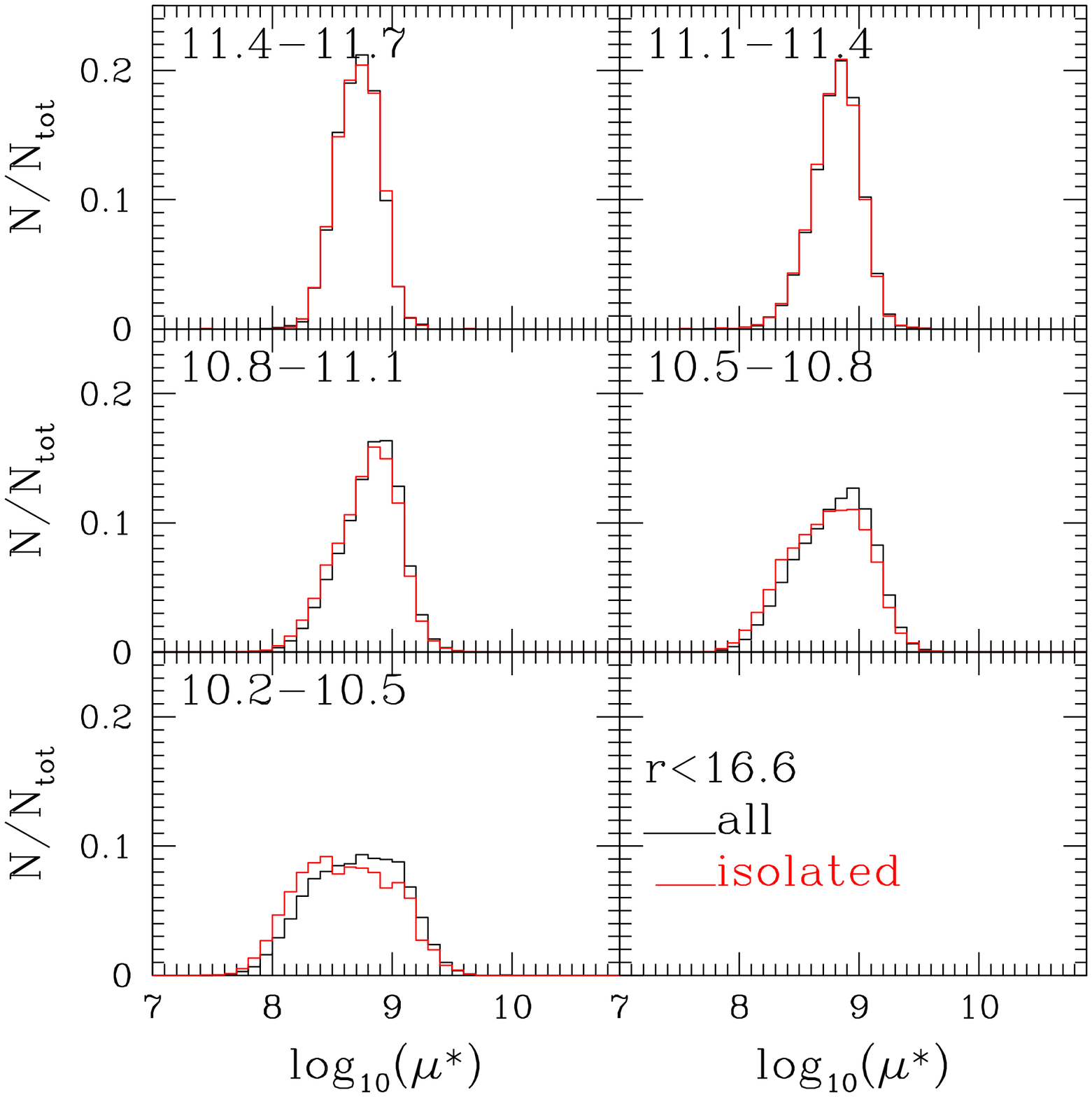,width=0.45\textwidth} }
\caption{Volume weighted distributions for the parent sample of SDSS/DR7
  galaxies with $r<16.6$ (black curves) and for our sample of isolated
  galaxies (red curves). Different panels within each set refer to different
  ranges of $\log M_\star/M_\odot$ as labelled.  Each set corresponds to a
  different property. {\bf Top left:} $^{0.1}(g-r)$ colour -- the vertical
  dashed line in each panel shows the colour separating red and blue
  populations. {\bf Top right:} Concentration $C=R_{90}/R_{50}$.  {\bf Bottom
    left:} Petrosian half-light radius $R_{50}$ (in kpc) -- the numbers at
  bottom right of each panel indicate the numbers of galaxies in the parent
  (black) and isolated samples (red) shown. {\bf Bottom right:}
  Stellar surface mass density $\mu_\star= M_\star/2\pi R_{50}^2$.
\label{fig:prop_dis}
}
\end{figure*}

\subsection{The mock catalogue}

In the analysis and interpretation of our results for satellite galaxy
populations we will make considerable use of mock catalogues built from the
galaxy formation simulations of \citet[][hereafter
  G11]{2011MNRAS.413..101G}. These are implemented on two very large dark
matter simulations, the Millennium Simulation \citep[MS;][]{2005Natur.435..629S} and
the Millennium-II Simulation \citep[MS-II;][]{2009MNRAS.398.1150B}. The MS follows the
evolution of structure within a cube of side $500h^{-1}\rm{Mpc}$ (comoving)
and its merger trees are complete for subhaloes above a mass resolution limit
of $1.7\times10^{10}h^{-1}\rm{M}_{\sun}$. The MS-II follows a cube of side
$100h^{-1}\rm{Mpc}$ but with 125 times better mass resolution (subhalo masses
greater than $1.4\times10^{8}h^{-1}\rm{M}_{\sun}$).  Both adopt the same
WMAP1-based $\Lambda$CDM cosmology \citep{2003ApJS..148..175S} with parameters
$h=0.73, \Omega_m=0.25, \Omega_\Lambda=0.75, n=1$ and $\sigma_8=0.9$. These
are outside the region preferred by more recent analyses (in particular,
$\sigma_8$ appears too high) but this is of no consequence for the issues we
study in this paper. For consistency, we will adopt this cosmology
when quoting numbers in the rest of this paper. 

In G11's galaxy formation model, the uncertain star formation and feedback
efficiencies are tuned to produce close fits to the stellar mass, luminosity
and autocorrelation functions of low redshift galaxies as inferred from the
SDSS.  Together with their high resolution (particularly for the MS-II) and
large volume (for the MS) this makes them ideal for our purposes in this
paper. Here we use the publicly available data from
http://www.mpa-garching.mpg.de/millennium.  We project the simulation boxes in
three orthogonal directions parallel to their $x$, $y$ and $z$ axes.  In each
projection we can assign each galaxy a redshift based on its ``line-of-sight''
distance and peculiar velocity, and we can select isolated primaries using
criteria which are directly analogous to those used for the SDSS (though we do
not need to worry about the complications due to completeness and boundary
issues).  In addition to observables like luminosities, colours, sizes and
morphologies, the simulation databases provide information which is not
directly accessible for real galaxies (e.g. halo mass, environment type and
full 3D position and peculiar velocity).  These can give insight into the
nature of the isolated primary sample we have selected.  

All the SDSS luminosities and colours we use in this paper are rest-frame
quantities K-corrected to the $^{0.1}r$ band.  The absolute magnitudes in our
mock catalogues are in the true $z=0$ $r$ band, because this is what is
directly provided by the database and the difference between the $^{0.1}r$
and $r$ bands is too small to significantly affect luminosities. We do,
however, transform the database $(g-r)$ colours to the $^{0.1}(g-r)$ band
using the empirical fitting formula of \cite{2007AJ....133..734B} because here
the shifts seem large enough to cause (minor) differences.

Structure in the MS and MS-II is characterized using Friends-of
Friends (FoF) groups partitioned into sets of disjoint self-bound
subhaloes. The subhalo populations at neighboring output times are
linked to build merger trees which record the assembly history of all
nonlinear structures and provide the framework for the galaxy
formation simulations. In these simulations galaxy evolution is
affected by environment in several ways. The galaxy at the centre of
the most massive subhalo of each FoF group (which usually contains
most of its mass) is considered the ``central galaxy'' and is the only
one to accrete material from the diffuse gas associated with the
group. When evolution joins two FoF groups, G11 continue to treat the
galaxy at the centre of the less massive subhalo as a central galaxy
until it falls within the nominal virial radius\footnote{We define
  this as $r_{200}$ the radius of the sphere centred on the
  gravitational potential minimum of the FoF group within which the
  mean density is 200 times the critical value. $M_{200}$ is then the
  mass within this sphere.} of the new FoF group. After this point the
infalling galaxy is considered as a ``satellite'', the mass of its
dark halo starts dropping as a result of tidal stripping, and its
diffuse gas is assumed to be removed in proportion to the subhalo dark
matter. Such satellites may later lose their subhaloes entirely
through tidal disruption.  At this point they are either disrupted
themselves or (more commonly) they become ``orphan satellites'' which
continue to orbit until dynamical friction causes a merger with their
central galaxy. In this paper we will follow G11, considering together
the two kinds of satellites (with and without a dark matter subhalo)
and the two kinds of centrals (in dominant and in newly accreted,
distant subhaloes).

\begin{figure*}
\centerline{             \epsfig{figure=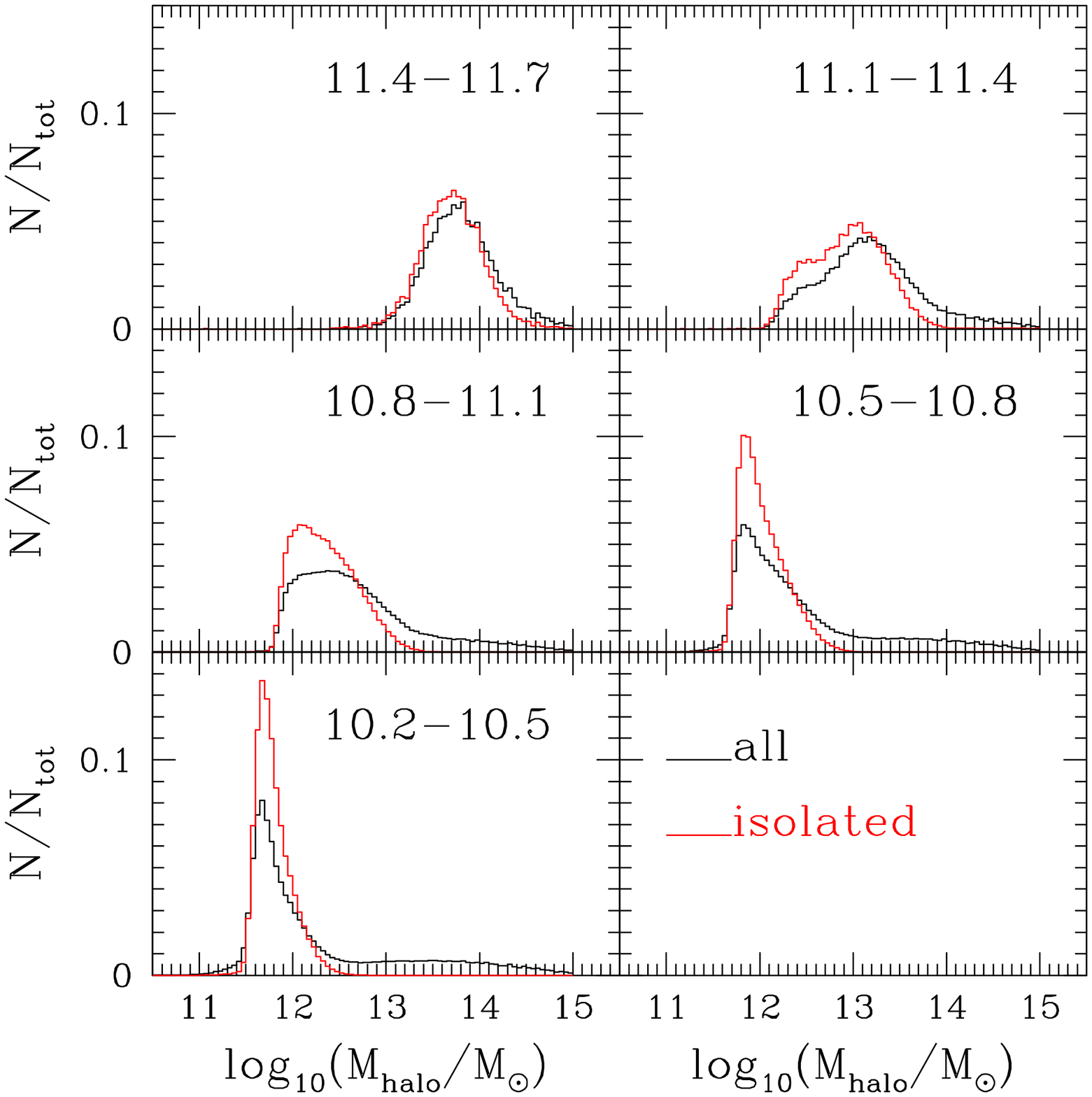,width=0.45\textwidth}
  \epsfig{figure=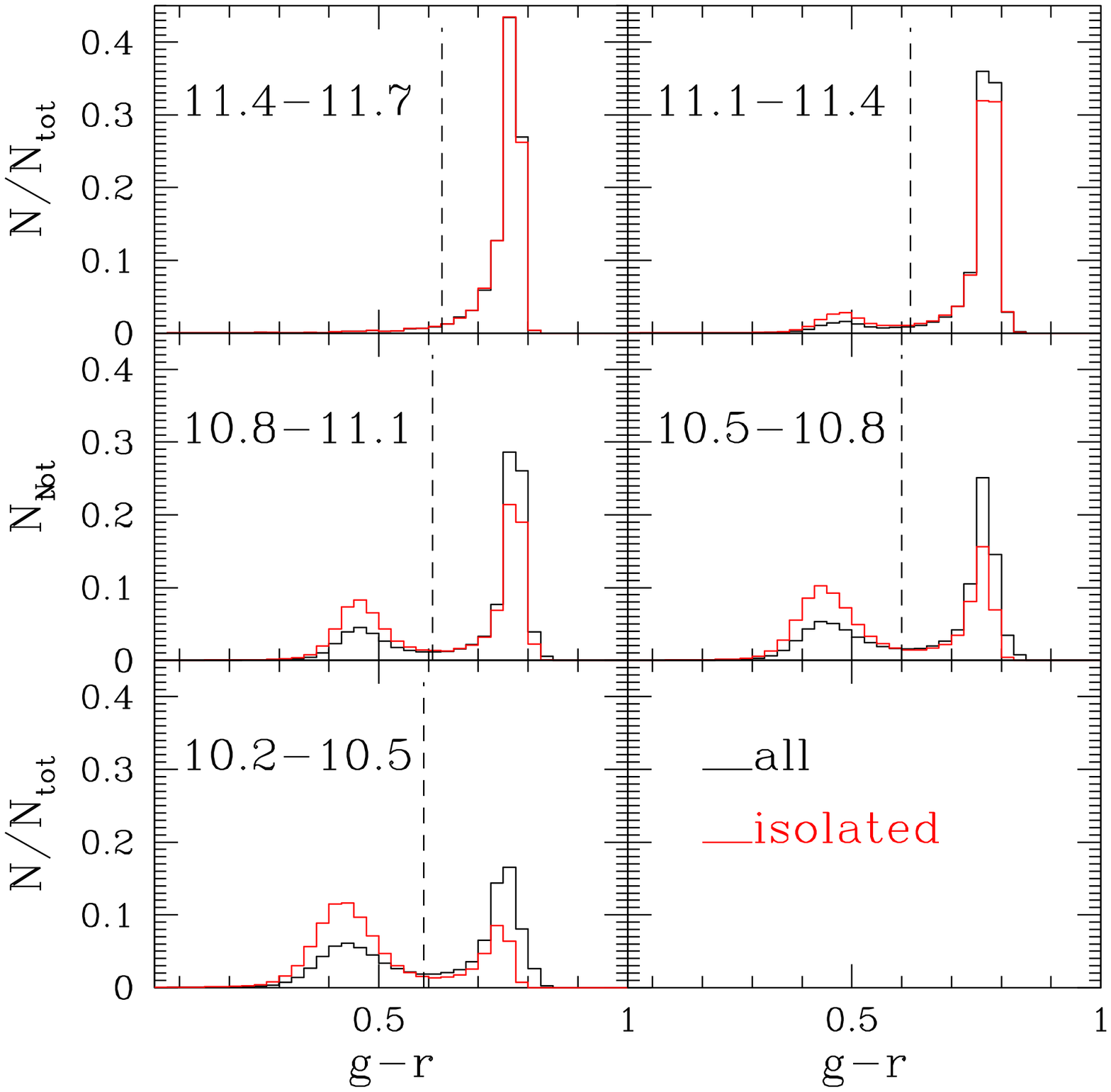,width=0.45\textwidth} }
\caption{{\bf Left:} Halo mass distributions for isolated galaxies (red
  curves) in five disjoint stellar mass bins and for the parent populations in
  the simulations of G11 from which they were drawn (black curves). {\bf
    Right:} Colour distributions for these same sets of simulated
  galaxies. Vertical dashed lines indicate the colour at which we separate red
  and blue populations.}
\label{fig:prop_dis2}
\end{figure*}

The left panels of figure~\ref{fig:prop_dis2} compare halo mass distributions
as a function of stellar mass for isolated simulation galaxies to those of
their parent population. Here, halo mass is $M_{200}$ of the FoF group for the
dominant central galaxy and all the satellites, and $M_{\rm inf}$ for the
other ``centrals'', where $M_{\rm inf}$ is the $M_{200}$ of the old FoF group
of a newly accreted central just prior to infall. Our isolation criteria have
a substantial effect on these mass distributions, eliminating a high-mass tail
which is particularly evident for lower stellar mass galaxies. This tail is
due to the satellites which, as we will show more explicitly below, are very
effectively excluded by our criteria.  

The right panels of figure~\ref{fig:prop_dis2} compare colour
distributions for these same simulated galaxy populations. Just as for
the real SDSS galaxies (figure~\ref{fig:prop_dis}) our isolation
criteria bias the distributions to bluer galaxies because central
galaxies have ongoing gas accretion and so are more actively
star-forming than satellites. Selection induces larger shifts for the
simulated distributions than for the real ones, however, and the
bimodal nature of the colour distributions is more obvious in the
simulation.  As already discussed by G11 and
\citet{2011MNRAS.416.1197W} this reflects the facts that the red and
blue sequences are more sharply defined in the simulation than in
reality and that satellite galaxies appear to be too uniformly
red. Note also that the red and blue populations separate at bluer
colours (indicated by dashed vertical lines) in the simulation than in
the SDSS data, particularly at high mass. This appears to be a
consequence of the stellar population synthesis models used in the
simulation, together with the fact that stellar metallicities are too
low for high-mass galaxies \citep[see][]{2010MNRAS.403..768H}.

\begin{figure*}
\centerline{             \epsfig{figure=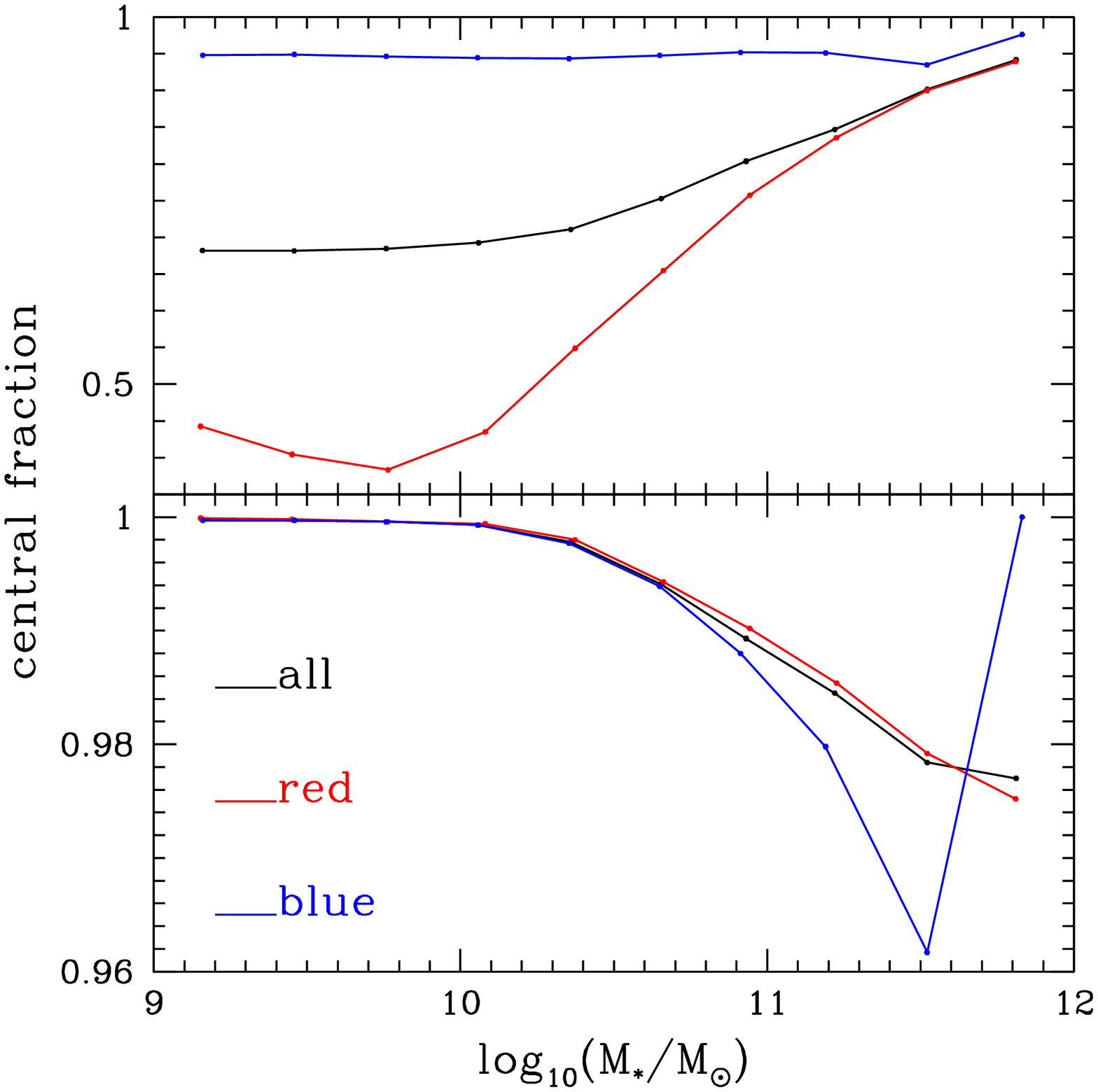,width=0.45\textwidth}
  \epsfig{figure=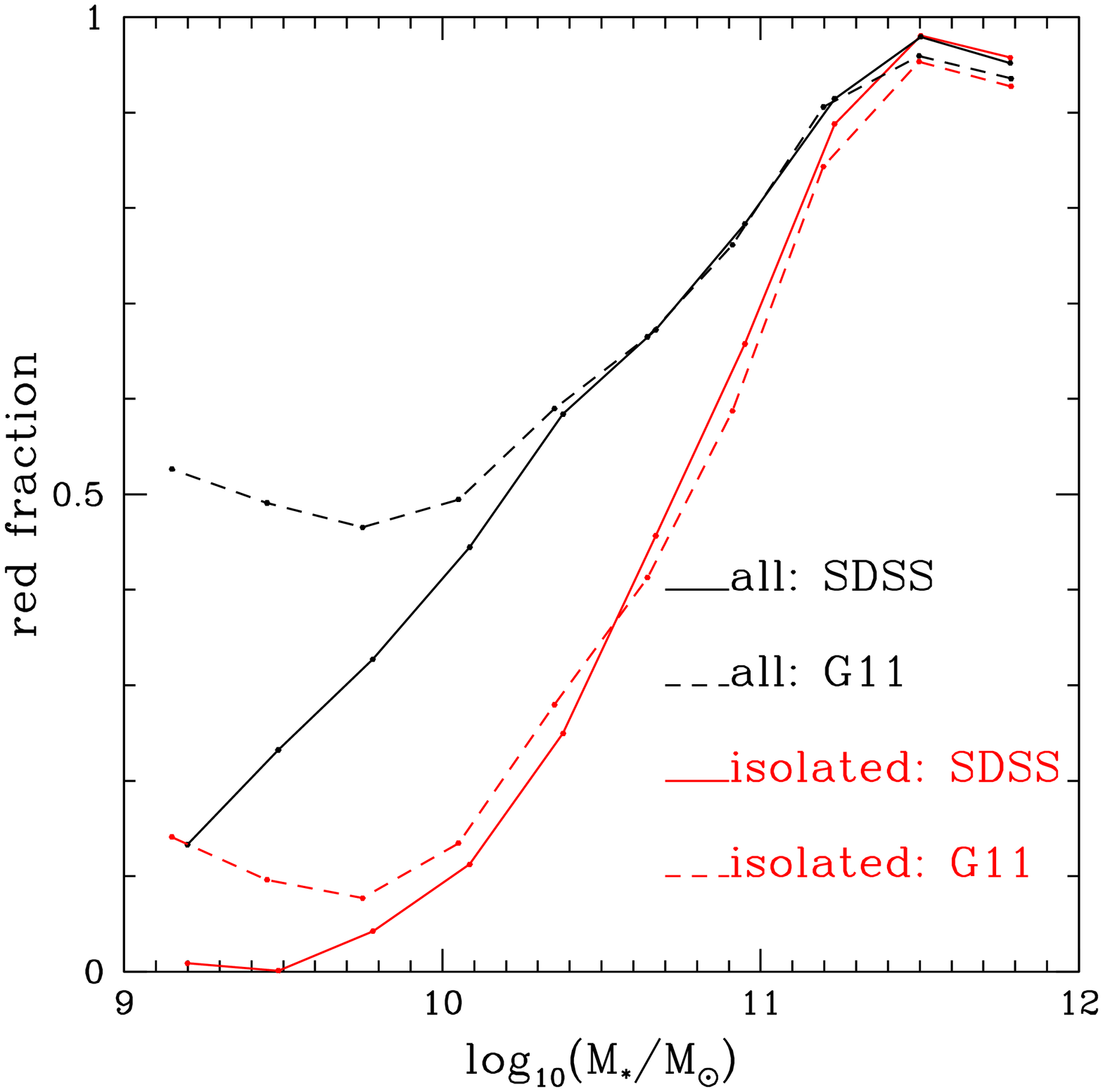,width=0.45\textwidth} }
\caption{{\bf Left:} The fraction of isolated galaxies (bottom panel) and of
  their parent sample of all galaxies (top panel) in the galaxy formation
  simulation of G11 which are classified as centrals rather than satellites
  according to the G11 criteria which we also adopt here (see the text for
  details). Black, red and blue lines indicate the fractions as a function of
  stellar mass for all, for red and for blue galaxies respectively.  {\bf
    Right:} The fractions of galaxies which are red (rather than blue) are
  shown as a function of stellar mass for isolated galaxies (red curves) and
  for the parent sample of all galaxies (black curves). The solid curves in
  each case represent observed results for the SDSS (after weighting to
  represent values for volume-limited samples) while the dashed curves are for
  the galaxies in the simulation of G11, Note the excellent agreement for
  $\log M_\star/M_\odot > 10.2$ the range of interest for this paper.}
\label{fig:frac}
\end{figure*}

The left panels of figure~\ref{fig:frac} illustrate how well our
isolation criteria select central galaxies, at least in the
simulation.  The fraction of centrals according to the definitions of
G11 is plotted as a function of stellar mass in the lower panel, with
the black curve referring to all isolated galaxies and the red and
blue curves referring to the red and blue subpopulations. For the
isolated population as a whole and for its red subpopulation, the
contamination by satellites {\it maximizes} at just over 2\%. Slightly
larger contamination occurs at high mass in the blue subpopulation,
but such massive blue galaxies are in any case very rare (see
below). In contrast, the curves in the upper panel show that only 65\%
of the parent population in the G11 model are centrals at $\log
M_\star/M_\odot = 10$, about 80\% are centrals at $\log
M_\star/M_\odot = 11$ and 88\% are centrals at $\log M_\star/M_\odot =
11.5$.  For blue galaxies the central fraction is above 90\% at all
masses, while for $\log M_\star/M_\odot < 10.3$ most red galaxies are
satellites. The application of our isolation criteria to the simulated
galaxy population of G11 thus results in an extremely pure sample of
central galaxies.

We cannot be sure, of course, that the elimination of satellite galaxies is as
effective in the SDSS samples as in the simulation. We can, however, check
that the separation into red and blue subpopulations matches as a function of
stellar mass, both for isolated galaxies and for their parent population.
This comparison is shown in the right panel of figure~\ref{fig:frac}. Here red
curves indicate the red fraction as a function of stellar mass for isolated
galaxies, while black curves indicate the same quantity for the parent
sample. In each case, solid curves are the observational result from SDSS and
dashed curves are for the simulation of G11. For the observations we have
again used $1/V_{\rm max}$ weighting to ensure that the plotted quantity is
appropriate for a volume-limited sample. For the primary stellar masses
relevant for this paper ( $\log M_\star/M_\odot > 10.2$) the agreement between
observation and simulation is almost perfect. At lower stellar masses there
are too many red galaxies in the simulation, again reflecting the problem
noted above and discussed in detail by \citet{2011MNRAS.416.1197W}: at low
masses the simulated satellite galaxies are too uniformly red.

\begin{figure}
\epsfig{figure=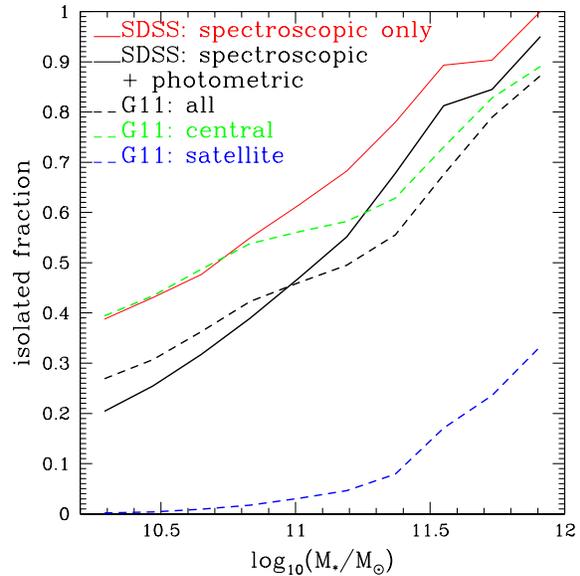,width=0.45\textwidth}
\caption{The fraction of galaxies that are selected by our isolation
  criteria as a function of stellar mass. The thick black solid line
  gives the result for our final sample of isolated SDSS galaxies,
  while the thick black dashed line is the corresponding result for
  simulated galaxies in the G11 model. These agree moderately well
  over the full mass range shown. The thin solid red line shows the
  fraction of retained SDSS galaxies after demanding isolation
  relative to the spectroscopic sample but before additionally
  requiring isolation relative to the photometric sample. Green and
  blue dashed curves indicate the fractions of central and satellite
  galaxies which remain after applying our isolation criteria to the
  G11 simulated galaxy sample.}
\label{fig:select}
\end{figure}

Another comparison of the effects of our isolation criteria on the
observed and simulated galaxy samples is shown in
Figure~\ref{fig:select}. The thick black solid line here indicates, as
a function of stellar mass, the fraction of galaxies from the parent
sample which remain in our final sample of isolated SDSS galaxies. The
thick black dashed line shows the result when analogous isolation
criteria are applied to our G11 mock catalogues. The agreement of
observation and simulation is again quite good, though not as perfect
as in the right panel of figure~\ref{fig:frac}. Note, however, that
perfect agreement is not expected, since the underlying SDSS sample is
magnitude-limited whereas the G11 sample is volume-limited.

This same plot provides a convenient way to summaries the relative effects of
the spectroscopic and photometric catalogues in defining our observed sample
of isolated galaxies. The thin red solid line shows the fraction of objects
which are isolated relative to the spectroscopic sample, but not necessarily
relative to the photometric sample.  At most masses, use of the photometric
catalogue increases the number of objects with an identified ``companion'' by
30 to 50\%. This has a relatively small effect on the isolated fraction at
high mass, since the great majority of massive galaxies are isolated by our
criteria. At $\log M_\star/M_\odot = 10.3$ it reduces the number of isolated
galaxies by almost a factor of two, however, because the majority of such SDSS
galaxies have a companion by these same criteria. These numbers suggest (and
we have checked in more detail) that our photometric rejection step is
conservative, in that a significant fraction of the photometric ``companions''
(about one third) are, in fact, galaxies at a significantly different redshift
which are projected on top of the primary. This does not matter for the
analysis of this paper -- it just causes us to end up with a slightly smaller
sample of isolated primaries than if we had been less conservative.

Finally we can use the simulated catalogue to see how our isolation criteria
affect the numbers of satellite and central galaxies in our samples. The green
and blue dashed lines in Figure~\ref{fig:select} show as a function of stellar
mass the fractions of centrals and satellites which pass our isolation
criteria. At high mass, most centrals are indeed isolated, but for $\log
M_\star/M_\odot < 10.6$ more than half of them are rejected because of
apparent companions. In most cases these companions are actually the central
galaxies of other haloes. On the other hand, for $\log M_\star/M_\odot < 11$
almost all satellite galaxies are found to have companions according to our
criteria. With increasing mass the number of apparently isolated satellites
grows, reaching 30\% at the highest mass. This is because these objects lie in
very massive galaxy clusters and so can be projected more than 1~Mpc from
their associated central galaxies. Note, however, that the actual number of
such massive satellites is very small (see figure~\ref{fig:frac}). It is the
interplay of these different effects which produces the very high purity
(i.e. central galaxy fraction) at all masses in our final isolated galaxy
sample.

\subsection{The photometric catalogue for satellites}
\label{subsec:photo}

Both the spectroscopic catalogue and the photoz2 photometric catalogue used in
the last section are based on DR7, the seventh release of SDSS data.  When
identifying and verifying the isolation of our isolated primary galaxies, only
objects with apparent magnitude brighter than $r=17.6$ had to be
considered. This is far above the magnitude limit of the SDSS photometry, and
the accuracy of the DR7 magnitudes/colours and of the derived photo-$z$
distributions is well tested at these magnitudes. When compiling
counts of faint satellite galaxies around these primaries, we need to go to
the SDSS limit for reliable photometry, however, which we take to be $r=21$.
At this limit, improvements to the photometry pipeline have continually
enhanced its reliability, identifying (and correcting when possible)
systematic artifacts which can have a significant influence on our
analysis. We therefore use the photometric catalogue from
SDSS/DR8\footnote{http://skyservice.pha.jhu.edu/casjobs/} \citep{2011ApJS..193...29A} when compiling
satellite counts. To be specific we created a reference photometric catalogue
by downloading objects that are classified as galaxies in the survey's primary
object list, and that do not have any of the flags BRIGHT, SATURATED,
SATUR\_CENTER or NOPETRO\_BIG set. This follows the selection criteria
for the DR7 photoz2 catalogue used above,

It is important to note that the magnitude we use for satellites is the
so-called SDSS model magnitude, which, at faint apparent magnitudes, is
claimed to have the highest signal-to-noise of all the alternatives
catalogued. In contrast, when defining our primary sample and checking its
isolation we used Petrosian magnitudes as listed on the NYU-VAGC website. We
continue to use these magnitudes for the primaries below. When we quote
colours, these are always measured within a well-defined aperture related to
the Petrosian radius, and are rest-frame quantities K-corrected to the
$^{0.1}(g-r)$ band for both primaries and satellites. All absolute magnitudes
quoted for SDSS galaxies are also in this same rest-frame band.

We have carried out a variety of tests for systematics in the SDSS
photometry, checking completeness and the quality of star-galaxy
separation by comparing with much deeper HST data, and quantifying
systematic biases in the magnitudes of faint images (e.g. the
satellites) in the neighborhood of substantially brighter images
(e.g. the primaries). Detailed descriptions of these tests are given
in an Appendix, along with details of tests of our procedures
(outlined in the next section) for counting apparent companions around
our primary galaxies and correcting for the (usually dominant)
contribution from unrelated foreground and background objects.

\section{Satellite counting methodology}
\label{sec:method}
We want to study the abundance of satellites as a function of their
luminosity, stellar mass and colour around our sample of isolated bright
primaries, and to see how these abundances depend on the stellar mass and
colour of the primary.  We will need to use the SDSS data down to their
reliable photometric limit (which we take to be $r=21$) and as a result the
great majority of the potential satellites do not have a spectroscopically
measured redshift.  Hence it is necessary to count {\it all} apparent
neighbours around our primaries and to correct statistically for unassociated
objects which happen to be projected near them. The number of such (mainly)
background objects can substantially exceed the number of true satellites, so
it is important to take considerable care in making these corrections.  We
adopt the following procedure.

For each isolated bright galaxy, we identify all photometric galaxies with
apparent projected separation $r_p<0.5$~Mpc and we accumulate counts in bins
of projected separation $r_p$, apparent magnitude $r$ and observed colour
$g-r$.  From the count in each ($r_p$,$r$,$g-r$) bin, we subtract the expected
number of background galaxies which we take to be $N(r,g-r) A(r_p,z)f/A_{\rm
  tot}$, where $N(r,g-r)$ is the total number of galaxies in the ($r$,$g-r$)
bin in the full photometric catalogue, $A_{\rm tot}$ is the solid angle of the
survey footprint, $A(r_p,z_{\mathrm{pri}})$ is the solid angle corresponding
to the annular $r_p$ bin at the redshift $z_{\mathrm{pri}}$ of the primary
galaxy, and $f$ is the incompleteness factor, the fraction of this annulus
which lies within the survey footprint (we estimate this using the random
points generated around the position of each primary during the selection
process -- see above). We then use the redshift of the primary galaxy to
convert observed apparent magnitudes and colours into rest-frame luminosities
and colours (in the $^{0.1}r$ and $^{0.1}(g-r)$ system)\footnote{We use the
  empirical fitting formula of \cite{2010PASP..122.1258W} which gives the
  K-correction as a function of redshift and observed colour.} and we transfer
the background-subtracted satellite counts from our narrow bins of $r$ and
$g-r$ into substantially broader bins of the rest-frame quantities.  Finally
we average these counts for each $r_p$ bin over the set of all primaries in
the desired range of stellar mass (and sometimes colour) and we sum the result
over the desired range in $r_p$.  Uncertainties in the resulting numbers are
estimated from the scatter among results for 100 bootstrap resamplings of the
set of primaries.

Some apparent companions are too red to be at the redshift of the primary galaxy. 
It is useful to exclude them when accumulating counts since they add noise without
adding signal.  Hence, we exclude all bins redder than
$^{0.1}(g-r)=0.032\mathrm{log}_{10}M_\star+0.73$, a fit to the upper envelope
of the distribution of rest-frame colour against stellar mass for galaxies of
measured redshift. The stellar mass $M_\star$ of the apparent companion is
estimated by assuming it to be at the primary's redshift and adopting
\begin{equation}
(M/L)_r=-1.0819^{0.1}(g-r)^2+4.1183^{0.1}(g-r)-0.7837
\label{eqn:masstolight}
\end{equation}

This empirical relation is a fit to a flux-limited ($r<17.6$) galaxy
sample from the NYU-VAGC website for which stellar masses were 
estimated from the K-corrected galaxy colours by fitting stellar population 
synthesis models assuming a \cite{2003PASP..115..763C} initial mass function 
\citep{2007AJ....133..734B}. For this sample the 1-$\sigma$ scatter in $(M/L)_r$ 
of this simple relation is about 0.1.

The photometric catalogue we use is complete down to an $r$-band apparent
model magnitude of 21. This limit corresponds, of course, to different
satellite luminosities and stellar masses for different primary redshifts and
different satellite colours. In order to ensure that our samples are complete
when compiling satellite luminosity functions, we allow a particular primary
to contribute counts to a particular luminosity bin only if the K-corrected
absolute luminosity corresponding to $r=21$ for a galaxy at the redshift of
the primary and lying on the red envelope of the intrinsic colour distribution
is fainter than the lower luminosity limit of the bin. Thus only the nearest
primaries will contribute to the faintest luminosity bins of our satellite
luminosity functions, and different numbers of primaries will contribute to
each bin.  We follow an exactly analogous procedure when compiling stellar mass
functions for satellites. 

For each individual satellite luminosity/stellar mass bin, this
treatment is equivalent to imposing an upper limit on primary
redshift.  Thus there is a maximum volume which is surveyed for
satellites in the $j$th luminosity/stellar mass bin which we
denote $V_{max,bin,j}$.

On the other hand, our primary sample is flux-limited at $r=16.6$, so
for brighter satellite bins where $V_{max,bin,j}$ is large,
intrinsically faint primaries will not be visible to the redshift
limit. The effective volume surveyed is then $V_{max,pri,i}$, the
total survey volume over which the $i$th primary would lie above the
flux limit.  Note that because of K-corrections, this volume depends
on the intrinsic colour of the primary as well as its intrinsic
luminosity. We present our final results in the form of the mean
number of satellites per primary for a volume-limited primary sample
\begin{equation}
N_{sat,j}=\frac{\Sigma_i N_{sat,i,j}/V_{max,ij}}{\Sigma_i 1/V_{max,ij}}, 
\label{eqn:numsat}
\end{equation}
where 
\begin{equation}\label{eqn:error}
V_{max,ij}= {\rm min}\big[V_{max,pri,i}, V_{max,bin,j}\big].
\end{equation}
Thus, for satellite luminosity/stellar mass bin $j$, we sum satellite
counts over all primaries $i$ that are within $V_{max,bin,j}$. At the
bright end of the satellite luminosity function, we expect
$V_{max,ij}=V_{max,pri,i}$ because satellites are less than 4.4
magnitudes fainter than their primaries and so can be seen around all
primaries.  For intrinsically faint satellites, however,
$V_{max,ij}=V_{max,bin,j}$ because primaries can be seen to well
beyond the distance at which $r=21$ for the satellites.
  
We apply similar selection criteria to our mock catalogues based on
G11. Since we know the absolute magnitude, rest-frame colour and
stellar mass of all galaxies, the background subtraction can be
carried out directly using rest-frame quantities. In addition, the
effective depth of the satellite catalogue is the same for all
primaries so that all primaries can contribute to all satellite
luminosity or stellar mass bins and we do not need any weighting in
order to obtain proper volume-weighted statistics. Note that we do not
use any information about the redshift difference between primary and
apparent companion, so projection effects occur over $\Delta z =
50,000$~km/s corresponding to the side of the Millennium Simulation
``box''. Both for the simulation and for the real SDSS data we use a
global background estimate based on the full survey in order to
minimize the statistical uncertainty in the correction.  

Previous work often estimated a background density locally for each
primary \citep[e.g.][]{1994MNRAS.269..696L,2011MNRAS.417..370G}. This not only
substantially increases the noise, it also introduces a significant
bias since galaxies are correlated on all scales and the
``background'' fields are, in fact, expected to have faint galaxy
densities significantly above the mean. The extent of the bias is, in fact,
strongly dependent on the isolation criteria and can be of either sign
\citep[see, for example,][]{1976ApJ...205L.121F}. The large-scale uniformity of
the SDSS photometry is such that there appears to be no advantage to
adopt a local estimate, provided all photometric quantities are
properly corrected for Galactic extinction. The accuracy of our method
is confirmed by the accurate convergence to unity at large angular
scale in the left panel of figure~\ref{fig:photo} and by a variety of tests
which we describe in detail in the Appendix.

\section{Luminosity and Mass Functions of Satellite Galaxies}
\label{sec:LFMF}

\subsection{Luminosity functions}
\begin{figure*}
\epsfig{figure=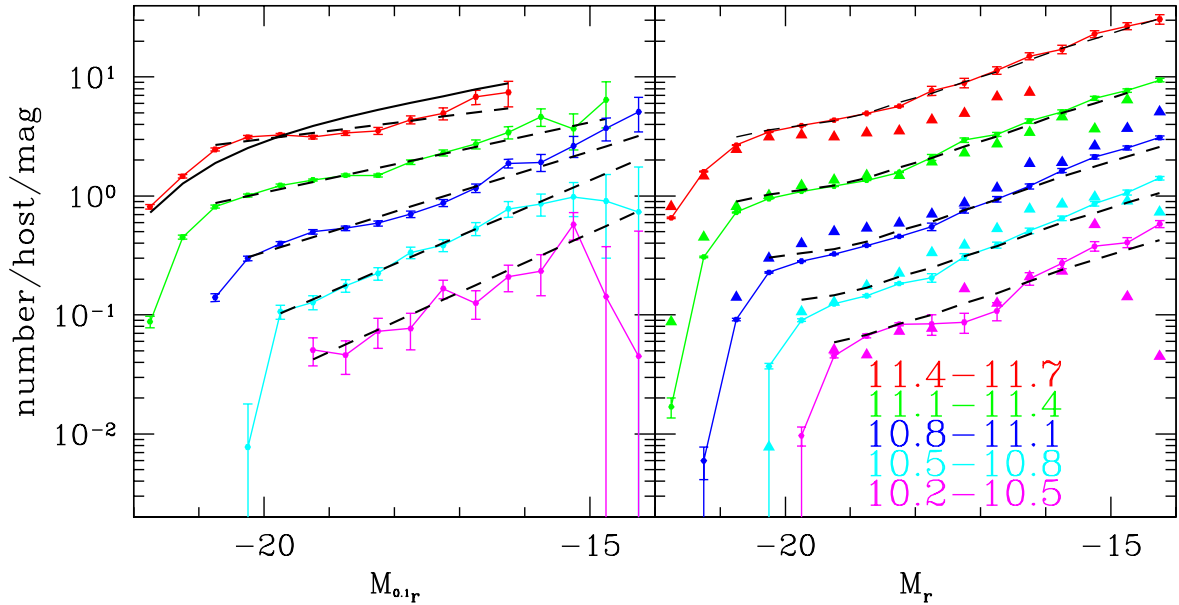,width=0.9\textwidth}
\caption{Luminosity functions in the $^{0.1}r$ band for satellites of primary
  galaxies in five disjoint ranges of $\log M_\star/M_\odot$, as indicated in
  the legend. Satellites are counted within a projected radius of 300~kpc,
  except for the lowest mass range where we count within 170~kpc. In the left
  panel, the data points connected by solid lines give observational results
  for SDSS/DR8 with error bars estimated by bootstrap resampling of the
  primary sample. The solid black line is a fit of a
  \citet{1976ApJ...203..297S} function to the most massive bin, with the
  characteristic luminosity and faint-end slope constrained to match those of
  the SDSS field luminosity function. Dashed black lines are power-law fits
  excluding one or two of the brightest points for each mass range, as
  indicated by the line extent (see the text).  In the right panel, the small
  symbols connected by solid lines show corresponding results for the G11
  simulated galaxy populations.  Black dashed lines here show the mean
  luminosity function for all simulated galaxies, renormalized to fit the
  satellite data as in the left panel. The SDSS results are overplotted as
  filled triangles for ease of comparison.}
\label{fig:LFall}
\end{figure*}

In figure~\ref{fig:LFall} we present $^{0.1}r$-band luminosity functions for
satellites projected within 300~kpc of their primaries, except for the
faintest bin, where the halo virial radius is much smaller than 300~kpc and we
estimate the luminosity function within 170~kpc in order to increase the
signal-to-noise. In the left panel, data points connected by solid lines show
our observational results for SDSS/DR8.  As indicated in the legend, colours
encode the range in $\log M_\star/M_\odot$ of the primaries contributing to
each luminosity function estimate.  Error bars are derived by bootstrap
resampling the primary sample. At the faint end we lose higher redshift
primaries because of the apparent magnitude limit of our photometric
catalogue. We do not plot data for bins with fewer than eight primaries.  It
is evident that satellite numbers increase strongly with primary mass.

The black solid line in the left panel shows a \cite{1976ApJ...203..297S}
function fit to the data for the most massive primaries. We have fixed the
characteristic luminosity and faint-end slope to be those of the SDSS field
luminosity function in $^{0.1}r$ \citep{2009MNRAS.399.1106M}. The result is a
moderately good fit to the satellite data, although these appear steeper than
the field luminosity function at the faintest magnitudes. The satellite
luminosity functions clearly become steeper for lower mass
primaries. Power-law fits to the faint-end data are shown as dashed lines in
the figure and give the slopes $\alpha$ listed in the first row of
table~\ref{tbl:slope}. The bright end of each function is cut off by our
requirement that every satellite be at least one magnitude fainter than its
primary. We therefore exclude one or two of the brightest points when making
these fits. Specifically, we find the median absolute magnitude
$M_{\mathrm{med}}$ for primaries in each mass range, and we include only
points for which the corresponding absolute magnitude bin lies entirely below
$M_{\mathrm{med}}+1$. The ranges fitted in each case are indicated by the
extents of the dashed black straight lines.\footnote{We have checked that the
remaining points are indeed unaffected by rebinning our data as a function
of the $r$-band absolute magnitude of the primaries. In this case we know
exactly which bins are unaffected by our isolation criterion. When
primary absolute magnitude and stellar mass are matched appropriately, the
resulting satellite luminosity functions match those of
figure~~\ref{fig:LFall} very closely over the full range used to determine
the faint-end slope and are unaffected by the isolation criterion over this
range.}  The faint-end slope decreases from $\alpha\sim-1.2$ for $\log
M_{\star,p}/M_\odot \sim 11.5$ to $\alpha\sim-1.6$ for $\log
M_{\star,p}/M_\odot \sim 10.3$. 

\cite{2011MNRAS.417..370G} divided their primaries into three
luminosity ranges ($M_r=-23.0\pm0.5$, $-23.0\pm0.5$ and $-23.0\pm0.5$)
and compiled satellite luminosity functions in bins of
satellite-primary magnitude difference rather than satellite absolute
magnitude. They quote faint-end slopes of -1.45, -1.725 and -1.96 for
these three sets of primaries, with the fainter primaries having
steeper luminosity functions. For the largest measured magnitude
differences ($\Delta m\sim 8$) they found similar numbers of
satellites independent of primary luminosity. These $\alpha$ values
are substantially more negative than ours, particularly for the
faintest primaries. In order to compare with their results, we adopt
similar isolation criteria, we take the same ranges of primary
luminosity, and we also accumulate satellite number as a function of
magnitude difference. Fitting the faint-end slope over the same
satellite magnitude range as in figure 7 of
\cite{2011MNRAS.417..370G}, we find $\alpha$ values of -1.189, -1.376
and -1.588, substantially shallower than those of
\cite{2011MNRAS.417..370G} and quite compatible with those we quote in
Table~\ref{tbl:slope}. Detailed tests show this inconsistency to be due 
partly to the local background subtraction scheme of \cite{2011MNRAS.417..370G}
which removes part of the signal {\footnote{\cite{2011MNRAS.417..370G} used 
photometry from SDSS/DR7  while our own tests, similar to those discussed in section 
A1 of the  Appendix, showed to suffer from substantially more serious systematics for 
faint images close to brighter ones than is the case for the SDSS/DR8 catalogues used 
here.}, but mainly to the fact that they use model magnitudes K-corrected to $z=0$ for 
their primaries, rather than the $^{0.1}r$ Petrosian magnitudes which we use here.

\begin{table*}
\caption{Exponent $\alpha$ of the  faint-end-slope for SDSS luminosity and mass functions}
\begin{center}
\begin{tabular}{lrrrrr}\hline\hline
Range in primary $\log M_\star/M_\odot$ & \multicolumn{1}{c}{11.7-11.4} & \multicolumn{1}{c}{11.1-11.4} & \multicolumn{1}{c}{10.8-11.1} & \multicolumn{1}{c}{10.5-10.8} & \multicolumn{1}{c}{10.2-10.5} \\ \hline
Luminosity function & -1.170 & -1.295 & -1.424 & -1.587 & -1.622 \\
Mass function       & -1.231 & -1.297 & -1.455 & -1.800 & -1.748 \\
\hline
\label{tbl:slope}
\end{tabular}
\end{center}
\end{table*}

In the right panel of figure~\ref{fig:LFall}, the small symbols joined by
solid lines show analogous results for the galaxy formation model of G11,
based on the Millennium and Millennium-II simulations.  Points brighter than
$M_r=-18$ are MS data. At fainter magnitudes, resolution effects cause the MS
to underestimate galaxy abundances and we take our data from the MS-II.  (The
two simulations agree very well in the range $-18>M_r>-20$.) To facilitate
comparison, we replot the SDSS data from the left panel as filled
triangles. Agreement of model and observation is fair but far from
perfect. The simulation overpredicts the number of satellites around the most
massive primaries by 25 to 50\% for $M_r>-19.5$. In the two lower primary
stellar mass ranges, the simulation underpredicts the number of satellites by
20 to 30\% . As we will see below, the latter discrepancy reflects a problem
with the colours of the simulated satellites rather than with their stellar
masses. The black dashed lines in this panel are the ``field'' luminosity
function for the full simulations renormalized to fit the satellite data for each
primary mass range. Here also there is a trend for the faint-end slope to be
steeper for satellites than in the field, but the effect is much less marked
than for the SDSS data. Furthermore the variation of faint-end slope with
primary mass seen in the SDSS is weak or absent in the simulation data.

Many previous papers have investigated the shape and faint-end slope of
group/cluster luminosity functions \citep[e.g.][]{2001A&A...367...59P,
  2002AJ....123.1807G,2003ApJ...591..764C,2003MNRAS.342..725D,2005MNRAS.360..727A,
2005A&A...433..415P,2006A&A...445...29P,2006ApJ...650..137Z,
2009ApJ...699.1333H,2010MNRAS.401..941A,2011MNRAS.414.2771D}. Unfortunately
there is little consensus. Some authors found large differences between
cluster and field luminosity functions, while others found the two to be quite
similar. For example, by stacking SDSS data around the centres of clusters
detected in the Rosat All Sky Survey, \cite{2006A&A...445...29P} obtained
cluster luminosity functions with a very obvious steepening at faint
magnitudes.  This faint-end upturn appeared to be contributed primarily by
early-type galaxies.  In our most massive primary stellar mass bin
($11.7>\mathrm{log}_{10}M_\star>11.4$), there is qualitatively similar
behaviour with an upturn at $M_r\sim-18$ as in \cite{2006A&A...445...29P} but
the steepening is much less dramatic in our data than in theirs. 
In contrast, no evidence of an upturn at the faint end was found by 
\cite{2010MNRAS.401..941A} and \cite{2011MNRAS.414.2771D} in galaxy clusters.

\subsection{Stellar mass functions}

\begin{figure*}
\epsfig{figure=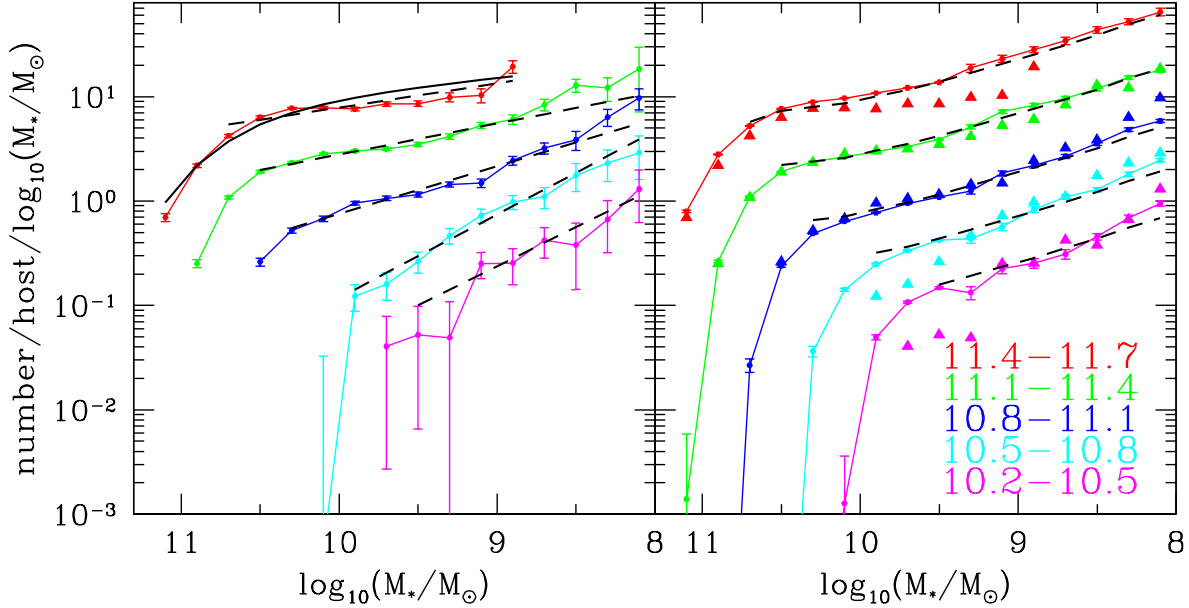,width=0.9\textwidth}
\caption{Similar to figure~\ref{fig:LFall} but showing stellar mass functions
  for satellites projected within 300~kpc (or 170~kpc) of their primaries,
  both for the SDSS (left panel) and for the G11 simulations (right panel with
  the SDSS data repeated as filled triangles). As in figure~\ref{fig:LFall},
  the primaries are grouped into five disjoint ranges of $\log
  M_\star/M_\odot$, as indicated by the colours.  A black solid line in the
  left panel is a renormalized version of the ``field'' stellar mass function
  of \citet{2009MNRAS.398.2177L} overplotted on the data for the highest mass
  primaries, while dashed black lines show a power law fit to each mass
  function estimate. In the right panel the dashed black lines are fits of the
  satellite data to renormalized versions of the stellar mass function of the
  simulation as a whole.}
\label{fig:MFall}
\end{figure*}

Figure~\ref{fig:MFall} is similar to figure~\ref{fig:LFall}, but shows
stellar mass functions for satellites both in the SDSS (left panel)
and in the G11 simulations (right panel with the SDSS data repeated as
filled triangles). The strong dependence of satellite number on
primary mass is again evident. The black solid line overplotted on the
most massive bin is a Schechter function fit with characteristic mass
and low-mass slope fixed to the ``field'' values of
\cite{2009MNRAS.398.2177L}. The observed satellite stellar mass
functions are again steeper than the corresponding field function at
the low-mass end. The black dashed lines in the left panel are
power-law fits to the observational data. The corresponding
faint-end-slopes are given in the second row of table~\ref{tbl:slope}.
Here also we have ignored the few brightest points in each
estimate. As before the extent of each dashed black line indicates the
points actually used in the fit.  The steepening of the satellite
stellar mass functions with decreasing primary mass is even stronger
than was the case for the satellite luminosity functions.

In the right panel of figure~\ref{fig:MFall}, the G11 points are
taken from the MS at $\log M_\star/M_\odot >9.5$ and from the MS-II at
lower mass. Fits to the field stellar mass function from the
simulations are shown as dashed lines and indicate no significant
shape difference between the two. Comparison with the SDSS data shows
that the overprediction of the satellite luminosity function for the
most massive primaries persists in very similar form in their stellar
mass function, The underprediction found for the two lower primary
mass ranges has gone away, however, indicating that the discrepancy
was due primarily to the colours of the simulated satellites, rather
than to their stellar masses. The simulation does not reproduce the
clear steepening of the SDSS satellite mass functions with decreasing
primary mass.  Any effect in this direction is very weak.

To investigate further the origin of the observed steepening with
decreasing primary mass, we take all spectroscopic galaxies brighter
than 16.6 in the $r$-band, without applying any isolation criterion,
and calculate the mass function of surrounding satellites within 300
(or 170)~kpc in exactly the same way as for our isolated primaries.
Results are shown as solid curves in figure~\ref{fig:background}. In
this case there is no steepening with decreasing primary mass -- the
different colour curves are more or less parallel to each other and to
the field stellar mass function.  Apparently, the steepening is caused
by our isolation criteria -- isolated galaxies have fewer lower mass
neighbours than typical galaxies.  The suppression is stronger for
more massive neighbours, steepening the satellite mass functions of
isolated primaries. Notice that for primaries with $\log
M_\star/M_\odot > 10.8$, the mean numbers of low-mass companions are
similar for isolated and for typical objects, but that lower mass
primaries have substantially fewer such companions if they are
isolated.  This is because many of the lower mass non-isolated
''primaries'' are, in fact, satellites themselves, and their low-mass
``companions'' are fellow satellites within the larger system.  It is
quite interesting that the abundance of relatively small satellites is
strongly affected by the presence or absence of a nearby galaxy
comparable in luminosity to the primary, particularly since the
observed trends are not fully reproduced by the simulation.  This suggests
that dwarf galaxy formation may be influenced by nearby giants in a
way which the simulation does not represent.

\begin{figure*}
\epsfig{figure=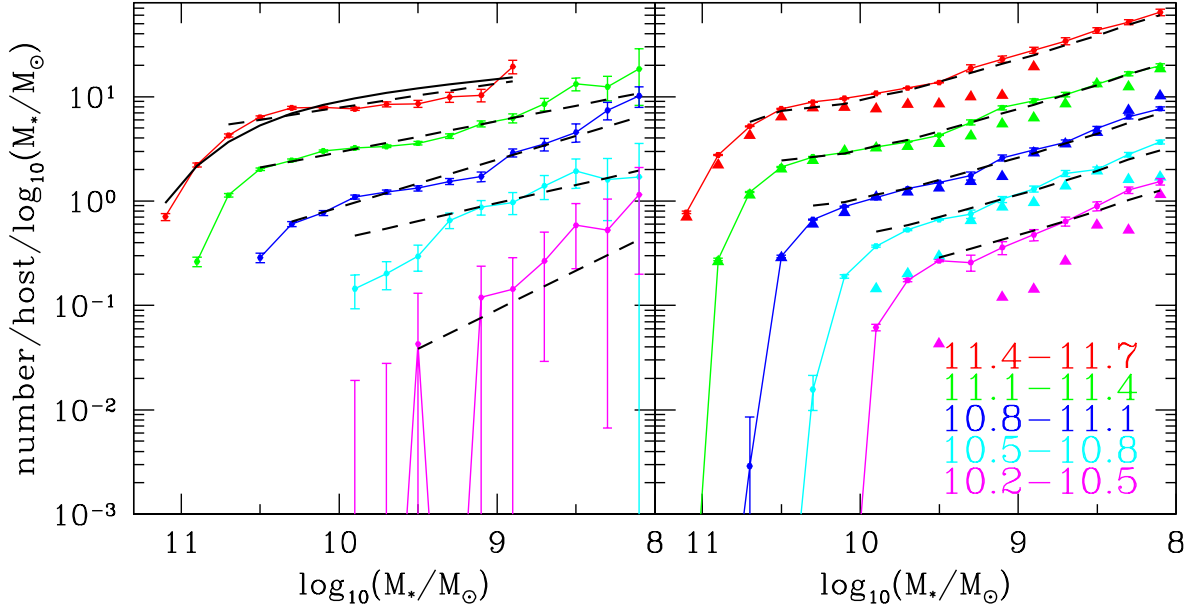,width=0.9\textwidth}
\caption{Similar to figure~\ref{fig:MFall}, but for red primaries.
  See figures \ref{fig:prop_dis} and \ref{fig:prop_dis2} for the
  colour cuts separating red and blue primaries.}
\label{fig:MFred}
\end{figure*}

\begin{figure*}
\epsfig{figure=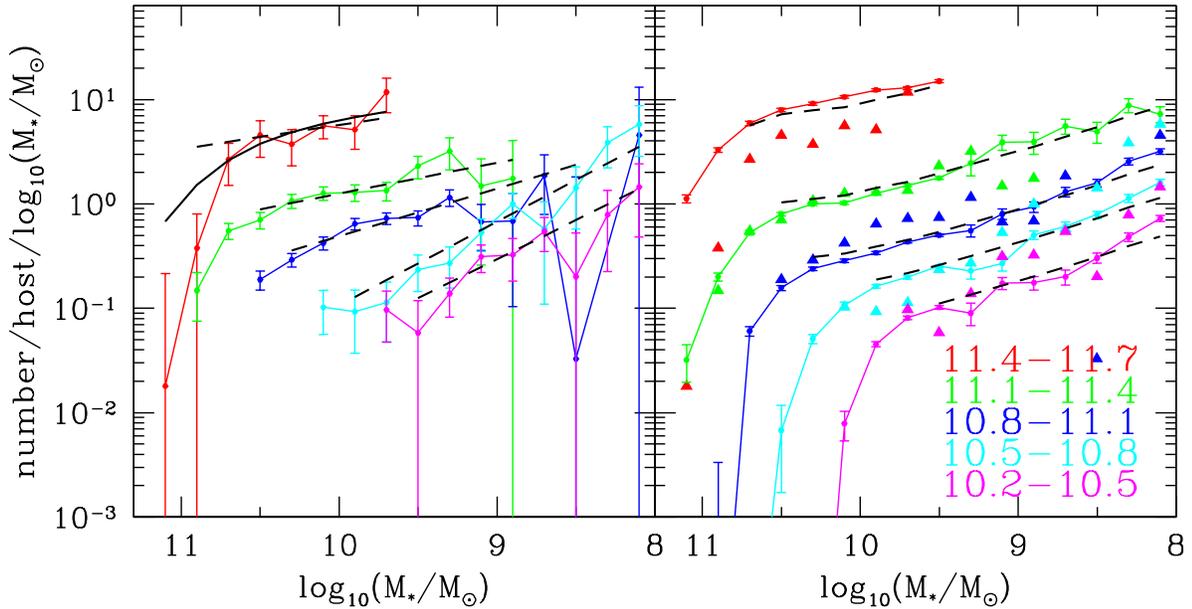,width=0.9\textwidth}
\caption{Similar to figure~\ref{fig:MFall}, but for blue primaries. In the
  right panel, the red dots connected by a curve stop at
  $\mathrm{log}_{10}M_\star=9.5$, because the Millennium-II Simulation has
  fewer than eight blue primaries more massive than $\log M_\star/M_\odot=11.4$.}
\label{fig:MFblue}
\end{figure*}

It is interesting to see whether satellite galaxy populations depend on the
colour of the primary galaxy as well as on its stellar
mass. Figures~\ref{fig:MFred} and \ref{fig:MFblue} show the satellite mass
functions surrounding red and blue primaries respectively, where the primary
populations have been split at the colours indicated in figures
\ref{fig:prop_dis} and \ref{fig:prop_dis2}. The black dashed lines in the left
panels are power-law fits, but in each case the slope is fixed to be that
found for all primaries in the relevant stellar mass bin
(table~\ref{tbl:slope}). In the right panel of figure~\ref{fig:MFblue} the red
curve stops at $\log M_\star/M_\odot=9.5$ because the MS-II contains fewer
than eight blue primaries more massive than $\log M_\star/M_\odot=11.4$ and
the MS population is affected by numerical resolution at lower satellite mass.
For red primaries the SDSS and G11 data agree quite well, apart from the slope
discrepancy and a residual overprediction of the abundance of faint satellites
around high-mass primaries.  For blue primaries, the differences are bigger
but in the highest mass bin this could reflect the small number of SDSS
primaries and the correspondingly large observational error bars. The
discrepancy for primaries in the stellar mass range $10.1 > \log
M_\star/M_\odot>10.8$ is smaller but more significant, particularly since
there is good agreement in this mass range for red primaries. Interestingly, a
comparison of figures \ref{fig:MFred} and \ref{fig:MFblue} shows that the
amplitude of the satellite luminosity function is higher around red primaries
than around blue primaries of the same stellar mass.  We analyze this result
in more detail in the following subsections.

\subsection{Satellite abundance as a function of primary
stellar mass and colour}

\begin{figure*}
\centerline{             \epsfig{figure=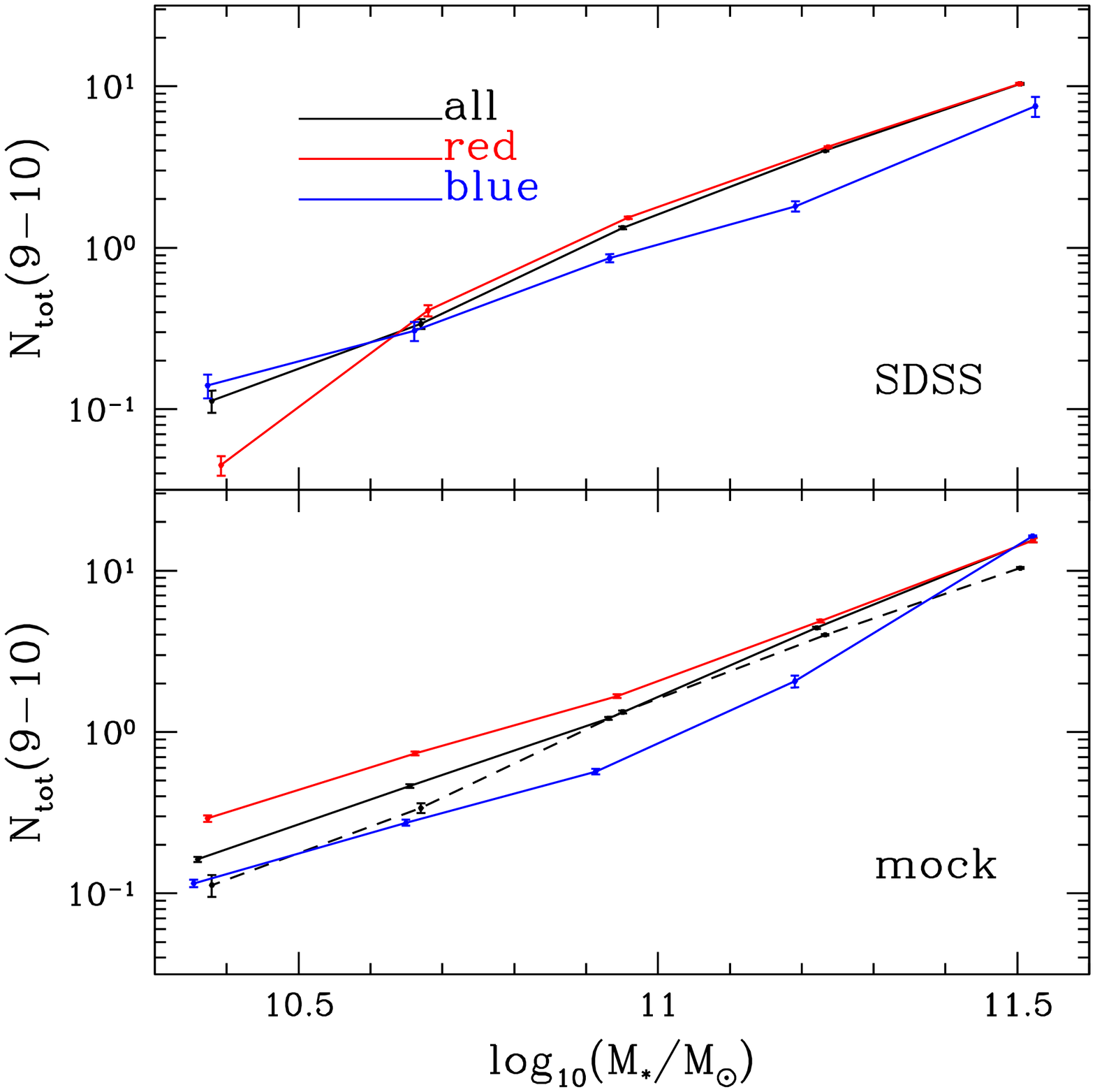,width=0.45\textwidth}
  \epsfig{figure=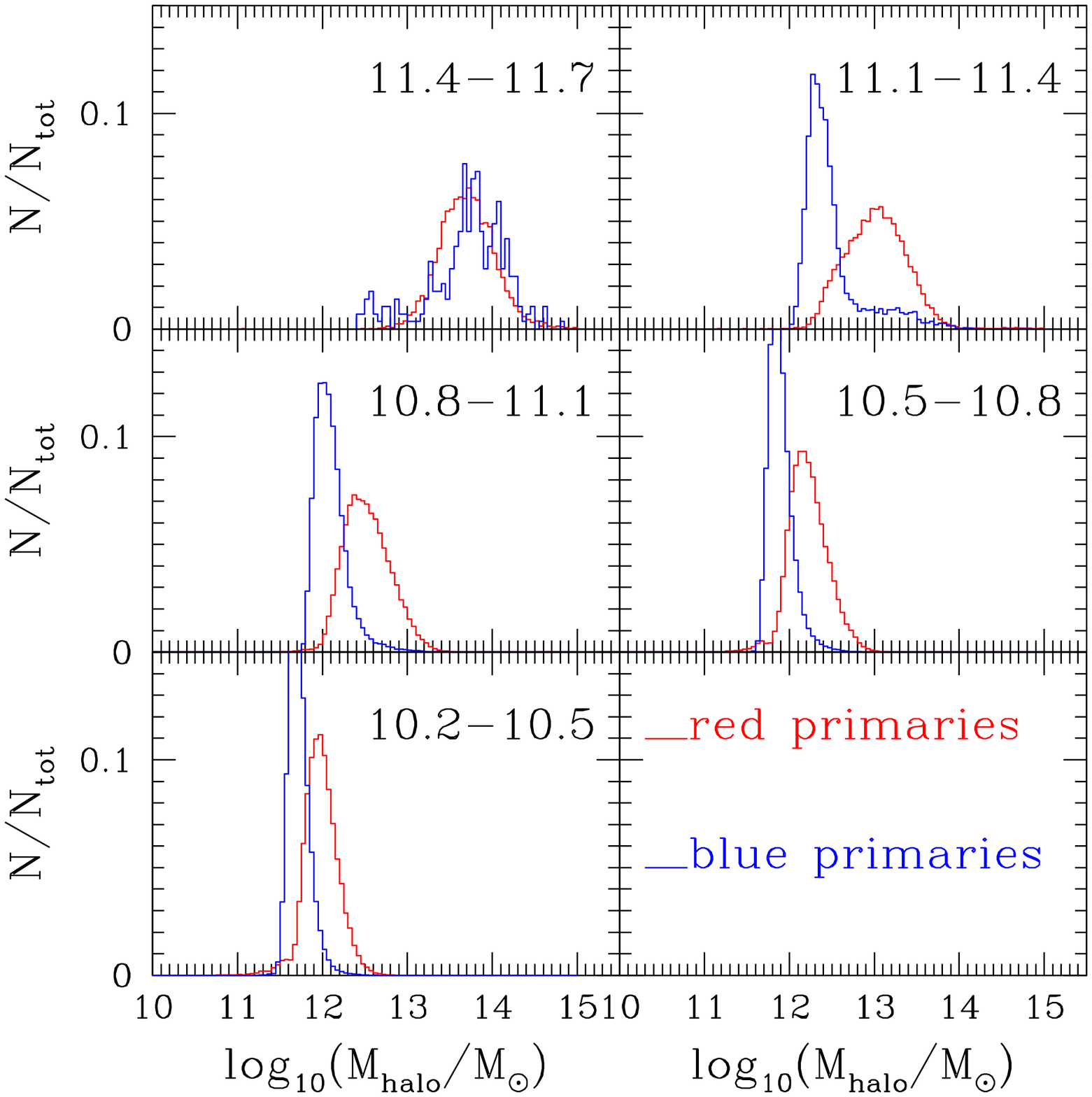,width=0.45\textwidth} }
\caption{{\bf Left:} Mean number of satellites in the stellar mass range $10.0
  >\log M_\star/M_\odot>9.0$ as a function of primary stellar mass.  Black,
  red and blue points refer to all, to red and to blue primaries,
  respectively. For the points at lowest primary mass, satellite counts were
  accumulated within 170~kpc, whereas for all other primary masses they were
  accumulated within 300~kpc.  The top panel gives observational results for
  the SDSS while the bottom panel gives corresponding results for the G11
  galaxy formation simulations.  The SDSS result for all primaries is
  re-plotted as a dashed curve in the bottom panel in order to facilitate
  comparison.  {\bf Right:} Host halo mass distributions for red and blue
  primaries in the simulated catalogues, split into the same primary stellar
  mass ranges as in the left-hand plots.  }
\label{fig:number}
\end{figure*}

In order to better display how the abundance of satellites depends on
the stellar mass and colour of the primary, we use the power-law fits
shown as dashed lines in figures~\ref{fig:MFall}, ~\ref{fig:MFred}
and~\ref{fig:MFblue} to predict the mean number of satellites per
primary in the stellar mass range $10.0>\log M_\star / M_\odot >9.0$
and at projected separation $r_p<300$~kpc ($r_p<170$~kpc for the
lowest mass primaries).  This has the advantage of producing a robust
measure of satellite abundance which is little affected either by
selection-induced cut-offs (most important for low-mass and red
primaries) or by incompleteness (most important for high-mass and blue
primaries). The results are shown in the left panels of
figure~\ref{fig:number}, where black dots and lines give results for
all primaries, while red and blue dots and lines give results for red
and blue primaries, respectively.  The top panel presents results for
the SDSS and the lower panel results for the G11 simulations.  The
SDSS result for all primaries is repeated in the lower panel, showing
that the simulation overpredicts the number of satellites in this mass
and projected radius range both for the highest mass and for the
lowest mass primaries. Note that because the low-mass slopes differ in
simulation and observation, the result for low-mass primaries depends
on the satellite mass range chosen for the comparison.

At high mass the black and red curves in figure~\ref{fig:number} are close to
each other, reflecting the fact that the fraction of red primaries is large
(see figure~\ref{fig:frac}). At the highest mass, the blue curve indicates a
consistent number of satellites around blue primaries, although with
considerable uncertainty because such primaries are rare. At somewhat lower
mass, however, blue primaries have significantly fewer satellites than red
primaries {\it of the same stellar mass}, both in the SDSS data and in the
simulation. This is a primary result of our paper. The effect is a factor of
two to three in satellite abundance for primaries with $\log
M_\star/M_\odot\sim 11$. In the SDSS data there is some indication that the
the colour dependence may get smaller again for lower mass primaries, but this
does not happen in the simulation, where there is still more than a factor of
two difference for $M_\star\sim 2\times 10^{10}M_\odot$. Overall, the
differences appear somewhat larger in the model than in the real data.  

The cause of this effect in the G11 simulation is easy to track down. In the
right-hand panels of figure~\ref{fig:number} we plot histograms of host halo
mass for isolated galaxies as a function of their stellar mass and colour. (As
shown in figure~\ref{fig:frac}, almost all isolated galaxies in the simulation
are the central galaxies of their haloes.)  For all except the highest stellar
mass range, red primaries have significantly more massive dark haloes than
blue ones. The shift between the peaks of the two distribution is an order of
magnitude for primaries with $11.4>\log M_\star/M_\odot\sim 11.1$, dropping to
a factor of two for $10.5>\log M_\star/M_\odot\sim 10.2$. In the simulation
red primaries have more satellites because they live in more massive haloes. A
direct indication that the same may hold for real galaxies comes from the
galaxy-galaxy lensing study of \cite{2006MNRAS.368..715M}. By combining their
SDSS lensing data with HOD modeling, these authors concluded that red
galaxies have more massive haloes than blue ones for $\log M_\star/M_\odot >
11$. At lower central galaxy masses their results appear consistent with no
offset, although the error bars are large (see their figure 4).

The fact that both the number of satellites and the mass of the
associated dark halo depend not only on the stellar mass of the
primary galaxy but also on its colour contradicts the assumptions
underlying many HOD or abundance matching schemes for interpreting
large-scale galaxy clustering. Such a dependence could be included in
more complex versions of at least the former, but would require
additional parameters and additional observational data to constrain
them
\citep[e.g.][]{2009MNRAS.398..807S,2009MNRAS.399..878R,2009MNRAS.392.1080S}.

\begin{figure}
\epsfig{figure=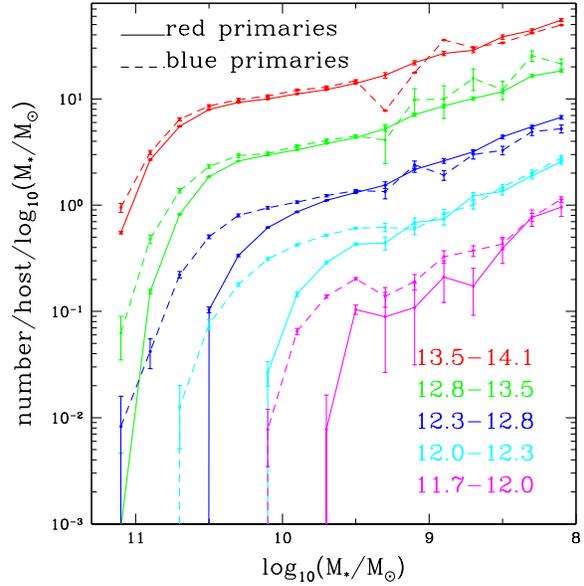,width=0.45\textwidth}
\caption{Satellite mass functions within $r_p=300$~kpc, split according to
  {\it halo} mass, for dark haloes in the G11 galaxy formation simulation. The
  colours refer to different ranges of $\log M_{\rm halo}/M_\odot$ as indicted
  in the legend. For each halo mass range, satellite stellar mass
  functions are shown separately for haloes with red (solid lines) and blue
  (dashed lines) central galaxies.}
\label{fig:halo}
\end{figure}

Within the simulation it is possible to check whether halo mass is the only
factor responsible for the difference in satellite abundance between red and
blue primaries. In figure~\ref{fig:halo} we present stellar mass functions for
satellites of isolated galaxies as a function of the {\it halo} mass of the
primary. As before these are compiled for satellites projected within
$r_p=300$~kpc, The halo mass ranges for this plot are chosen to correspond
roughly to the primary stellar mass ranges in previous figures.  For each halo
mass range, the mass functions are also split according to the colour of the
primary galaxy, with solid and dashed lines referring to results for red and
blue primaries respectively.  At low satellite mass there is excellent
agreement between the solid and dashed curves, indicating that satellite
abundance does not depend on central galaxy colour at fixed halo mass.  For the
most massive haloes there are few blue primaries in the MS-II, so the red
dashed curve is quite noisy below $\log M_\star/M_\odot=9.5$. For massive
satellites there are obvious discrepancies between the dashed and solid
curves, but these result from our sample definition. At given halo mass, red
primaries have smaller stellar masses and so substantially lower luminosities
than blue ones. Our isolation criteria then imply a correspondingly lower
upper limit on the stellar mass of satellites for the red primaries. The
effect is largest for the lowest mass haloes.

\cite{2007MNRAS.382.1901S} used the galaxy formation simulation of
\cite{2006MNRAS.365...11C}, also based on the Millennium Simulation,
to study the relation between primary luminosity/stellar mass and
satellite velocity dispersion, which should be a good diagnostic of
halo mass. Although they found a strong dependence of velocity
dispersion on galaxy colour at fixed primary luminosity, this
dependence almost vanished at fixed primary stellar mass. This appears
to contradict our results from the G11 simulation. At $\log
M_\star/M_\odot\sim 11$, where we find the biggest difference in
satellite abundance between red and blue primaries, almost a factor of
three, the difference in velocity dispersion in their figure 13 is at
most 20\%, corresponding to a factor of at most 1.7 in halo mass
(since $M_h\propto \sigma^3$).  The discrepancy could result from the
different isolation criteria adopted in the two studies, from
departures from a straightforward relation between 3-D velocity
dispersion within $r_{200}$ (the quantity considered by
\cite{2007MNRAS.382.1901S}), halo mass and projected satellite count
within 300~kpc (the quantity considered here), or from differences
between the two galaxy formation models.

So far we have characterized the abundance of satellites by the count
within $r_p = 300$~kpc (or $r_p=170$~kpc for the lowest mass bin in
figure~\ref{fig:number}) In order to better understand the relation
with halo mass, it is useful instead to consider the count within the
virial radius of the haloes, which we define as $r_{200}$, the radius
of a sphere within which the mean mass density is 200 times the
critical value.  We obtain such a measure using the equation,
\begin{equation}
N(r<r_{200})=N_p(r_p < 300~{\rm kpc})\times
\frac{M(r<r_{200})}{M_p(r_p< 300~{\rm kpc})},
\label{eqn:abunvir}
\end{equation}
where $M_p$ and $N_p$ are projected quantities and $M$ and $N$ the
corresponding 3-dimensional quantities. The projected abundance $N_p(r_p <
300~{\rm kpc})$ is taken directly from figure~\ref{fig:number}, and we
calculate the mass ratio on the {\it rhs} of equation~\ref{eqn:abunvir}
assuming an NFW profile . The
halo radius $r_{200}$ is known for each simulated galaxy, but can only be
inferred indirectly for the SDSS objects. We assume that the mean halo mass at
given stellar mass is the same in the observations as in the G11
simulations. The mass ratio also depends weakly on halo concentration which we
take from the model of \cite{2009ApJ...707..354Z}.

For each primary stellar mass bin in figure~\ref{fig:number} we can
then estimate a mean halo mass $\langle M_{200}\rangle$. We take the
mean satellite number per primary given by equation~\ref{eqn:abunvir}
and divide by this mean halo mass to obtain the abundance per unit
halo mass of satellites in the stellar mass range $10.0>\log
M_\star/M_\odot > 9.0$. Finally, we divide this quantity by the
abundance per unit (total) mass of galaxies in this same stellar mass
range in the Universe as a whole, taken from
\cite{2009MNRAS.398.2177L} for the SDSS data and from the G11
simulation as a whole for the mock data. The result is a measure of
the formation efficiency of low-mass galaxies as a function of their
present environment, as characterized by mass of the halo in which
they live.  Figure~\ref{fig:num_vir} shows this efficiency as a
function of mean halo mass for all primaries in the five stellar mass
bins of figure~\ref{fig:number}. There are three significant points to
take from this plot: (i) the formation efficiency of low-mass
satellite galaxies varies rather little with the mass of the halo in
which the galaxies are found today; (ii) in massive haloes this
efficiency is about 50\% larger than the efficiency for forming such
galaxies in the universe as a whole (remember that, globally, about
half the galaxies in this stellar mass range are satellites); (iii)
finally there is fair agreement between the formation efficiencies in
the simulation and in the real universe, although this is in part due
to our use of the simulation to assign halo masses to the observed
galaxies.

\begin{figure}
\epsfig{figure=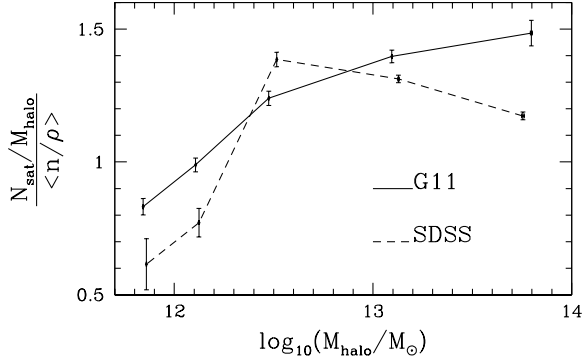,width=0.45\textwidth}
\caption{Mean number of satellites in the stellar mass range
  $10.0>\log M_\star/M_\odot>9.0$ per unit total halo mass, relative
  to the mean abundance per unit total mass of such galaxies in the
  Universe as a whole. Isolated primary galaxies are grouped into the
  same five stellar mass bins used in figure~\ref{fig:number}. Results
  for the G11 simulations and for the SDSS are shown by solid and
  dashed curves respectively. The mean halo mass has been calculated
  directly for each bin in the simulation data. Each SDSS primary is
  assigned a halo mass using the simulation relation between mean halo
  mass and primary stellar mass for all primaries. This plot hence
  shows the efficiency of low-mass galaxy formation as a function of
  present-day halo mass in units of the overall efficiency in the
  Universe as a whole.}
\label{fig:num_vir}
\end{figure}

\section{Satellite colour distributions}
\label{sec:colour}
So far we have studied the abundance of satellites as a function of the
stellar mass and colour of their primary and as a function of their own
luminosity and stellar mass. In this section we study how the {\it colours} of
satellite galaxies depend on the properties of their primaries.
Figures~\ref{fig:colour_obs} and~\ref{fig:colour_mock1} show cumulative colour
distributions for satellites in the SDSS and in the G11 catalogues
respectively, as a function of the stellar mass and colour of their
primary. The distributions are for satellites in two different stellar mass
ranges, as indicated by the labels above the relevant panels, and refer to all
satellites projected within 300~kpc.  The top, middle and bottom panels in
each column refer to satellites of all, of red and of blue primaries
respectively, while the different colours of the curves in each panel encode
the stellar mass range of the primary galaxies. The black curves which repeat
in all the panels of each column give the colour distributions for field
galaxies in the same stellar mass range as the satellites (calculated for SDSS
from all galaxies in the NYU-VAGC with $r<17.6$, and for G11 from all galaxies
within the simulation volume). The dashed horizontal line is merely a
reference to facilitate identification of the median colour.

A number of systematic trends are evident in these plots. Concentrating first
on the observational results in figure~\ref{fig:colour_obs}, we see that more
massive primaries have redder satellites within 300~kpc (in every panel the
curves are ordered cyan-blue-green-red from top to bottom), that low-mass
satellites are bluer than high-mass ones (the curves in the right panels are
always bluer than the corresponding curves in the left panels), satellites are
systematically redder than field galaxies of the same mass, except possibly
for the lowest mass primaries (the coloured curves almost always lie below the
corresponding black curves), and red primaries have redder satellites than
blue primaries of the same stellar mass (every coloured curve in the lowest
panels is bluer than the corresponding curve in the middle panel). This last
trend is the ``galactic conformity'' effect pointed out by
\cite{2006MNRAS.366....2W}.

If we now compare with the simulation results in figure~\ref{fig:colour_mock1}
we see that the same four systematic trends are present. More massive
primaries have redder satellites; lower mass satellites are bluer; satellite
galaxies are redder than field galaxies of the same stellar mass; and red
primaries have redder satellites than blue primaries of the same stellar mass.
However, there is an obvious discrepancy in that simulated satellites are
systematically redder than observed satellites. This is true for all primary
and satellite masses, but is particularly marked for lower mass and red
primaries, and for lower mass satellites. Clearly, the theoretical model of
G11 suppresses star formation much more effectively in such satellites than is
the case in the real universe. This echoes the conclusions of
\cite{2006MNRAS.366....2W} about the earlier models of \cite{2007MNRAS.375....2D}.  
The excessive reddening of the simulated satellite population reduces but does
not eliminate all the other trends mentioned above.

The galactic conformity phenomenon has been discussed in a number of previous
publications
\citep[e.g.][]{2006MNRAS.366....2W,2008MNRAS.389...86A,2010MNRAS.409..491K,
2011MNRAS.417.1374P}. \cite{2006MNRAS.366....2W} showed that, among groups of
given luminosity (which they considered a proxy for halo mass), those with an
early-type central galaxy have a larger fraction of early-type satellites.
They considered several physical processes which might be responsible for this
(halo and/or galaxy mergers, ram-pressure stripping, strangulation,
harassment...), focusing on whether these processes could alter galaxy
morphology. However, as discussed in some detail by \cite{2004MNRAS.353..713K}
and re-emphasized in the context of galactic conformity by
\cite{2010MNRAS.409..491K}, it is important to separate star formation
activity, stellar mass and galaxy structure when analyzing the influence of
environment on galaxy properties. Typical classifications into ``early'' and
``late'' types mix aspects of all of these.  The conformity effects we see
here are for given central galaxy stellar mass (rather than luminosity), are
within a fixed projected radius (300~kpc), and refer specifically to the {\it
  colours} of satellites and primaries. It seems possible that they could be
due at least in part to the tendency for red centrals to have more massive
haloes than blue ones, together with the trends for satellites to get redder
with increasing halo mass and decreasing $r/r_{200}$.

This can be checked directly for the simulated galaxy catalogues of G11.  In
figure~\ref{fig:colour_halo} we again show cumulative satellite colour
distributions for two different satellite mass ranges and for all, for red and
for blue primaries, but now in bins of host halo mass rather than of primary
stellar mass.  In this plot, the colour distributions are calculated for all
satellites projected within the halo virial radius rather than for a fixed
projected radius of 300~kpc. The halo mass ranges have been chosen to
correspond approximately to the primary stellar mass ranges we have been using
in previous plots. For the two high-mass bins, the colour distributions show
no significant dependence on the colour of the central galaxy, but for the two
low-mass bins the dependence on primary colour, while smaller than in
figure~\ref{fig:colour_mock1}, is clearly still present.  Thus there must be
physical processes in the model which contribute to the ``galactic
conformity'' phenomenon in addition to those which result in red primaries
having more massive haloes than blue primaries of the same stellar mass.

We have analyzed our simulation to identify candidates for these additional 
processes. In figures~\ref{fig:bh} and~\ref{fig:time} we plot distributions 
for the two which show the strongest trends. 
Figure~\ref{fig:bh} shows the distributions of black hole
mass and of hot gas mass for haloes in these same four mass ranges, split
according to the colour of the central galaxy. Figure~\ref{fig:time} shows
similar plots for the cumulative distributions of infall redshift (defined as
the redshift when a satellite last entered the virial radius $r_{200}$ of the
main progenitor of its halo) for satellites with stellar mass above
$10^9M_\odot$. In all cases the distributions depend strongly on central
galaxy colour for haloes in the two lower mass ranges where galactic conformity
effects are substantial, but at most weakly for higher mass haloes where these
effects are absent.  For $\log M_{\rm h}/M_\odot < 12.8$, haloes with red
central galaxies have more hot gas, more massive central black holes and
earlier satellite infall redshifts than haloes of the same mass with blue
central galaxies. Notice that the black hole mass distribution for blue
centrals is bimodal in the lower left panel of figure~\ref{fig:bh}.  Clearly,
there is a transition in this halo mass range between blue centrals with
high-mass black holes (about 0.3\% of their stellar mass, as in the upper
panels) and blue centrals with low-mass black holes (roughly an order of
magnitude smaller as a fraction of stellar mass, as in the lower right
panel). The hot gas fractions are similar to the overall cosmic baryon
fraction ($\sim 17\%$) for haloes with red central galaxies and about a factor
of two smaller in the lower mass haloes with blue centrals. The transition
halo mass is approximately the mass at which both observational and simulated
samples of isolated galaxies become dominated by red objects (see
figure~\ref{fig:frac}).

Given the modeling assumptions of G11, these systematic differences mean
that, for $\log M_{\rm h}/M_\odot < 12.8$, haloes with red central galaxies
have gas cooling rates which are twice but ``radio mode'' heating rates which
are 20 times those of similar mass haloes with blue central galaxies. Thus,
their central galaxies are clearly red because feedback has quenched their
growth, and this is why their stellar masses are smaller than those of the
blue centrals which have continued to grow to the present day. The redness of
the satellite population in haloes with red centrals can be traced to the
facts that the satellites are accreted earlier and orbit through a denser hot
gas medium.  Thus both tidal and ram pressure stripping processes are more
effective, and the satellites have had longer to exhaust any remaining
star-forming gas. Finally, the earlier assembly indicated by the higher
satellite infall redshifts presumably explains why these particular haloes
have red centrals, bigger black holes and more hot gas.

Thus, at least in the models, it is clear why there is galactic conformity at
given halo mass. The more easily observable galactic conformity at given
central stellar mass is predicted to be stronger as a result of the
combination of these effects with the tendencies for for red centrals to have
more massive haloes at given stellar mass than blue ones, and for more massive
haloes to have redder satellites. Yet stronger conformity effects are
expected at given central galaxy luminosity, since blue central galaxies have
smaller stellar masses (and thus even smaller halo masses) than red central
galaxies of the same luminosity.

\begin{figure}
\epsfig{figure=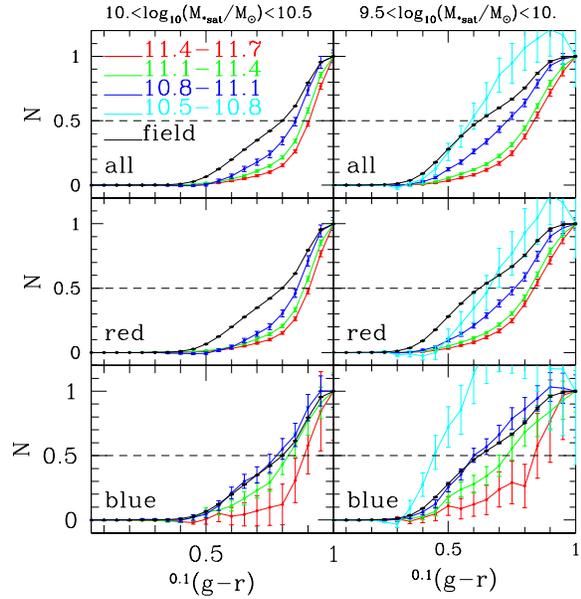,width=0.45\textwidth}
\caption{Cumulative colour distributions for satellites projected within
  300~kpc of their primary as a function of primary stellar mass (indicated by
  line colour) for two ranges of satellite mass (left and right
  columns with the range indicated by the label above each column) and for
  all, for red and for blue primaries (upper middle and lower panels in each
  column).  Black lines show the cumulative colour distribution for a
  volume-limited sample of field galaxies derived from the SDSS DR7
  spectroscopic sample with $r<17.6$.}
\label{fig:colour_obs}
\end{figure}

\begin{figure}
\epsfig{figure=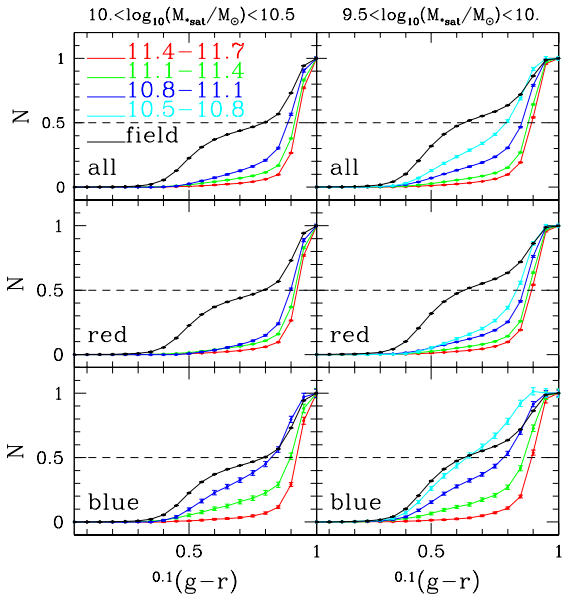,width=0.45\textwidth}
\caption{Similar  to figure~\ref{fig:colour_obs},  but for the simulated
  galaxy catalogues of G11. The black lines here are the colour distributions
  for all galaxies in the specified stellar mass ranges within the full
  simulation volume.}
\label{fig:colour_mock1}
\end{figure}

\begin{figure}
\epsfig{figure=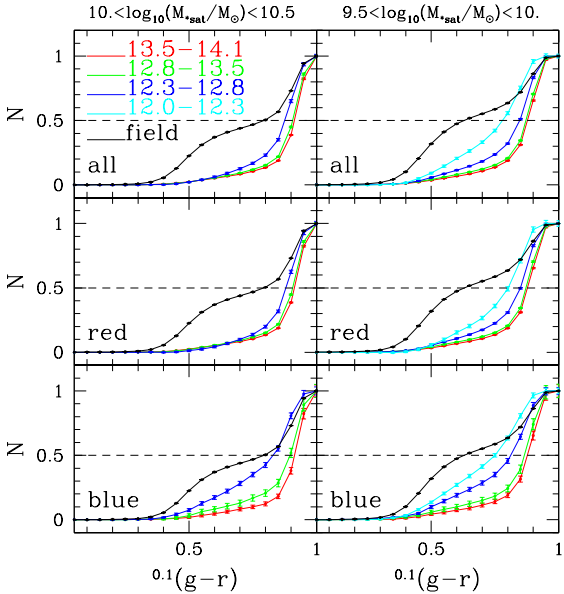,width=0.45\textwidth}
\caption{Colour distribution of satellites projected within the halo virial
  radius $r_{200}$ of all primaries, of red primaries and of blue primaries
  (top, middle and bottom rows) as a function of host halo mass (indicated by
  line colours corresponding to the ranges of $\log M_{\rm h}/M_\odot$ given
  in the legend) in the simulated galaxy catalogues of G11. Results are again
  shown for two ranges of satellite stellar mass.}
\label{fig:colour_halo}
\end{figure}

\section{Summary and Conclusions}
\label{sec:conclusion}
We have used a photometric catalogue of SDSS/DR8 galaxies brighter
than $r=21$ to study the satellite populations of 41271 
 isolated galaxies with $r<16.6$ selected from the
SDSS/DR7 spectroscopic catalogue.  In particular, we have studied how
the abundance of satellites as a function of luminosity, stellar mass
and colour depends on the stellar mass and colour of the central
galaxy.  Our study differs from other recent SDSS-based studies of
satellite galaxies \citep{2011AJ....142...13L,2011MNRAS.413..101G} in
the size of the sample analyzed, in our primary focus on systematics
as a function of stellar mass, and in our detailed comparison with the
predictions of simulations of the evolution of the galaxy population
in the concordance $\Lambda$CDM cosmology. In general, our results
confirm and extend those obtained earlier, but the comparison with
simulations allows us to identify the likely physical cause of most of
the effects we see, and to isolate those which do not have a natural
explanation within our current theory of galaxy formation.

Our observational samples and analysis procedures allow us to measure the
properties of the satellite population in an unbiased way down to absolute
magnitudes $M_{^{0.1}r}\sim-14$ and stellar masses $\log M_\star/M_\odot
\sim8$. Our main observational conclusions are as follows:

\begin{itemize}

\item Satellite luminosity and stellar mass functions have shapes consistent
  with those of the general field galaxy population only around the highest
  stellar mass primaries $\log M_\star/M_\odot > 11.4$. These are all
  brightest cluster galaxies. For lower mass primaries, these functions become
  progressively steeper, even after accounting for the bright-end cut-off
  induced by our isolation criteria. This steepening is more marked for the
  stellar mass functions than for the luminosity functions because observed
  satellites get bluer as their stellar mass decreases. 

\item The mean abundance of satellites increases strongly with primary 
  stellar mass, approximately as expected if the number of satellites is 
  proportional to dark halo mass.

\item For $\log M_\star/M_\odot > 10.8$, red primaries have more
  satellites than blue primaries of the same stellar mass. The effect exceeds
  a factor of two for $\log M_\star/M_\odot \sim 11.2$. This is reminiscent of
  the result of \cite{2006MNRAS.368..715M} who showed that at high stellar
  mass, red central galaxies have more massive haloes than blue ones.  This 
  trend could in part be due to colour dependent errors in deriving 
  stellar masses from the photometry, but such errors would need to be quite 
  large and to depend on primary mass.

\item Satellite galaxies are systematically redder than field galaxies of the
  same stellar mass except around blue primaries with $\log M_\star/M_\odot <10.8$
  where the satellites can have similar colours or even be systematically
  bluer than the field (i.e. the galaxy population within a large
  representative volume).

\item The satellite population is systematically redder around more massive
  primaries, for more massive satellites and around red primaries. The first
  effect reflects the fact that cluster galaxies are systematically redder
  than field galaxies, the second echoes the trend found in the general
  field, and the third is the galactic conformity effect pointed out by
  \cite{2006MNRAS.366....2W} but measured here for fixed central stellar mass
  rather than fixed central luminosity.

\end{itemize}

We used criteria directly analogous to those employed on the SDSS to construct
an isolated galaxy sample from the $z=0$ output of the publicly available
galaxy formation simulations of \citet[][G11]{2011MNRAS.413..101G}. These are based on the
Millennium and Millennium-II Simulations. The mock catalogue contains similar
magnitude, stellar mass, colour and position/velocity information to the real
catalogue, but also contains information about dark haloes and the location of
the galaxies within them. Based on the mock sample, we conclude that $\sim
98\%$ of our isolated galaxies are the central objects of their dark
haloes. Both in the SDSS and in the mock catalogue, the distributions of
intrinsic properties for the isolated and parent populations are very similar.
Only the colour distributions shift slightly, with the isolated galaxies being
systematically bluer than the full population. A detailed comparison of the
mock and real samples leads to the following conclusions

\begin{itemize}

\item At all primary masses, the luminosity and stellar mass functions of G11
  satellites are quite similar both in shape and in normalization to those
  measured for SDSS.  However, in the simulation there is only a weak tendency
  for the satellite functions to be steeper than those of the field, or to be
  steeper for lower mass primaries. This disagrees with the SDSS where the
  steepening with decreasing primary mass is quite marked.

\item In the mock catalogues the abundance of satellites increases with
  primary stellar mass almost in proportion to mean halo mass. For high-mass
  haloes, the abundance per unit mass of satellite galaxies is about 1.5 times
  the value for the universe as a whole.
 
\item For $\log M_\star/M_\odot <11.4$, red simulated primaries have more
  satellites then blue ones of the same stellar mass. The effect is similar in
  strength to that seen in SDSS but continues to lower primary mass. In the
  simulation it is due entirely to red primaries having more massive haloes
  than blue ones. At fixed {\it halo} mass the abundance of faint satellites is
  independent of the colour of the primary, but red primaries have lower
  stellar masses because of their truncated star formation histories.

\item Satellite galaxies in the G11 simulations are systematically redder than
  in the SDSS. The effect is particularly marked for lower mass
  satellites and around lower mass primaries. The modeling improvements
  introduced by G11 to address this issue are apparently insufficient to
  fully solve it. Star formation is still terminated too early when galaxies
  become satellites.

\item Despite this overall shift, the colours of simulated satellites depend
  on the colour and stellar mass of their primary and on their own stellar
  mass in very similar ways as in the SDSS. Satellites are systematically
  redder if they are more massive, if their primary is more massive, and if
  their primary is red. The first trend echoes that found for field galaxies,
  while the second reflects the fact that more massive haloes contain redder
  satellites. The third trend, the galactic conformity effect, is caused by
  redder primaries having more massive haloes at fixed stellar mass, and
  denser hot gas atmospheres (hence more effective ram-pressure stripping) at
  fixed halo mass.

\end{itemize}

Satellites go red in the G11 simulations because they lose their source of new
cold gas. Once they fall within the virial radius of their host, their hot gas
reservoirs are gradually removed by tidal and ram-pressure stripping and no
new material is added by infall. Star formation uses up the remaining cold
gas and then switches off. Clearly this happens too quickly (see
\cite{2011MNRAS.416.1197W} and also \cite{2007MNRAS.377.1419W} for an explicit
demonstration of how lengthening the relevant timescales can cure the problem)
so improving the model will require changing the star formation assumptions to
increase the time to gas exhaustion (for example, by removing the threshold
gas surface density for star formation) or providing new sources of fuel (for
example, by including the gas return from stellar evolution). Once this is
fixed, it seems likely that the other trends of satellite colour with
environment will be well matched.

The other clear discrepancy between the SDSS data and the simulation is the
steepening of satellite mass and luminosity functions as one goes to fainter
primaries. This is a relatively strong effect in the real data (and is visible
also in the analysis of \cite{2011MNRAS.413..101G}) but does not occur at a significant
level in the simulations. Thus, it must reflect not merely the statistics of
hierarchical clustering in a $\Lambda$CDM cosmology, but in addition some
difference in the star-formation histories of satellite galaxies living in
different mass host haloes. This may be related to the discrepancy noted by
G11 -- comparisons with high-redshift data suggest that low-mass galaxies form
too early in their simulations -- but this can only be checked by further
simulation work. It is clear that the detailed and relatively precise
statistical information provided by large-sample studies of satellite galaxies
is useful for testing and refining our understanding of how galaxies form.
In a follow-up paper we will extend our current study by considering 
satellite galaxy properties as a function of distance from the primary,

\begin{figure*}
\centerline{                \epsfig{figure=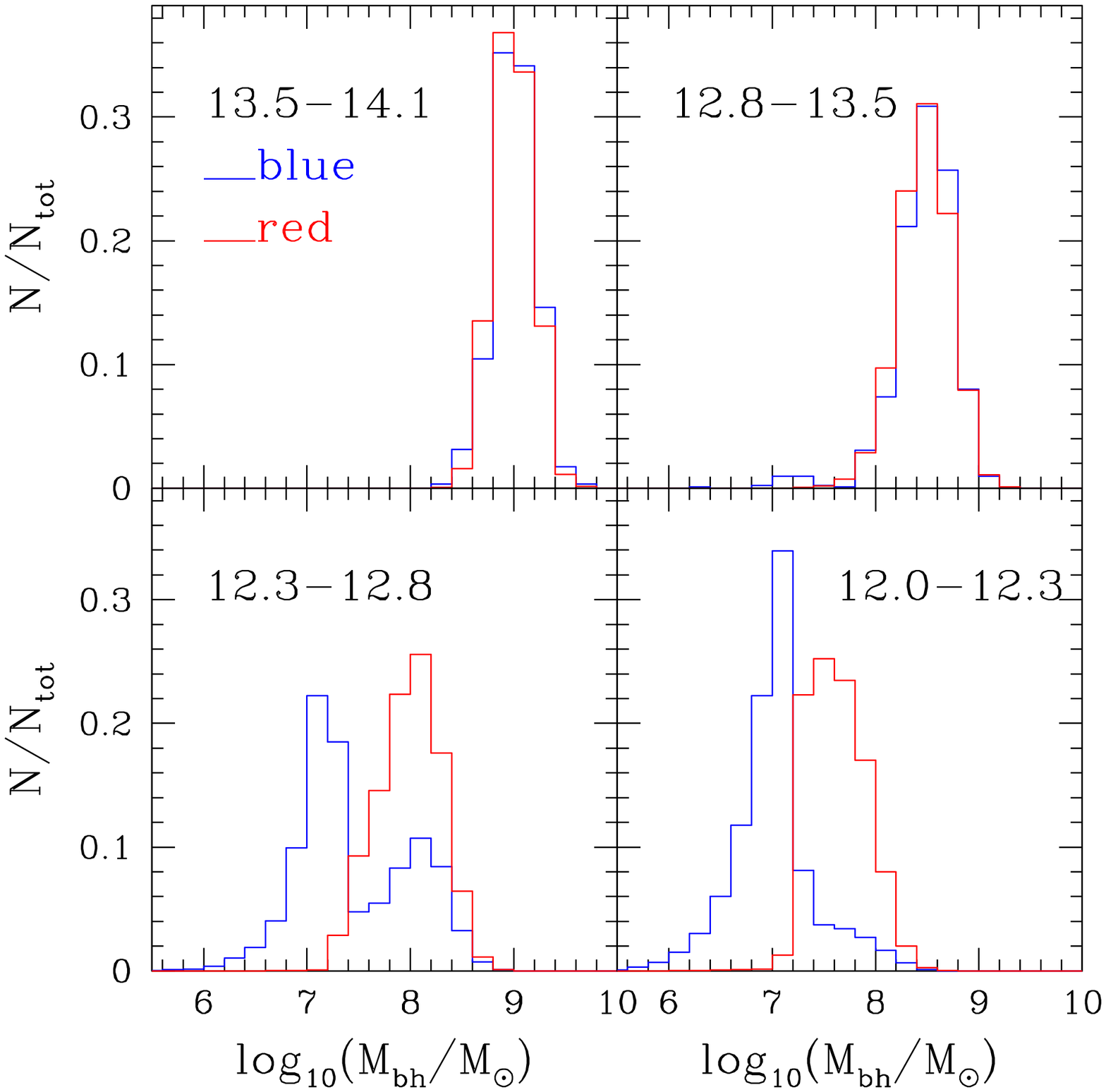,width=0.45\textwidth}
  \epsfig{figure=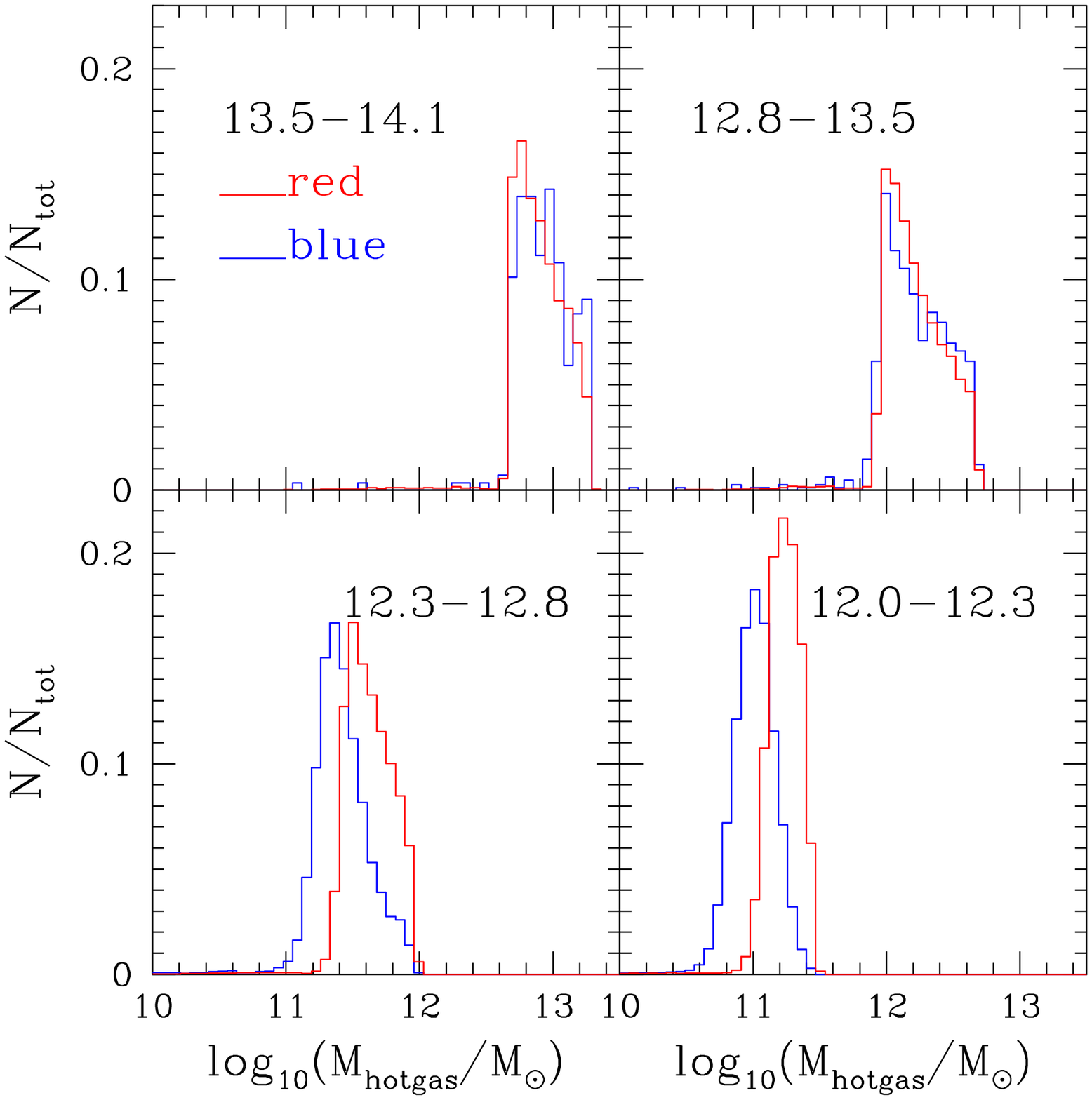,width=0.45\textwidth} }
\caption{{\bf Left:} Central black hole mass distributions for the same four
  ranges of halo mass used in figure~\ref{fig:colour_halo} and split according
  to the colour of the central galaxy.  {\bf Right:} Hot gas mass
  distributions for the same four sets of haloes and again split according
  to the colour of the central galaxy.}
\label{fig:bh}
\end{figure*}

\begin{figure}
\epsfig{figure=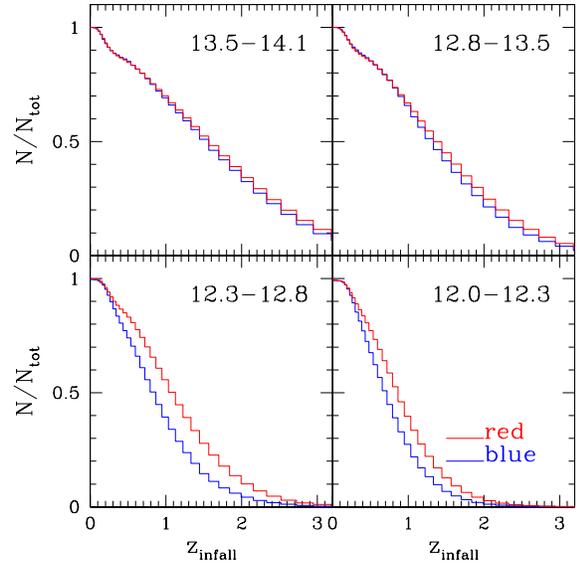,width=0.45\textwidth}
\caption{Infall time distributions for satellites in the same four ranges of
  halo mass used in figure~\ref{fig:colour_halo} and again split according to
  the colour of the central galaxy.}
\label{fig:time}
\end{figure}

\section*{Acknowledgements}
We gratefully thank Rachel Mandelbaum for useful discussions about SDSS systematics, Xu Kong for supplying 
the COSMOS mask and Cheng Li for discussions about details of the NYU-VAGC and photoz2 catalogues. 
Wenting Wang is partially supported by NSFC (11121062, 10878001, 11033006,11003035), and by the CAS/SAFEA 
International Partnership Program for Creative Research Teams (KJCX2-YW-T23).

\appendix

\section{Data Systematics and Completeness}
\subsection{SDSS Photometry Systematics}
\label{appendix:a}

\begin{figure*}
\centerline{             \epsfig{figure=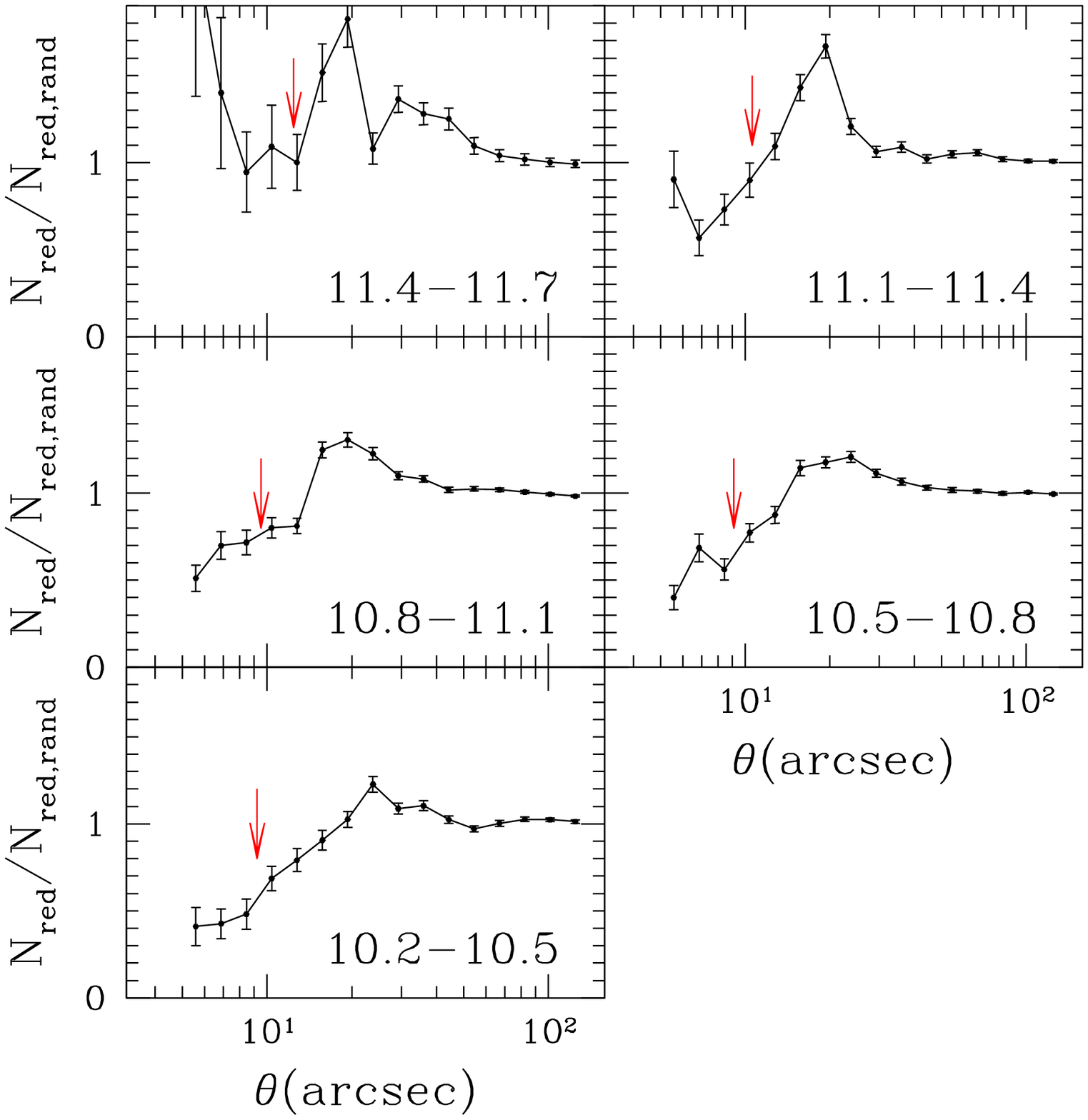,width=0.45\textwidth}
  \epsfig{figure=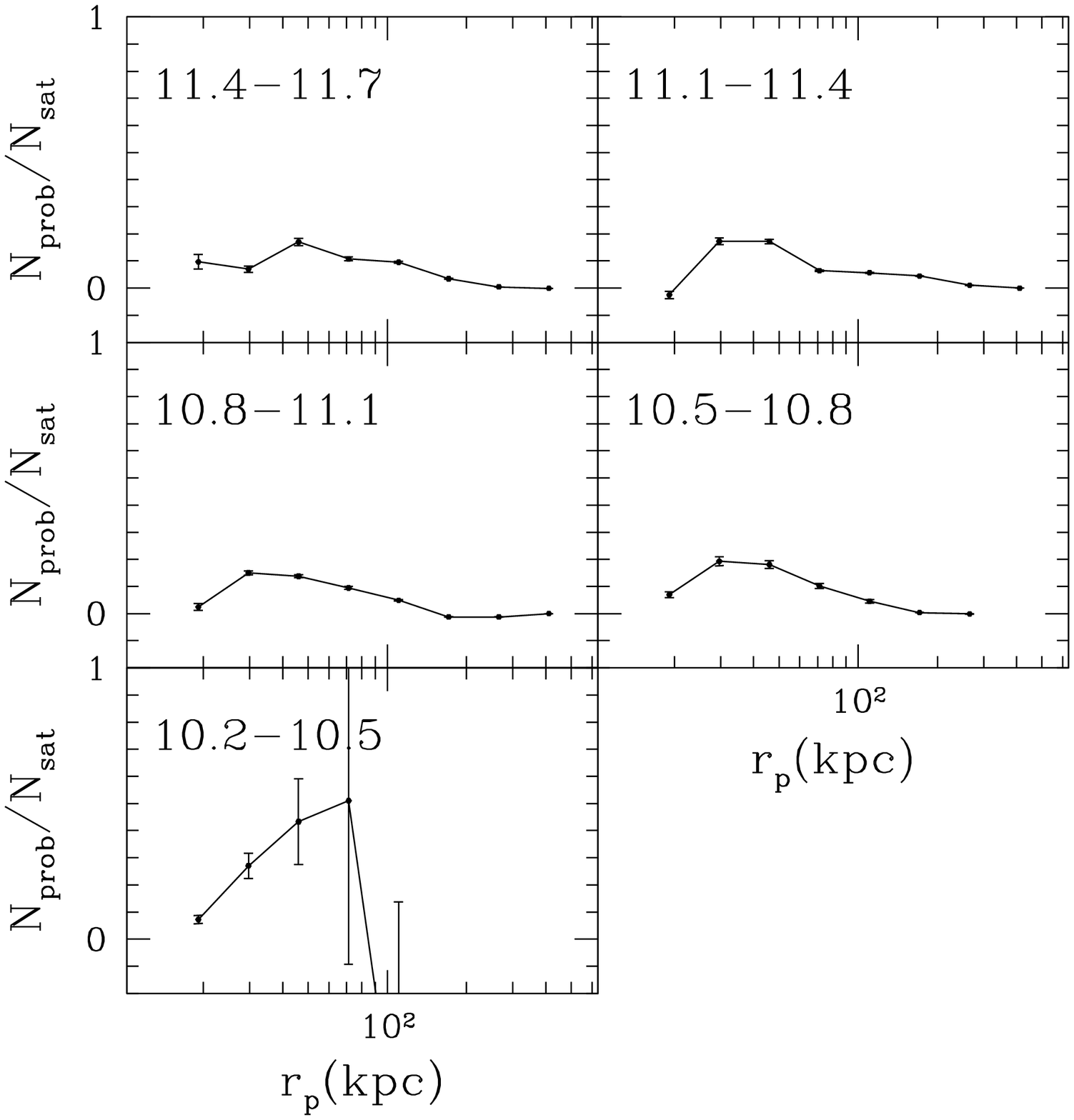,width=0.45\textwidth} }
\caption{{\bf Left:} The ratio of the number profile of ``red'' background
  galaxies with $r<21$ around our primary sample to a similar profile for the
  same primaries made after randomizing the positions of the photometric
  galaxies within the survey footprint. The five panels refer to our five
  ranges of $\log M_\star/M_\odot$ as indicated by the labels. The arrows show
  $\langle\theta_{90}\rangle$, the mean angular size of the primaries in each
  panel.  {\bf Right:} The ratio of our estimate of the number of ``false''
  satellites brighter than $r=21$ resulting from systematic errors in
  background subtraction to our estimate of the total number of satellites
  above this same apparent magnitude limit. The error bars are obtained by
  propagation of those in the left panel, together with bootstrapping the
  primary sample when estimating the satellite+background counts.}
\label{fig:photo}
\end{figure*}

As mentioned by \cite{2005MNRAS.361.1287M, 2006MNRAS.372..758M} and also
discussed in the SDSS/DR8 paper \citep{2011ApJS..193...29A}, the magnitudes of
faint objects close to brighter galaxies are affected by systematic problems
related to sky subtraction and deblending. This results in an apparent over-
or underabundance of faint galaxies close to bright ones.  This is potentially
a critical problem for our study, since we evaluate the mean number of
satellites surrounding some set of primaries by counting all faint neighbours
in the photometric catalogue and then subtracting the number of background
objects that should be projected onto these same regions ``at random''. We
therefore test explicitly for such affects in the DR8 photometry we will use.

We take our set of isolated primary galaxies and use the techniques
described in the main body of our paper to count ``red'' neighbours
brighter than $r=21$ in a set of annuli of fixed angular scale. We
also make similar counts using a fake photometric catalogue built by
randomizing the positions of the actual photometric galaxies within
the survey footprint.  We then accumulate both sets of counts for each
annulus and for all primaries within each of the five stellar mass
bins used in figure~1, and we divide the real counts by the random
counts to get a normalized profile. Error bars are estimated by
bootstrapping the sample of primaries. Our definition of ``red'' is
that the colours of the photometric galaxy should be too red for it to
be at the same redshift as its primary. We estimate a stellar mass
$M_\star$ for the photometric galaxy {\it assuming} it to be at the
redshift of the primary (see section~\ref{sec:method}).  We then consider 
any photometric galaxy with $^{0.1}(g-r)>0.032\mathrm{log}_{10}M_\star+0.73$ 
to be a background object. (This is based on the upper envelope of a
rest-frame colour versus stellar mass scatter plot for SDSS galaxies
with measured redshift.)  Our normalized profile is thus expected to
be unity to within its statistical errors. The actual results are
shown in the left panel of figure~\ref{fig:photo} for our five primary
stellar mass bins.

All the profiles go nicely to unity at angular separations of an arcminute or
more, This shows that our colour cuts have indeed isolated a population of
background galaxies which has no significant spatial correlation with the
primaries. On the other hand, there are substantial deviations from unity at
smaller scale which can have either sign and are similar for the different
primary mass ranges. These most likely reflect photometric errors caused by
the effect of the bright galaxy on the local sky background estimate used when
measuring the apparent magnitude of its faint ``red'' neighbour
\citep{2006MNRAS.372..758M}. At $r=21$ the logarithmic slope of the integral
$r$-band number counts is 0.34 dex per magnitude, so a change in the
apparent surface density of background objects by a factor of 1.5 requires an
effective shift of about 0.5 mag in the apparent magnitude
scale.\footnote{Gravitational lensing could also change the apparent density
  of background objects, but simple estimates suggest the effect is too small
  to be relevant.}  Note that the typical angular size of the primaries
(characterized by $\langle\theta_{90}\rangle$, where $\theta_{90}$ is the
angular scale containing 90\% of the Petrosian luminosity, and indicated by
the arrows in the individual panels of figure~\ref{fig:photo}) is quite
similar for each of our stellar mass bins and is comparable to the separations
where we see substantial systematic effects.

These problems will affect the SDSS photometry both of true satellites and of
background galaxies, presumably by similar amounts. We will neglect the shifts
in satellite magnitudes in the following since they are only substantial at
separations below 20 arcsec, and so will affect very few objects. A more
serious issue is whether the apparent excess (or deficit) of background
galaxies will cause our background subtraction to fail, since this assumes the
number density of background galaxies to be statistically uniform and
independent of position on the sky relative to the primary. We estimate the
size of the problem as follows.  We construct a random sample by taking the
full photometric catalogue and randomizing the positions of the galaxies
within the survey footprint, keeping all their other properties fixed. We then
accumulate the total count $N_{\mathrm{tot}}$ of random galaxies with $r<21$
in each radial annulus for all the primaries in each mass bin.  We multiply
$N_{\mathrm{tot}}$ by the factor $N_{\rm red}/N_{\rm red,rand}-1$.  (This is
the quantity on the $y$-axis in the left panel of figure~\ref{fig:photo} minus
unity.)  The result is our estimate $N_{\rm prob}$ of the number of apparent
satellites brighter than $r=21$ resulting from incorrect background
subtraction.  This systematic error will cause us difficulties if $N_{\rm
  prob}$ is comparable to the total number of satellites.

In the right panel of figure~\ref{fig:photo} we show the ratio of this
estimated number of spurious satellites to the total number of satellites we
find with $r<21$ when the techniques discussed in section~\ref{sec:method} are applied to the real
data.  It seems that this systematic error is not a problem for the four
massive primary bins where $|N_{\mathrm{prob}}|/N_{\mathrm{sat}}$ is less than
20\% on all scales. In the least massive bin, however, 
$|N_{\mathrm{prob}}|/N_{\mathrm{sat}}$ seems to be around 40\% to 50\% at $r_p>30$kpc, 
though with pretty large error-bars. For this primary mass range the number of 
satellites is very small, and thus even a small misestimation of the number of 
background galaxies can cause a relatively big effect. Looking at the scale 
range corresponding to $r_p>30$kpc in the left panel of that bin (this is 
roughly $\theta>30\arcsec$ for the relevant mean redshift $z\sim0.06$) we find 
that the background density modulation required is smaller than $10\%$, corresponding 
to an apparent magnitude error of $\sim 0.1$~mag. A shift of this order in magnitude 
due to background subtaction problems does not seem implausible.

We conclude that our results are insensitive to these small-scale photometric
problems except for the lowest stellar mass range we consider, where satellite
counts may be significantly affected. We do not attempt to correct this
systematic because we are unable to estimate sufficiently accurately how the
effect depends on the photometric properties (apparent magnitude, colour,
profile shape...) of the brighter and fainter images. One might hope that the
effects would become smaller at brighter apparent magnitudes, but we have been
unable to demonstrate this convincingly. We note that we made similar tests
using DR7 photometry, finding substantially larger effects than presented here
for DR8. This is our primary reason for choosing DR8 photometry when searching
for satellites, even though our primary sample was defined using DR7 data.

\subsection{Completeness in SDSS}
\label{appendix:b}

\begin{figure}
\epsfig{figure=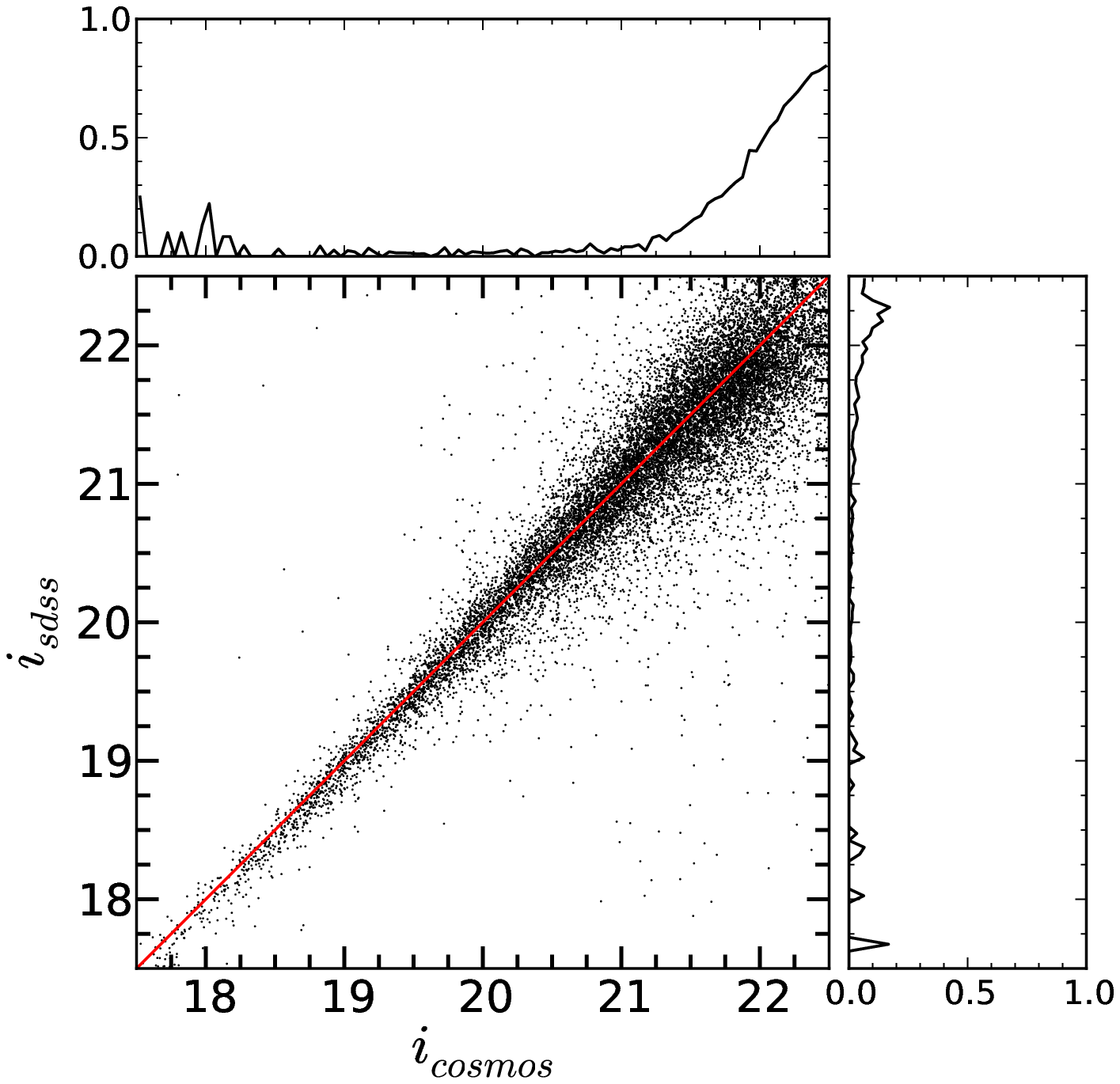,width=0.45\textwidth}
\caption{Scatter plot comparing SDSS and COSMOS i-band apparent
  magnitudes (including both galaxies and stars). Panels on the top
  and on the right show the fraction of COSMOS objects unmatched in
  SDSS and of SDSS objects unmatched in COSMOS, respectively.}
\label{fig:cosmos}
\end{figure}

In this paper we study the properties and the distribution of
satellite galaxies down to faint apparent magnitudes where
identification, classification and magnitude calibration could be
affected by a variety of observational factors such as seeing, sky
brightness and extinction, in addition to the issues of photometric
accuracy in the neighborhood of brighter objects which we addressed in
the previous section. In this section we focus on estimating
quantitatively the fraction of galaxies that might have been missed in
the SDSS catalogues. We test the accuracy of star-galaxy separation in
section~\ref{appendix:c}, of our background galaxy subtraction methods
in \ref{appendix:d} and the influence of additional observing factors
in section~\ref{appendix:e}.

We investigate the completeness of the SDSS/DR8 photometric catalogue
using the much deeper HST data available for the 2 squ.deg. COSMOS
survey.  We cross-match galaxies and stars in the COSMOS
field \footnote{We use 150.7~$>ra>$~149.5 and 2.72~$>dec>$~1.65 to
  exclude the irregular edge of the full survey.} to galaxies and
stars in SDSS/DR8 and examine the failed matches in both
directions. The COSMOS survey was initially conducted by HST in broad
$I$-band using the Advanced Camera for Surveys (ACS) Wide-Field Channel
(WFC) detector. The current COSMOS database also includes later
observations from a variety of telescopes. It reaches a limiting
magnitude of $I_{AB}=28$ for point source, and $I_{AB}=26$ for
galaxies of diameter $\sim0.5\arcsec$ \citep{2007ApJS..172...38S}.
This is much deeper than SDSS. The specific COSMOS catalogue used here
is the Zurich Structure and Morphology
catalogue\footnote{http://irsa.ipac.caltech.edu/data/COSMOS/tables/morphology/},
which contains the measurements presented in
\cite{2007ApJS..172..406S} and \cite{2007ApJS..172..434S}.

We further excluded COSMOS and SDSS objects which are outside the
region common to both the COSMOS and the SDSS masks. The SDSS mask
used here is made up of the ``spherical polygons'' provided on the
NYU-VAGC website, and is the same as used in the main body of our
paper when evaluating and correcting for edge effects. The COSMOS mask
was kindly provided by Xu Kong (private communication). Objects in the
Zurich COSMOS catalogue which have flag ACS$\_$CLEAN=0,
ACS$\_$MU$\_$CLASS=3 or JUNKFLAG=1 are also excluded. We consider
objects (galaxies or stars) to be matched in COSMOS and SDSS if their
angular separation is smaller than 1.5$\arcsec$.

Figure~\ref{fig:cosmos} is a scatter plot comparing the COSMOS and
SDSS $i$-band magnitudes of matched objects. For about $4\%$ of SDSS
objects with $r<21$ there are two matched candidates in COSMOS.  This
reflects the much higher resolution of the COSMOS data which allows
substantially better separation of close pairs of images, double stars,
merging galaxies and chance superpositions, than is possible with
SDSS. In such cases, we sum the flux of all the matched COSMOS objects
and compare to that of the single SDSS object.

The red straight line is a reference with
$\mathrm{i}_{cosmos}=\mathrm{i}_{sdss}$ to guide the eye. In general,
$i$-band magnitudes in COSMOS and SDSS agree with each other quite
well, with a bigger scatter for fainter objects. SDSS $r$-band
magnitudes of $r=19, 20, 20.5$ and 21 correspond roughly to $i$-band
magnitudes, $i=18.6, 19.6, 20.1$ and 20.6. For these SDSS values the
scatter in COSMOS magnitudes in figure~\ref{fig:cosmos} is 0.277,
0.293, 0.360 and 0.367 magnitudes, respectively.

The histograms at the top and right of figure~\ref{fig:cosmos} show
the fraction of objects in each survey which is unmatched in the other as
a function of $i$-band magnitude. It is expected that SDSS objects
should be lost at the faint end, because COSMOS is much deeper. At the
bright end, there are a few spikes, but the total number of objects at
the bright end is small. For i~$<20.5$, we fail to match about $1\%$ of
SDSS objects in the COSMOS catalogue. This is because the COSMOS mask
we are using is not identical to that used for the Zurich
catalogue. In particular, it eliminates smaller regions around bright
stars so we find a few SDSS objects in such regions which are not in
the Zurich catalogue.

At $i\sim 20.5$, there are also about $1.5\%$ of COSMOS objects which are unmatched in
SDSS. This is only 71 objects in total, and we have looked at both SDSS and
COSMOS images of these fields. We found that about 31 of these 71 objects
failed to match because the angular separation between
their centroid coordinates was bigger than 1.5$\arcsec$. In some cases the
objects are quite extended or have a close neighbour. Given the relatively poor
seeing in SDSS, the determination of the coordinates of faint objects can be
uncertain and so can result in failure to match to COSMOS.

On the other hand, about 40 of the 71 objects really do not have a
match in SDSS. Of these, about 1/3 do indeed have an object at or
close to their coordinates in the COSMOS image which SDSS failed to
recognize. The remaining 2/3 do not, however, have an object at or
close to their coordinates even in the COSMOS image.  The Zurich
Structure and Morphology catalogue also provides a cross-match with
the ground based catalogue of \cite{2007ApJS..172...99C}. Using the
cross-match information provided, we found 15 out of these 40 objects
do have a match in the \cite{2007ApJS..172...99C} catalogue, with 12
having angular separation between the two matched objects bigger than
1.5$\arcsec$. The remaining objects do not have a match in the
ground-based \cite{2007ApJS..172...99C} catalogue. We are not clear
what produced these seemingly spurious objects in the COSMOS
catalogue, but the fraction is indeed small ($<1\%$).

The overall conclusion of this exercise is that incompleteness in the
SDSS/DR8 photometric catalogue is at or below the 1\% level at the
apparent magnitude limit relevant for our analysis.  This is small
enough to be negligible.

\subsection{Star-Galaxy Separation in SDSS}
\label{appendix:c}

\begin{figure}
\epsfig{figure=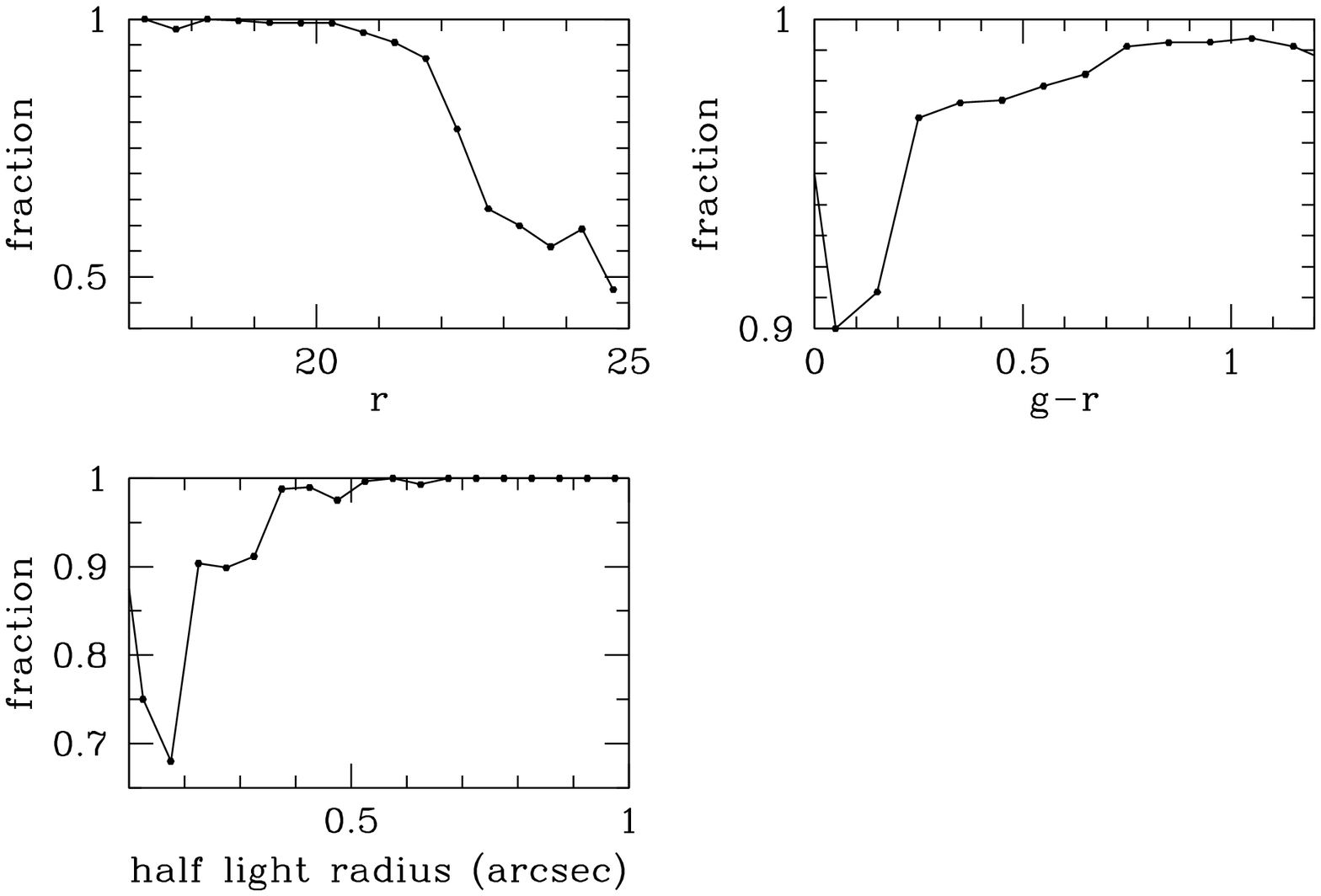,width=0.45\textwidth}
\caption{{\bf Top left:} The fraction of COSMOS galaxies which have a
  matched SDSS galaxy (as opposed to an SDSS ``star'') as a function
  of $r$-band apparent magnitude in SDSS. {\bf: Top right:} The
  fraction of COSMOS galaxies which have a matched SDSS galaxy with
  $r<21$, reported as a function of observed SDSS $g-r$ colour. {\bf
    Bottom left: } Similar to the top right panel, but showing the
  matched fraction as a function of angular size as measured in
  COSMOS, and again limited to objects with $r<21$ in the SDSS
  catalogue.}
\label{fig:cosmos2}
\end{figure}

SDSS carries out star-galaxy separation using the measured difference
between psf (Point Spread Function) and cmodel (Composite Model) magnitudes.  If psfMag-cmodelMag~$>$~0.145, an
object is classified as a galaxy, otherwise it is called a star.  The
quality of SDSS star-galaxy separation is thus unavoidably dependent
on observational conditions such as seeing and sky background, and is
also expected to be a function of galaxy apparent magnitude and
angular size. On the other hand, COSMOS should give much more reliable
star-galaxy separation because of its substantially greater depth, its
much smaller point-spread function and its more uniform observing
conditions. In this section we again use COSMOS data to evaluate the
quality of star-galaxy separation in the SDSS.

We take galaxies in the COSMOS field and cross-match them to SDSS
images, both galaxies and stars. In the match, we again used objects with
150.7~$>ra>$~149.5 and 2.72~$>dec>$~1.65 in order to avoid the COSMOS edge, and
we again excluded objects which are obscured by the mask of either survey.

For COSMOS galaxies that have a match in SDSS (either a ``galaxy'' or
a ``star''), we show in figure~\ref{fig:cosmos2} the fraction which
are classified as galaxies by SDSS as a function of SDSS $r$-band
apparent magnitude, SDSS $g-r$ colour and COSMOS GIM2D 
\citep[Galaxy Image Two-Dimensional][]{1998ASPC..145..108S} half-light
radius. The COSMOS GIM2D half-light radius is obtained by fitting a
\cite{1968adga.book.....S} model to the galaxy image
\citep{2007ApJS..172..434S}.

In the top left panel of figure~\ref{fig:cosmos2}, the fraction of
``correctly-classified'' SDSS galaxies is close to 1 at $r<20$ if we
take COSMOS as reference, and drops significantly beyond $r=21.5$. At
$r=21$, the fraction is about 95\%. There is some discussion about
star-galaxy separation on the official SDSS
website\footnote{http://www.sdss3.org/dr8/imaging/other\_info.php}. SDSS/DR1
was compared against the COMBO-17 survey (Classifying Objects by Medium-Band 
Observations in 17 Filters). This much older completeness
curve also drops significantly beyond $r=21.5$, and the fraction at
$r=21$ is about 95\%, in excellent agreement with our COSMOS test
here.

We now consider all COSMOS galaxies with matched SDSS apparent
magnitude $r<21$, and we investigate classification success as a
function of colour and angular size. The top right panel of
figure~\ref{fig:cosmos2} suggests that blue galaxies are more likely
to be misclassified than red ones, although the correct fraction is
always above 90\%. Only about 1\% of the reddest objects are
misclassified. We note that the number of bluish objects ($g-r<$~0.7) 
is only about one-fifth of the number of red objects. 

In the bottom panel of figure~\ref{fig:cosmos2}, we report galaxy
classification success as a function of angular size (Sersic
half-light radius from COSMOS in units of arcseconds.) The fraction
starts to drop for satellite half-light radii smaller than
0.4$\arcsec$, and is below $90\%$ at 0.2$\arcsec$, though never
dropping below 70\%. We emphasize that fewer than 100 objects in
COSMOS have $r_e <0.2\arcsec$, which is $\sim 2\%$ of all galaxies
with $r<21$. Only about 300 have $r_e<0.4\arcsec$. 

In summary, these results together imply that not more than about 1\%
of galaxies brighter than $r=21$ are misclassified as stars by SDSS,
which is a remarkably small number for a 2.5m telescope working on a
less than perfect site. They also imply that misclassification is a
negligible problem for our purposes.

\subsection{Background Subtraction Tests}
\label{appendix:d}

\begin{figure}
\epsfig{figure=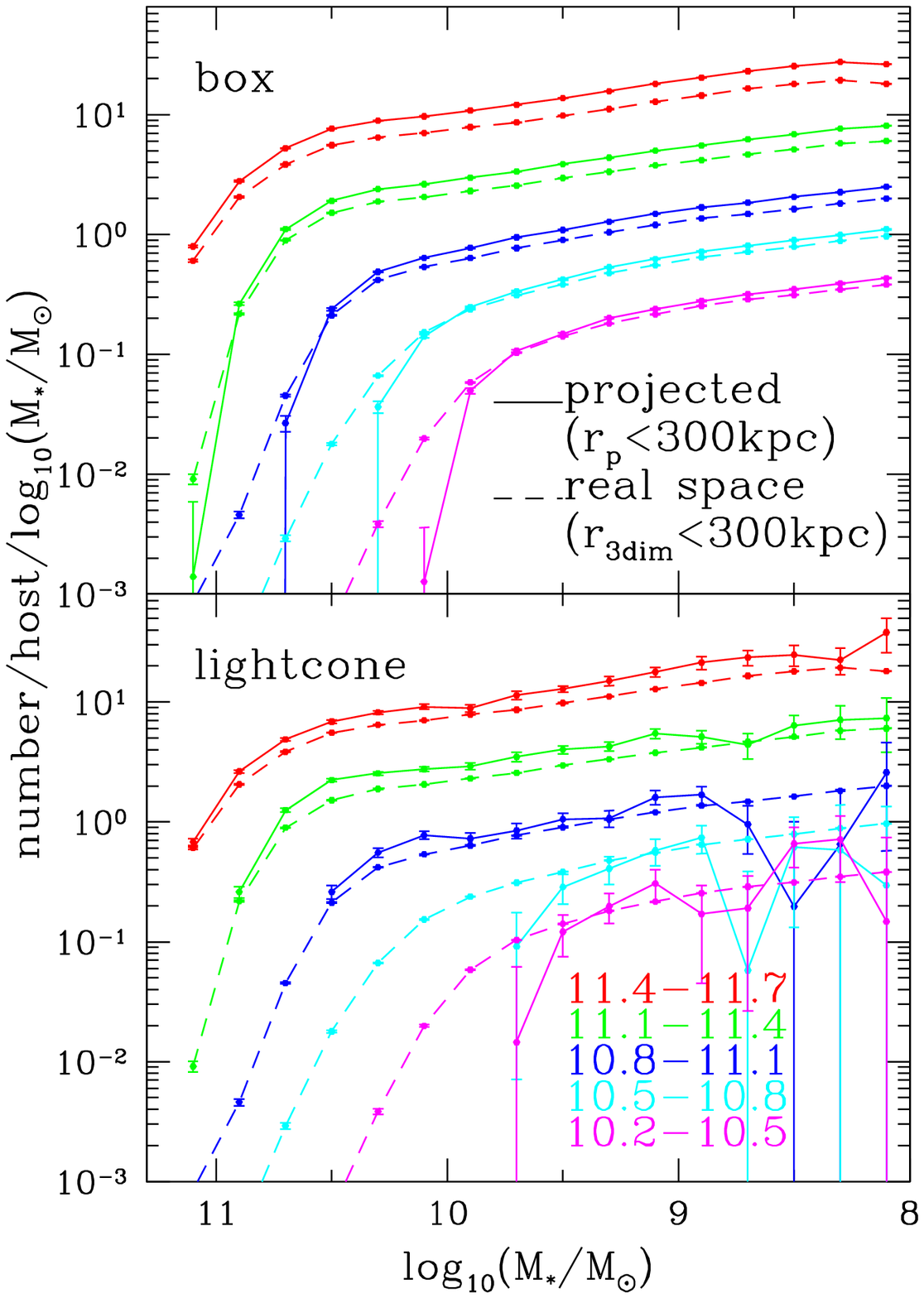,width=0.45\textwidth}
\caption{Tests of our background subtraction algorithms using stellar
  mass functions for satellites of isolated primaries in the G11
  simulation separated into five ranges of primary stellar mass as
  indicated by colour in the legend.  {\bf Top:} Results for isolated
  primaries defined in a projection of the full simulation as
  described in the main text and shown previously in
  figure~\ref{fig:MFall} (the solid curves) are compared with the
  ``true'' stellar mass functions defined as the mean count of
  satellites within 300 (or 170)~kpc of these same primaries in 3D
  (dashed curves). {\bf Bottom:} Results for isolated primaries defined
  in a full lightcone mock catalogue of SDSS/DR8 using exactly the same 
  criteria as for the real SDSS surveys and estimated using the identical
  code used for the SDSS data (solid curves). The dashed curves repeat the
  ``true'' answer from the upper panel. } 
\label{fig:bckgrd}
\end{figure}

All our satellite abundance measurements depend critically on our
ability to construct an unbiased estimate of the expected count of
unrelated foreground and background galaxies projected close to each
primary.  This background count often exceeds the count of true
satellites by a substantial factor, and the ratio of the two depends
strongly on distance from the primary and on the magnitudes of the primary
and satellite galaxies. As a first test of our procedures, we randomized
the positions of all galaxies in our SDSS/DR8 photometric catalogue within
the SDSS footprint, while retaining all their other properties. We then
repeated the analysis of luminosity and stellar mass functions which led
to figures~\ref{fig:LFall} and \ref{fig:MFall} keeping exactly the same sets of
spectroscopic primary galaxies and using the identical code. This null test
produced results which were consistent with zero within the estimated 
uncertainties for all satellite luminosity and stellar mass functions and 
around primaries in all five stellar mass ranges.

For our next tests, we used the fact that for the G11 simulation we
know the true properties of the satellite distribution. We took all
central galaxies in the model with stellar mass in the five mass 
ranges we use for our SDSS primaries and we compiled luminosity and
stellar mass functions for their true satellites, defined to be
companions within 300~kpc of the primary in 3-D.  We then compared
with the functions given in the right panels of
figures~\ref{fig:LFall} and \ref{fig:MFall} which were estimated in
projection after correction for background contamination. The result
for the stellar mass function is shown in the upper panel of
Fig.~\ref{fig:bckgrd}.  Although the normalizations differ slightly,
as expected because of the different geometry of the regions within
which satellites are counted in the two cases, the shapes of the
functions agree closely in all cases, demonstrating that our
``global'' correction does indeed remove the background properly from
our isolated galaxy sample.

Notice that this does not test aspects of our analysis concerned with
converting counts from the magnitude-limited SDSS catalogues to the
volume-limited statistics which we quote, for example, incompleteness
corrections due to the SDSS mask, effective volume corrections, and
K-corrections.  We therefore took an all-sky ``light-cone'' mock
catalogue constructed from the MS as described in
\cite{2012MNRAS.421.2904H}\footnote{This catalogue is available at
  http://www.mpa-garching.mpg.de/millennium.}. We applied our SDSS
mask and a magnitude limit of $r<21$ in order to create a mock
SDSS/DR8 catalogue. From this we created a mock SDSS/DR7 spectroscopic
catalogue including a model for incompleteness due to fibre collisions
and other observational problems. We then defined a sample of
primaries using exactly the same criteria as for the real SDSS data
(including a simple model of our use of photo-$z$'s to deal with
spectroscopic incompleteness when checking for isolation) and we
estimated satellite luminosity and stellar mass functions using the
same code as for the real data.  The results are shown in the lower
panel of Fig.~\ref{fig:bckgrd}.  Again agreement with the true answer
is good within the now somewhat larger error bars. This agreement 
provides a full end-to-end test of our code.

Nevertheless, for real data, there could be additional factors which
affect the accuracy of background subtraction.  For example, the
surface density of real galaxies on the sky could be modulated by
extinction, seeing, sky brightness or proximity to other objects (see
above), and thus affect our estimation of background from the whole
survey which assumes background galaxies to be distributed
uniformly. Thus it is important to test the quality of our background
subtraction directly on the real data.

To do so, we compare measurements of satellite mass functions for
projected distances between 300kpc and 1Mpc to measurements for
projected distances below 300kpc. In order to avoid complications
induced by our isolation criteria which are imposed at 500~kpc and
1.0~Mpc (see \cite{1976ApJ...205L.121F} for an example of the strong
effects of such criteria), when doing this test we use all SDSS
spectroscopic galaxies with r$<16.6$ without any further
selection. From our SDSS mock catalogue we estimate the mean surface
densities of true satellites in these two regions to differ by factors
ranging between 5.34 and 2.62 from the highest to lowest mass
primaries. Since the background level should be (and is assumed to be)
the same in the two regions and varies strongly with satellite mass,
any error in background subtraction is expected to show up as a
distortion in the shape of the satellite stellar mass function which
will be much more marked in the outer region.

Figure~\ref{fig:background_noiso} shows results for this test. Solid curves
represent satellite mass functions measured within $r_p=300$kpc, while
triangles are measurements for 1Mpc~$>r_p>$~300kpc. We renormalize the
1Mpc~$>r_p>$~300kpc results and replot them as dashed curves in order
to compare their shape with the $r_p<300$kpc results. The agreement
between the solid and dashed curves is quite good despite the fact
that the signal-to-background ratio changes by a factor of 3.5 on average. This
is both encouraging and interesting. Not only does it suggest that our
background subtraction is working well, the close similarity in shape
also demonstrates a significant regularity of the observed HOD, since
the $r_p<$300kpc results are dominated by satellites in the same halo
as the primary and the 1Mpc~$>r_p>$~300kpc results by companions in
other haloes, at least for the lower mass primaries.

\begin{figure}
\epsfig{figure=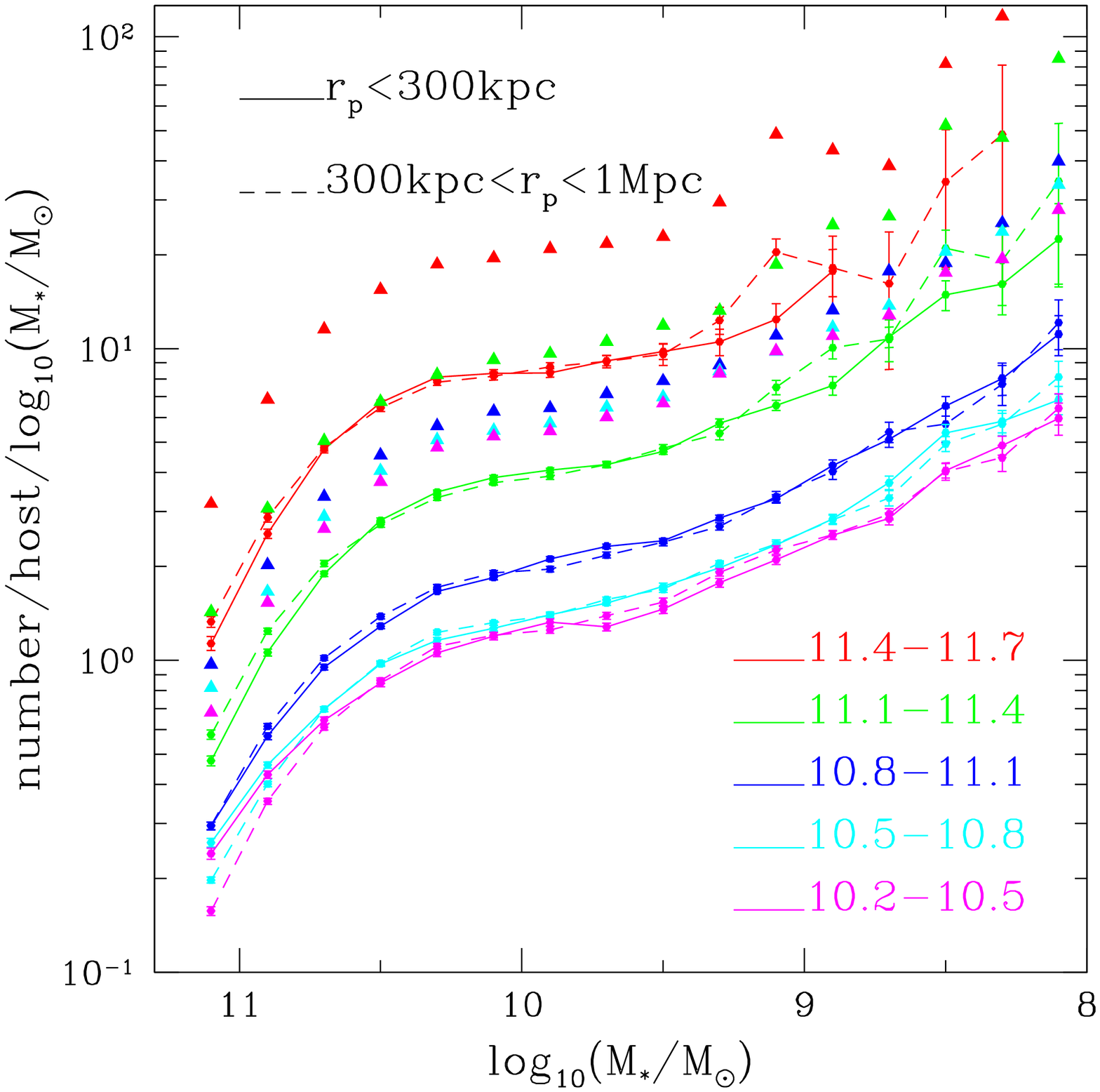,width=0.45\textwidth}
\caption{Stellar mass functions for companions projected within 300kpc
  (solid curves) and in the range 1Mpc~$>r_p>$~300kpc (triangles) of
  primary galaxies in five disjoint stellar mass ranges. The primaries
  in this plot are spectroscopic galaxies with $r<16.6$ but with no
  isolation criteria applied.  Curves and symbols of different colour
  correspond to different stellar mass ranges for the primaries, as
  indicated by the legend. Dashed curves show the 1Mpc~$>r_p>$~300kpc
  results renormalized in amplitude in order to compare their shape to
  the $<300$kpc results. }
\label{fig:background_noiso}
\end{figure}

\begin{figure}
\epsfig{figure=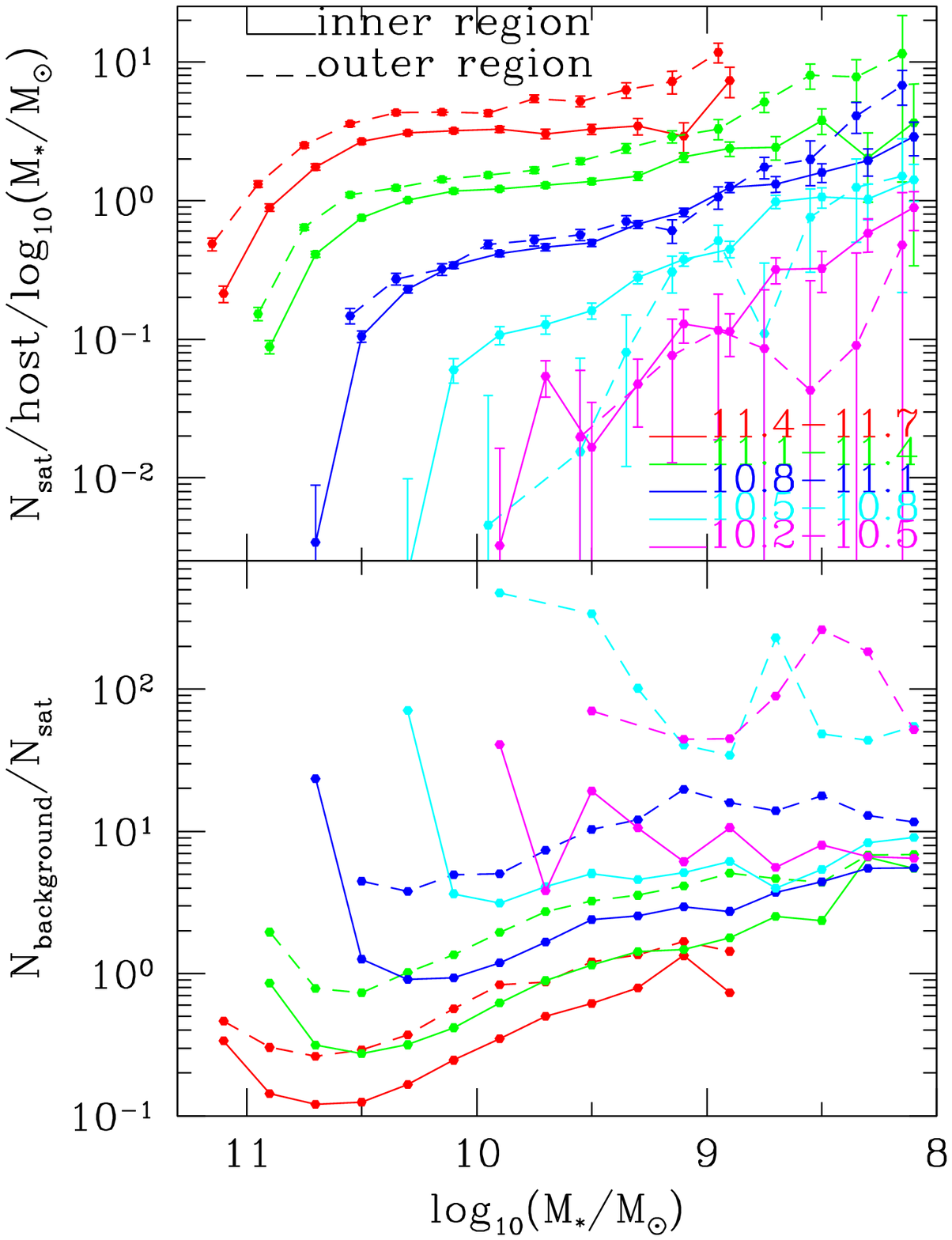,width=0.45\textwidth}
\caption{{\bf Top: }Stellar mass functions for satellites of isolated
  primaries separated into five ranges of primary stellar mass as
  indicated by colour in the legend. The results shown before in
  figure~\ref{fig:MFall} are split into disjoint inner (solid curves)
  and outer (dashed curves) regions as listed in
  table~\ref{tbl:radii}.  {\bf Bottom:} The ratio between the mean
  numbers of background and of satellite galaxies as a function of
  satellite stellar mass. Curves with different colour and line type
  have the same meaning as in the top panel. }
\label{fig:background}
\end{figure}

\begin{table*}
\caption{Boundaries of the subregions used for testing background subtraction.}
\begin{center}
\begin{tabular}{lccccc}\hline\hline
Range in primary $\log M_{\ast}/M_{\odot}$  & \multicolumn{1}{c}{11.7-11.4} & \multicolumn{1}{c}{11.1-11.4} & \multicolumn{1}{c}{10.8-11.1} & \multicolumn{1}{c}{10.5-10.8} & \multicolumn{1}{c}{10.2-10.5} \\ \hline
Inner region  & 15-150 (kpc) & 15-135 & 15-120 & 15-110 & 15-75 \\
Outer region  & 150-300 (kpc) & 135-300 & 120-300 & 110-300 & 75-170 \\
\hline
\label{tbl:radii}
\end{tabular}
\end{center}
\end{table*}

It is clearly important to carry out a similar test for the sample of
isolated galaxies which we analyses in this paper. For these objects,
we need to examine the accuracy of background subtraction within
300~kpc, since this is the outer radius to which we count companions
and lies well inside the smallest radius at which we apply an
isolation criterion (500~kpc).  We therefore split the region within
300~kpc into two disjoint subregions by radius. We choose the radius
separating the two subregions so that they would be expected to
contain equal numbers of satellites if these followed the projected
NFW profile inferred for the dark matter. We list in
table~\ref{tbl:radii} the radial boundaries we adopt for each of our
five different primary mass ranges.

The top panel of figure~\ref{fig:background} shows satellite stellar
mass functions as in figure~\ref{fig:MFall} but now split to display
results separately for the two subregions. Solid and dashed curves
give results for the inner and outer regions, respectively. Their
normalizations do not agree exactly, showing that the radial profiles
are not consistent with the NFW profiles we adopted when setting the
boundaries. For primaries in the two highest stellar mass ranges (the
red and green curves) the amplitude in the outer annulus is higher
than expected relative to that near the centre. For primaries in the
two lowest stellar mass ranges (the cyan and pink curves) the opposite
appears true, although the measurements are quite noisy. Thus, the
inner radial distribution of satellites is shallower than NFW for
massive primaries, and it could be steeper than NFW through the main
body of the halo for low mass primaries. We will leave detailed
discussion of satellite radial profiles to \cite{PaperII} which is in
preparation.

Despite the difference in amplitude, the dashed and solid red, green
and blue curves agree in shape quite well, except possibly at the
lowest masses where statistics are relatively poor. The dashed cyan and
pink curves are very noisy, but within their uncertainties they are
also consistent in shape with the solid curves.

The bottom panel of figure~\ref{fig:background} shows the ratio of the
mean number of background galaxies to the mean number of satellites as
a function of satellite stellar mass for each of the primary samples
shown in the top panel. Solid and dashed curves again refer to the
inner and outer regions.  For the most massive primaries (the red
curves) satellites substantially outnumber background galaxies at high
mass, and are comparable in number at low mass. For the green curves,
the background count already exceeds the number of satellites at
almost all masses, while for the blue, cyan and pink curves, it is
substantially larger, particularly for lower mass satellites and in
the outer annulus.  For these lower mass primaries the ratio of
background to satellites in the outer annulus is almost an order of
magnitude higher than in the inner region. For the blue curves, where
the mass functions have high signal-to-noise over the widest satellite
mass range, the background-to-satellite ratio ranges from unity to
about 5.5 in the inner region and from 4 to 20 in the outer
annulus. Give the large values of these corrections and their strong
variation with satellite mass and radius, it is gratifying that the
corresponding stellar mass functions agree in shape as well as they
do.

\subsection{Other Observational Factors}
\label{appendix:e}

In this section we investigate whether other observational factors
such as Galactic latitude, extinction, seeing and sky brightness 
introduce systematic effects in our results.  We start with our sample
of isolated primaries with $11.1>\log M_\star/M_\odot>10.8$ and
subdivide it into pairs of disjoint approximately equal subsamples
based on Galactic latitude, $r$-band extinction, seeing, sky
brightness at the time of observation of each object and redshift. In
each case we estimate satellite stellar mass functions for the two
subsamples separately and then compare the results.

In the top two panels of figure~\ref{fig:test6}, we compare stellar
mass functions for primaries with Galactic latitude $b~>60^\circ$ and
with $b<60^\circ$ as indicated by the legends. (Numbers in brackets
give the mean Galactic latitudes of the two subsamples.)  The two
panels differ in that at top left the background was estimated
globally using the survey as a whole and then applied in the same way
to primaries in each subsample, while at top right the background for
each subsample was estimated separately from its own half of the SDSS
footprint. A clear difference at low mass is seen between the two
stellar mass functions in the top left panel, but is almost entirely
eliminated in the top right panel. This reflects a difference in mean
surface density of about 3.5\% between the two regions for galaxies
with (extinction-corrected) apparent magnitude in the range
21~$>r>$~20, with the low Galactic latitude region having lower mean
surface density.

It is unclear what causes this problem. Clearly there are more stars
and greater dust optical depths at low Galactic latitude. However,
misclassification of stars as galaxies would {\it increase} the
apparent galaxy density at low latitudes and there is no obvious
reason why galaxies would be more likely to be misclassified as stars
at low latitude. Moreover, unless the COSMOS field analyzed above is
atypical, star-galaxy separation in the SDSS is too accurate for
errors to produce such a change in mean ``galaxy'' density. Effects
due to inappropriate correction for reddening also seem to be excluded
by the explicit extinction test we discuss next.

The central and lower left panels of figure~\ref{fig:test6} show that
our satellite stellar mass function results do not differ
significantly for regions observed through greater extinction, in
poorer seeing, or against a brighter sky. In all three cases the two
halves of the sample give results which appear in good agreement. Only
for the lowest mass objects where te uncertainties become substantial
is there any hint of a disagreement between the red and the black
curves, and even here no consistent pattern emerges. Thus none of the
observational factors discussed here appears to significantly bias
our results.

\subsection{A bias with redshift?}
\label{appendix:f}

In the bottom right panel of figure~\ref{fig:test6}, we split the
primary sample according to redshift. There is a significant
difference in satellite stellar mass function between two resulting
subsamples, corresponding to a difference in amplitude by about a
factor of 1.5 with the nearer primaries having more satellites. The
mean redshifts of the two subsamples are 0.0379 and 0.0713, and we
have checked in the G11 simulation that negligible evolution in
satellite abundance is expected over such a small redshift
interval. Further, the distributions of stellar mass and intrinsic
colour are indistinguishable for the two sets of primaries, so a
significant ``cosmic variance'' difference between the samples appears
excluded. The discrepancy cannot be a consequence of our isolation
criteria, since it persists for stellar mass functions like those of
figure~\ref{fig:background} which count companions around
spectroscopic galaxies without imposed isolation criteria.  An
identical analysis of isolated galaxies from the mock SDSS survey we
constructed from the MS light-cone finds no difference between 
the satellite stellar mass functions when split by redshift in the same way. 
Thus the discrepancy must reflect a property of the real data which is not
included in the simulation and in construction of the mock catalogue.

\begin{figure}
\epsfig{figure=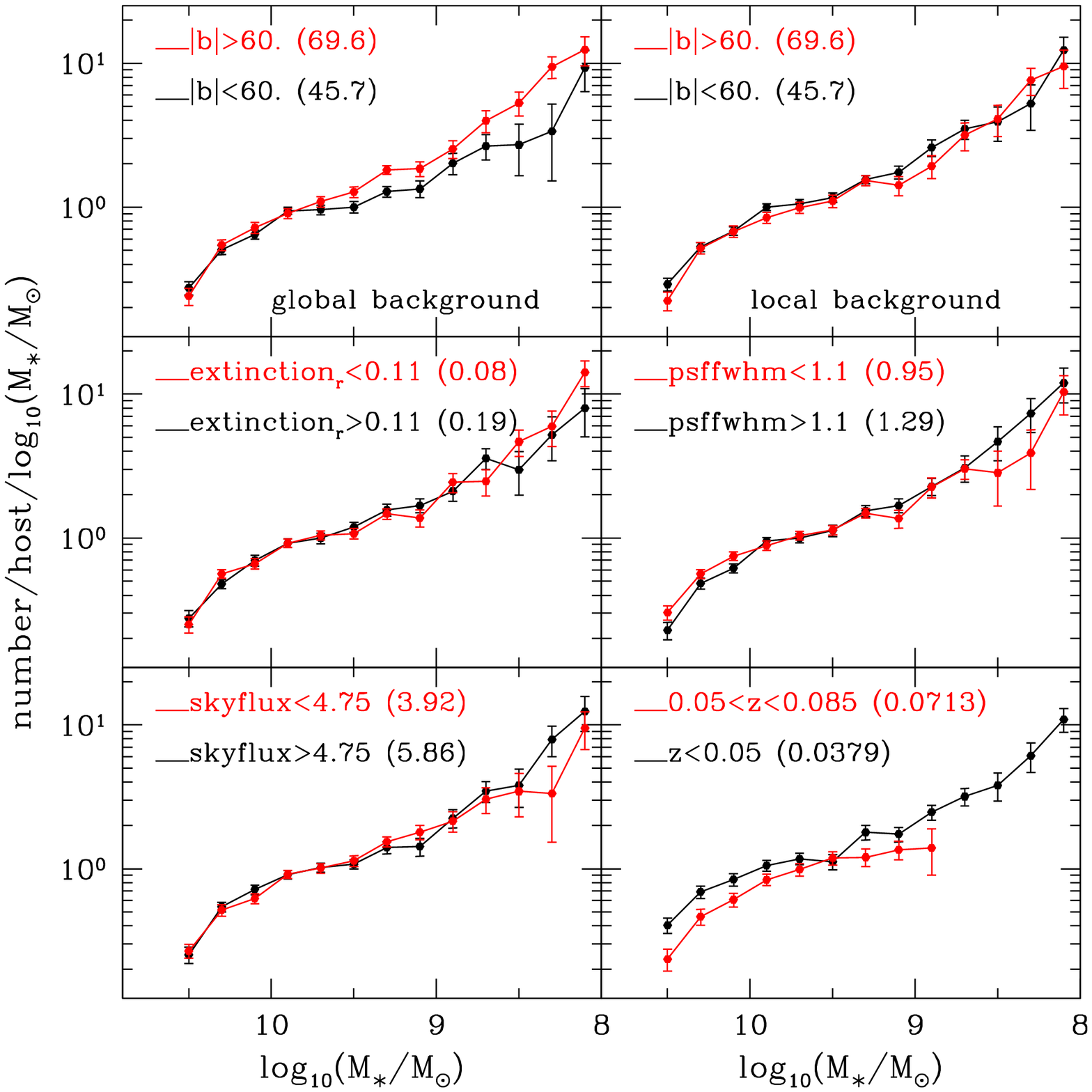,width=0.45\textwidth}
\caption{Stellar mass functions for isolated primaries with $11.1>\log
  M_\star/M_\odot>10.8$, split into subsamples of approximately equal
  size based on Galactic latitude, $r$-band extinction, seeing, sky
  brightness and redshift, as indicated by the legends. The top two
  panels show the same subsamples split according to Galactic latitude
  but corrected using a global background estimate on the left and a
  background estimated separately for each subsample on the
  right. Numbers in the parentheses following the legends in each panel
  give mean values of the separating quantity for each of the two
  subsamples.}
\label{fig:test6}
\end{figure}

In order to search for clues to the origin of this bias, we compare
the distributions of rest-frame colour and size in these two redshift
bins for satellites with $9.5>\log M_\star/M_\odot>8.75$ which are
projected within 300~kpc of isolated primaries with $11.1>\log
M_\star/M_\odot>10.8$.  Results are presented in
figure~\ref{fig:colorsize}. From the satellite colour distributions in
the left panel, it appears that the higher redshift subsample is
missing predominantly red satellites. The right panel indicates,
however, that these ``missing'' galaxies are similar in size to the
remaining satellites. The colours, angular sizes and surface
brightnesses of such objects are in the range where SDSS observations
detect and correctly classify galaxies without difficulty. Hence there
is no obvious reason why they should be missed if, in reality, they
are present.

Currently, we have no explanation for this apparent dependence on
redshift.  We have verified that it appears in very similar form for
other primary mass ranges. Given that the luminosity/stellar mass
functions presented in the main body of this paper are a combination
of counts around primaries at different redshifts, and that higher
redshift primaries contribute to the counts of brighter/more massive
satellites, this apparent decrease in satellite abundance with
redshift will result in measurement of a steeper slope for the full
sample as compared to either the black or the red curve in the bottom
right panel of figure~\ref{fig:test6}. The measured low-mass slopes
for these two subsamples are -1.444 (black) and -1.390 (red). The
corresponding low-mass slope for the stellar mass function in
figure~\ref{fig:MFall} is -1.455 (see table~\ref{tbl:slope}), which is
indeed steeper, as expected.  The difference is less than 5\%,
however, so this residual unidentified systematic should have only a
small effect on our conclusions.

\begin{figure}
\epsfig{figure=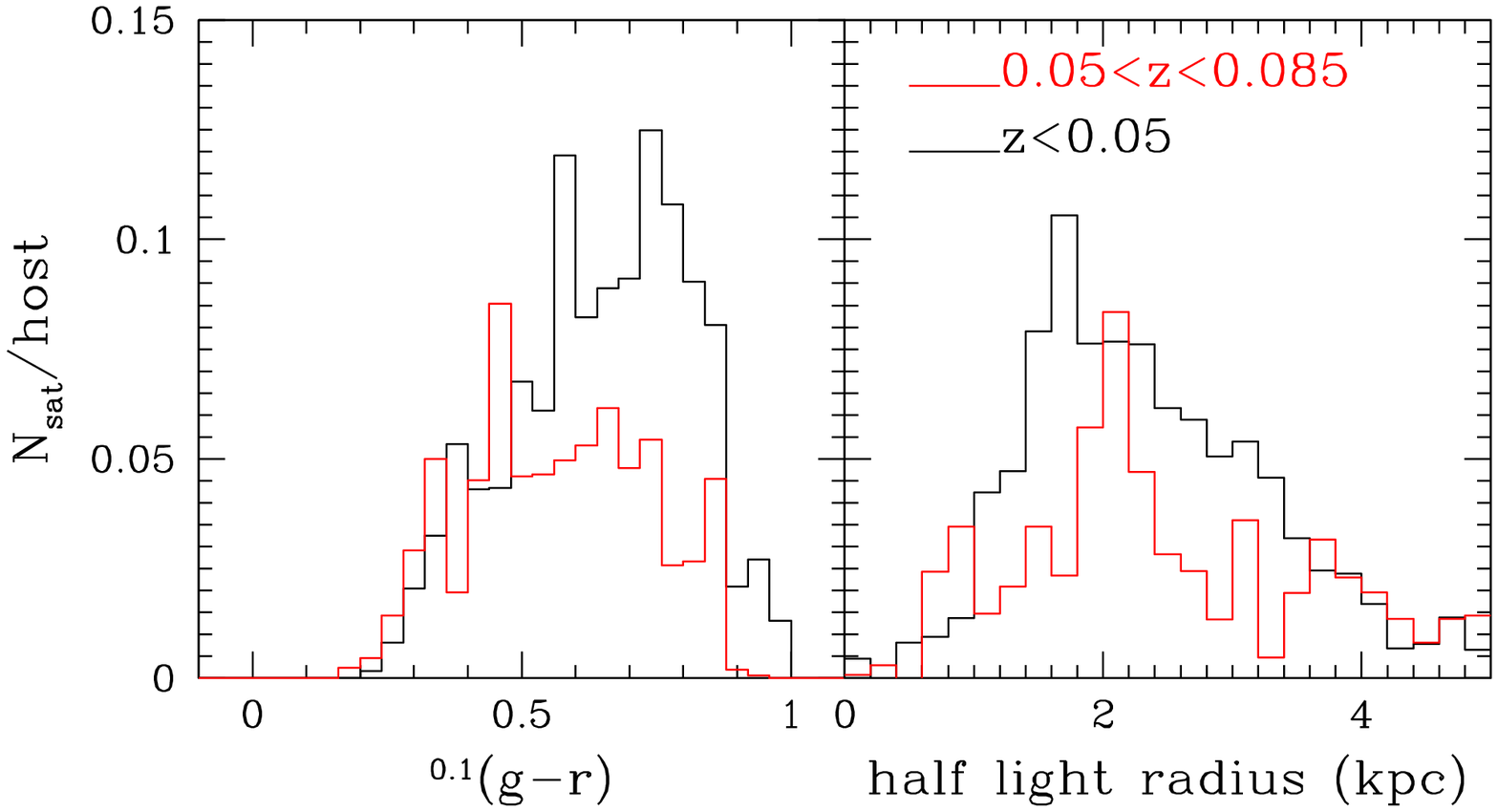,width=0.45\textwidth}
\caption{{\bf Left:} Rest-frame colour distributions, normalized per
  primary, for satellites with $9.5>\log M_\star/M_\odot>8.75$
  projected within 300~kpc of isolated primaries with
  $11.1>\log M_\star/M_\odot>10.8$ and split according to redshift
  at $z=0.05$.  {\bf Right:} Size distributions, also normalized per
  primary, for the same two sets of satellites.}
\label{fig:colorsize}
\end{figure}

\end{document}